\documentclass[notitlepage,twocolumn,letterpaper,natbib,aps,prd,amsmath,amsfonts,nofootinbib,preprintnumbers,superscriptaddress,secnumarabic]{revtex4-1}
\pdfoutput=1
\usepackage{amssymb,amsmath,latexsym,mathrsfs}
\usepackage{url}
\usepackage{enumitem}
\usepackage{graphicx}
\usepackage{booktabs}
\usepackage[usenames,dvipsnames]{color}
\usepackage[breaklinks,colorlinks,urlcolor=magenta,citecolor=magenta,linkcolor=magenta]{hyperref}
\usepackage{multirow}
\usepackage{float}
\usepackage{cases}
\usepackage{blindtext}
\usepackage{pifont}
\usepackage{hhline}
\usepackage[table,xcdraw]{xcolor}
\definecolor{linkcolor}{rgb}{0.0, 0.47, 0.75}
\definecolor{citecolor}{rgb}{1.0, 0.5, 0.0}
\definecolor{linkcolor}{rgb}{0.390625,0.5607843137,0.99609375}
\hypersetup{
  linkcolor  = linkcolor,
  citecolor  = linkcolor,
  urlcolor   = linkcolor,
  colorlinks = true
}

\begin{document}

\title{Peregrine: Sequential simulation-based inference for gravitational wave signals}

\author{Uddipta Bhardwaj}
\email{u.bhardwaj@uva.nl}
\affiliation{GRAPPA Institute, Anton Pannekoek Institute for Astronomy and Institute of High-Energy Physics,\\
University of Amsterdam, Science Park 904, 1098 XH Amsterdam, The Netherlands}

\author{James Alvey}
\email{j.b.g.alvey@uva.nl}
\affiliation{GRAPPA Institute, Institute for Theoretical Physics Amsterdam,\\
University of Amsterdam, Science Park 904, 1098 XH Amsterdam, The Netherlands}

\author{\\ Benjamin Kurt Miller}
\affiliation{GRAPPA Institute, Institute for Theoretical Physics Amsterdam,\\
University of Amsterdam, Science Park 904, 1098 XH Amsterdam, The Netherlands}
\affiliation{AMLab, Institute for Informatics, LAB42, Science Park 900, 1098 XH Amsterdam}
\affiliation{AI4Science, Institute for Informatics, LAB42, Science Park 900, 1098 XH Amsterdam}

\author{Samaya Nissanke}
\affiliation{GRAPPA Institute, Anton Pannekoek Institute for Astronomy and Institute of High-Energy Physics,\\
University of Amsterdam, Science Park 904, 1098 XH Amsterdam, The Netherlands}
\affiliation{Nikhef, Science Park 105, 1098 XG Amsterdam, The Netherlands}

\author{Christoph Weniger}
\affiliation{GRAPPA Institute, Institute for Theoretical Physics Amsterdam,\\
University of Amsterdam, Science Park 904, 1098 XH Amsterdam, The Netherlands}

\preprint{}


\begin{abstract}

\noindent The current and upcoming generations of gravitational wave experiments represent an exciting step forward in terms of detector sensitivity and performance. For example, key upgrades at the LIGO, Virgo and KAGRA facilities will see the next observing run (O4) probe a spatial volume around four times larger than the previous run (O3), and design implementations for \textit{e.g.} the Einstein Telescope, Cosmic Explorer and LISA experiments are taking shape to explore a wider frequency range and probe cosmic distances. In this context, however, a number of very real data analysis problems face the gravitational wave community. For example, it will be critical to develop tools and strategies to analyse (amongst other scenarios) signals that arrive coincidentally in detectors, longer signals that are in the presence of non-stationary noise or other shorter transients, as well as noisy, potentially correlated, coherent stochastic backgrounds. With these challenges in mind, we develop \texttt{peregrine}, a new sequential simulation-based inference approach designed to study broad classes of gravitational wave signal. In this work, we describe the method and implementation, before demonstrating its accuracy and robustness through direct comparison with established likelihood-based methods. Specifically, we show that we are able to fully reconstruct the posterior distributions for every parameter of a spinning, precessing compact binary coalescence using one of the most physically detailed and computationally expensive waveform approximants (\texttt{SEOBNRv4PHM}). Crucially, we are able to do this using \textbf{only 2\% of the waveform evaluations} that are required in \emph{e.g.} nested sampling approaches. Finally, we provide some outlook as to how this level of simulation efficiency and flexibility in the statistical analysis could allow \texttt{peregrine} to tackle these current and future gravitational wave data analysis problems. 

\vspace*{5pt} \noindent \textbf{\texttt{GitHub}}: The \texttt{peregrine} analysis and inference library will be made available \href{https://github.com/PEREGRINE-GW/peregrine}{here}.
\end{abstract}

\maketitle
\hypersetup{
  linkcolor  = linkcolor,
  citecolor  = linkcolor,
  urlcolor   = linkcolor
}

\renewcommand*{\thefootnote}{\arabic{footnote}}
\setcounter{footnote}{0}

\section{Introduction}\label{sec:intro}

\begin{figure*}[t]
    \centering
    \includegraphics[width=\linewidth,trim={0.1cm 0.1cm 0.1cm 0.1cm},clip]{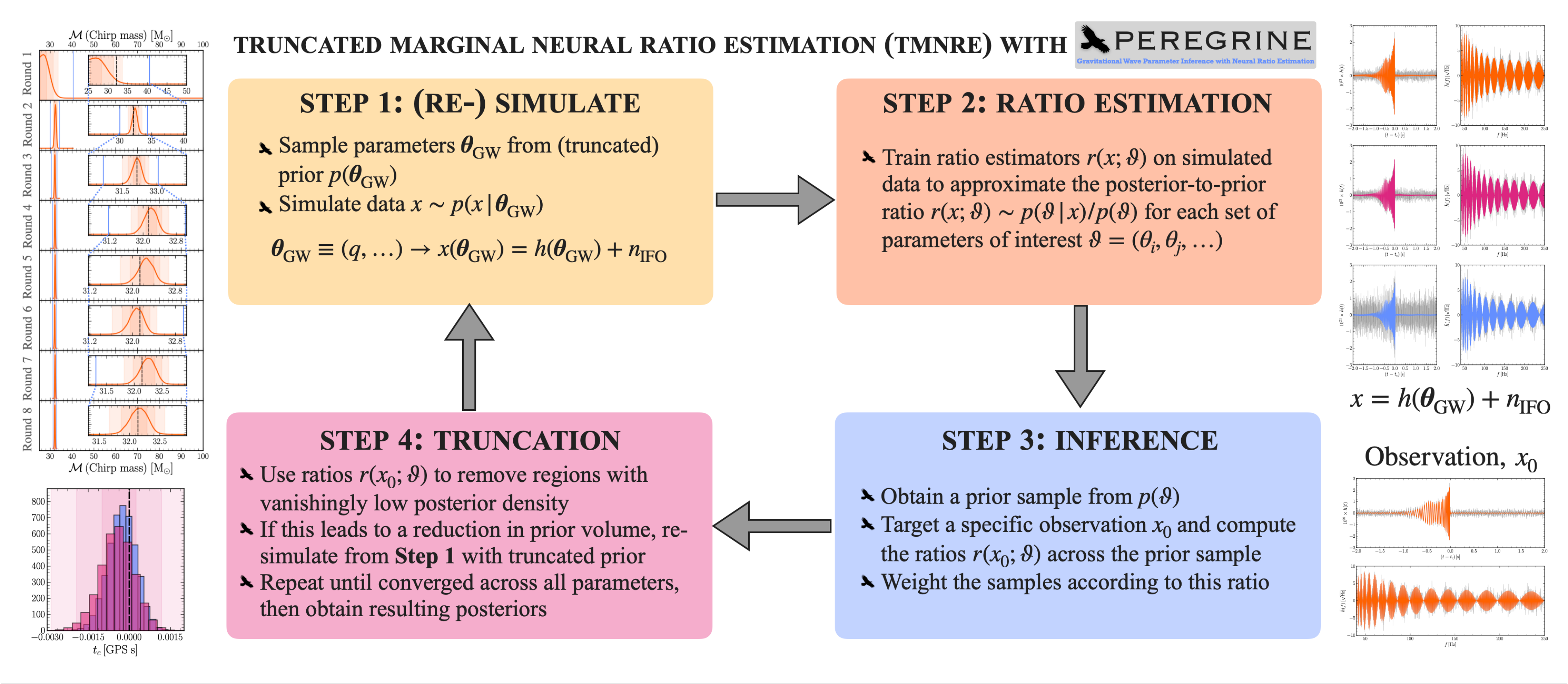}
    \caption{A schematic illustration of the data analysis method developed in this paper. We use Truncated Marginal Neural Ratio Estimation (TMNRE) to perform full parameter estimation on gravitational wave signals. The algorithm is implemented in our new code \texttt{peregrine}.}
    \label{fig:summary}
\end{figure*}

\noindent \emph{Observational Status.} The gravitational wave (GW) sky is now much louder than it was after the first detection of a gravitational wave signal from the merger of two black holes in 2015~\cite{LIGOScientific:2016aoc}. Indeed, source catalogs are now complete and varied enough to be used for studying questions in gravitational theory~\cite{LIGOScientific:2020tif}, cosmology~\cite{LIGOScientific:2019zcs,Nissanke:2009kt,Nissanke:2011ax}, and even the formation and internal properties of black holes and neutron stars~\cite{LIGOScientific:2020kqk, LIGOScientific:2018cki}. To give some current context, the LIGO-Virgo collaboration (LVKC) has now reported on 90 compact binary coalescence events\footnote{Additional catalogs such as the deep extended catalog~\cite{LIGOScientific:2021usb}, the 3-OGC catalog~\cite{Nitz:2021uxj} and others~\cite{Venumadhav:2019aa,Venumadhav:2020aa} report several other sources with different analysis pipelines or higher false alarm rate thresholds.}
that are believed to be of astrophysical origin~\cite{LIGOScientific:2020ibl,LIGOScientific:2021djp}.

Along with more efficient search pipelines~\cite{Allen:2005fk,Isi:2018vst,Venumadhav:2019tad}, one of the major reasons for the increase in size of the gravitational wave catalog is the large increase in search volume probed by the detectors. For example, in the previous LVKC observing run (O3), the projected comoving search volume for binary black hole (BBH) mergers was $V_\mathrm{BBH} = 3.4 \, \times 10^8 \, \mathrm{Mpc}^3$~\cite{KAGRA:2013rdx}. In comparison, however, the next observing run (O4) that is planned for May 2023 has a projected search volume of around $V_\mathrm{BBH} = 1.5 \, \mathrm{Gpc}^3$~\cite{KAGRA:2013rdx,Petrov:2021bqm}, representing over a 400\% increase in volume and the corresponding expected event rate\footnote{The equivalent volumes for binary neutron star (BNS) coalescence events are $V_\mathrm{BNS} = 3.3 \, \times 10^6 \, \mathrm{Mpc}^3$ and $V_\mathrm{BNS} = 3.4 \, \times 10^8 \, \mathrm{Mpc}^3$ for O3 and O4 respectively~\cite{KAGRA:2013rdx,Petrov:2021bqm}.}. This observing run is predicted to have the advanced LIGO (aLIGO) detectors running close to their design sensitivities (O5), and therefore, represents a key benchmark for the current generation of gravitational wave observatories. Such an improvement in detector sensitivity and corresponding event rate induces a new era of computational challenges when it comes to gravitational wave data analysis.

\emph{Gravitational Wave Data Analysis.} In general, the data analysis pipeline for gravitational waves splits into two parts: detection and parameter inference. Whilst studying the optimality of search pipelines~\cite{Allen:2005fk,Isi:2018vst}, especially for new classes of signal, is an interesting and important problem, in this work we will focus on the follow up step of high precision parameter inference. The traditional (and current) approach to gravitational wave parameter inference follows a Bayesian framework where the goal is to infer either the full ($N$-dimensional) joint posterior over all parameters or marginal posterior distributions over a (sub-)set of relevant intrinsic (\emph{e.g.} source properties) and extrinsic parameters (\emph{e.g.} the source-detector geometry)~\cite{Cutler:1994ys,Thrane:2019aaa}. In addition, one must make some informed choice (ideally using some astrophysically motivated range, or making a maximally uninformative choice in the absence of this) of the prior distributions for the parameters of interest. For current state-of-the-art waveform approximants/generators, which are now capable of describing complex scenarios such as fully precessing and spinning compact binary inspirals~\cite{Ossokine:2020kjp,Khan:2015jqa,Pratten:2020fqn,Edwards:2023sak}, the number of relevant parameters is around $N_\mathrm{params} \sim 12$--$17$. At a purely practical level, this means that approaches such as profiling the gravitational wave data likelihood are computationally infeasible, and it is common therefore to use some form of stochastic sampling method to investigate the parameter space.

Traditional stochastic algorithms such as Markov Chain Monte Carlo (MCMC)~\cite{Mackay:2003aaa,Foreman-Mackey:2012any} and nested sampling~\cite{Skilling:2006gxv,Handley:2015aaa,Ashton:2022grj,Speagle:2020aaa} are both implemented within the standard open source analysis software such as LALInference~\cite{Veitch:2014wba}, PyCBC~\cite{Biwer:2018osg}, and bilby~\cite{Ashton:2018jfp,Romero-Shaw:2020owr,Ashton:2021anp,Williams:2023ppp}. Both MCMC and nested sampling algorithms function by generating samples from the full joint posterior (although they achieve this in different ways), and then typically marginalise to obtain posteriors over individual, or pairs of parameters. The computational cost in these cases comes about since these algorithms carry out numerous likelihood evaluations for each individual sample point. Indeed, a general rule of thumb for current gravitational wave parameter inference is that generating around $10^4$ independent posterior samples requires at least $10^6$ likelihood evaluations\footnote{This is especially true for spinning and precessing compact binary inspirals.} (and corresponding forward evaluations of the waveform generation model)~\cite{Ashton:2021anp,Williams:2023ppp}. Depending on the signal duration and complexity of the waveform approximant, this can result in inference times of hours to weeks for a single events, see e.g.~\cite{Ashton:2021anp,Williams:2023ppp}. This is the context in which concrete data analysis challenges will arise as, amongst other things, \emph{(i)} the detection rate increases significantly~\cite{KAGRA:2013rdx}, \emph{(ii)} more realistic waveform approximants that incorporate \emph{e.g.} higher order modes become more expensive to evaluate~\cite{Iacovelli:2022bbs,Pratten:2020fqn}, and \emph{(iii)} future gravitational wave detections involving multiple overlapping waveforms~\cite{Pizzati:2021apa,Antonelli:2021vwg,Samajdar:2021egv,Janquart:2022nyz,Langendorff:2022fzq}, non-stationary noise distributions, or some stochastic background, see \emph{e.g.}~\cite{Caprini:2018mtu,Christensen:2018iqi,LIGOScientific:2019vic}, will render explicit likelihood evaluations increasingly complicated and costly. 

\emph{Simulation-based Inference.} Recently, there have been significant advances in the field of simulation-based inference (SBI)~\cite{Cranmer:2019eaq,Brehmer:2020cvb,Lueckmann:2021aaa} (sometimes also known as implicit likelihood inference), partly as a result of rapid developments in machine learning, but also as a response to emerging data analysis challenges such as those described above. The field of simulation-based inference is now wide and varied, with numerous approaches available as fully implemented open-source software~\cite{Miller:2021hys,Miller:2022shs,Tejero-cantero:2020aaa,Alsing:2019xrx}. It has been shown to be successful in a number of different physics settings, such as CMB analyses~\cite{Cole:2021gwr}, strong lensing image analysis~\cite{Montel:2022fhv}, point source searches~\cite{AnauMontel:2022ppb}, and field-level cosmology~\cite{Makinen:2021nly}, along with other examples~\cite{Dimitriou:2022cvc,Gagnon-Hartman:2023soa,Delaunoy:2020zcu,Karchev:2022xyn,Lin:2022ayr}. Broadly, the key advantages of this set of methods are \emph{(i)} they can be highly simulation-efficient compared to traditional methods, and \emph{(ii)} they are ``simulation-based" in the sense that they do not require an explicit likelihood to be written down, only a realistic forward simulator to be provided. Combined, this opens up the exciting possibility of using higher fidelity forward models to take full advantage of even the most complex data. In the context of gravitational waves, a particular version of simulation-based inference known as neural posterior estimation (NPE) has already been successfully applied to perform fully amortised inference\footnote{Amortized inference refers to training a neural network to approximate the posterior distribution over model parameters which allows for fast inference on any input data.} on compact binary mergers~\cite{Dax:2021myb,Wildberger:2022agw}. This leads to impressive inference performance across \emph{e.g.} the LVKC catalog, where after the complex training process (requiring millions of waveform evaluations) has been carried out, almost real-time inference can be done~\cite{Dax:2021tsq}. It is worth noting in this regard that the focus of our analysis pipeline is different. We look to perform analysis on individual gravitational wave signals, where this sort of global amortisation becomes infeasible. The prototypical example of this would be the overlapping waveform scenario~\cite{Pizzati:2021apa,Antonelli:2021vwg,Samajdar:2021egv,Langendorff:2022fzq,Janquart:2022nyz}, where the number of waveform calls required to perform reliable amortised inference would likely be at least an order of magnitude larger, if not more. In this current work, we however focus on single BBH mergers to develop our method because this is where clear, robust quantitative comparisons can be performed to validate our approach.

\emph{Key Contributions.} In this work, we develop a new data analysis method that is applicable to wide classes of gravitational wave signals. In particular, we implement a particular, sequential simulation-based inference algorithm known as Truncated Marginal Neural Ratio Estimation (TMNRE)~\cite{Miller:2022shs} based on the \texttt{swyft} software~\cite{Miller:2021hys}. We demonstrate the applicability, accuracy and robustness of this approach by studying two case studies -- a highly spinning, precessing binary black hole system at distances of $200 \, \mathrm{Mpc}$ (extremely high signal-to-noise ratio (SNR) of $\sim 100$ to act as a stringent test of the precision we can achieve) and $900 \, \mathrm{Mpc}$ (low SNR of $\sim 20$ to test our ability to reliably distinguish between signal and noise). Using a state-of-the-art waveform approximant \texttt{SEOBNRv4PHM}~\cite{Ossokine:2020kjp} (although in practice, any waveform approximant for \textit{e.g.} \texttt{IMRPhenomXPHM}~\cite{Pratten:2020ceb}, \texttt{IMRPhenomPV2} or \texttt{IMRPhenomD}~\cite{Hannam:2013oca,Khan:2015jqa}, can easily be used), we demonstrate that we can fully reproduce the posterior distributions generated by current likelihood-based methods. Importantly, we show that we can achieve this agreement \emph{with only 2\% of the waveform evaluations} that are needed in standard likelihood-based approaches. As far as the computational efficiency of the method is concerned, we provide a discussion of the design choices relevant for our TMNRE-based algorithm in Sec.~\ref{sec:sbi}, in particular focusing on the details of the network training and optimisation. We highlight the role that truncation plays in achieving efficient inference in Fig.~\ref{fig:truncation}, where we use an active learning approach to lower the training data variance with respect to the observation of interest. We also discuss the robustness tests that can be carried out to validate our results independent of any other method. In addition, we will release our implementation, which we call \texttt{peregrine}, as open-source, extendable analysis software available to the community.

\emph{Structure of the Work.} The remainder of the work is structured as follows: we begin in Sec.~\ref{sec:sbi} by briefly reviewing simulation-based inference approaches before giving a detailed explanation of TMNRE and its application to gravitational wave analysis. We apply this method in Sec.~\ref{sec:results} to the two case studies detailed above. Finally, in Sec.~\ref{sec:conclusion}, we provide some discussion and outline our conclusions regarding the application of our pipeline to current and future gravitational wave data analysis challenges.

\section{Simulation-based Inference for Gravitational Waves}\label{sec:sbi}

\noindent In this section, we will give a very brief overview of the general class of simulation-based inference methods, before focusing our attention on the particular implementation we use in this work. Finally, we will discuss some specific design choices that are relevant to the analysis of gravitational waves.

\subsection{Overview of Simulation-based Inference}\label{sec:sbi_overview}

\noindent Over the last three to four years, there has been a significant increase in the usage and development of so-called simulated-based inference approaches to data analysis~\cite{Cranmer:2019eaq,Brehmer:2020cvb}\footnote{As discussed in the introduction, this is arguably for two main reasons, the first is that the huge computational improvements in machine learning (ML) methods allow for density estimators, classifiers and neural networks to be efficiently trained on very high dimensional data. The second reason is then that utilising these developments opens up the possibility to tackle new data analysis challenges~\cite{Pizzati:2021apa,MockLISADataChallengeTaskForce:2009wir,Schafer:2022dxv}.}. Whilst there are now multiple options in terms of particular implementations and algorithms, every approach has a common setup. In particular, the main question driving all SBI algorithms is: \emph{can we still do robust Bayesian inference if we are only given some generative model?}~\cite{Cranmer:2019eaq}. In other words, given some forward model $p(x, \theta)$ that takes some underlying set of parameters $\theta$ and produces some simulated data $x$, can we construct meaningful posteriors $p(\theta | x)$ for all, or some subset, of the parameters $\theta$. In practice, how this is done is by studying Bayes' theorem applied to the posterior,
\begin{equation}\label{eq:sbi_bayes}
    p(\theta | x) = \frac{p(x | \theta) p(\theta)}{p(x)},
\end{equation}
where $p(\theta | x)$ is the posterior of $\theta$ given some observed or simulated data $x$, $p(x | \theta)$ is the likelihood of a given set of data $x$ given some input parameters $\theta$, $p(\theta)$ is the (Bayesian) prior over the parameters $\theta$, and $p(x)$ is the Bayesian evidence. The key functionality of simulation-based inference is to realise that having a forward generative model $p(x, \theta) = p(x | \theta) \, p(\theta)$ is equivalent to being able to sample from the (simulated) likelihood. This is the origin of the terms ``likelihood-free" or ``implicit likelihood" inference~\cite{Cranmer:2019eaq}.

At this point, the various SBI methods diverge somewhat in terms of how they use this capability of sampling from the likelihood to construct posterior densities or posterior samples. Broadly, they can be categorised in terms of how they split up Bayes' theorem in Eq.~\eqref{eq:sbi_bayes}. There are three main ways to do this given a prior $p(\theta)$ (which is explicitly known/chosen) and a generative model $p(x, \theta) = p(x | \theta) \, p(\theta)$.
\begin{itemize}[leftmargin=*]
    \item \emph{Neural Posterior Estimation (NPE).} The first class of methods attempts to directly estimate the posterior density $p(\theta | x)$, typically through the use of some flexible density estimator such as a normalising flow~\cite{Papamakarios:2016aaa,Dax:2021myb,Tejero-cantero:2020aaa,Zeghal:2022aaa}. This has been used in a number of contexts, including for the amortised analysis of gravitational waves~\cite{Dax:2021tsq,Wildberger:2022agw,Dax:2022pxd}.
    \item \emph{Neural Likelihood Estimation (NLE).} In the second case, given some training data generated from the forward model, one attempts to create a well-controlled proxy for the simulated data likelihood $p(x | \theta)$~\cite{Papamakarios:2016aaa,Alsing:2019xrx,Lin:2022ayr}. This can then be used directly in traditional algorithms such as MCMC to generate posterior samples.
    \item \emph{\textbf{Neural Ratio Estimation (NRE).}} The final class of methods takes a different approach and estimates the likelihood-to-evidence ratio $p(x | \theta)/p(x)$~\cite{Miller:2021hys,Rozet:2022aaa,Montel:2022fhv,Karchev:2022xyn,AnauMontel:2022ppb,Gagnon-Hartman:2023soa,Delaunoy:2022aaa,Cole:2021gwr,Dimitriou:2022cvc,Miller:2022haf,Hermans:2019aaa,Durkan:2020aaa}. In this work, we will be using a particular version of this known as Truncated Marginal Neural Ratio Estimation (TMNRE)~\cite{Miller:2022shs}. Ratio estimation ultimately works by translating the Bayesian inference problem into a binary classification task on joint and marginal samples. We provide more details of this mapping below in the specific context of gravitational waves.
\end{itemize}

\subsection{The TMNRE Algorithm}\label{sbi:tmnre}

\noindent In this work, we use a particular simulation-based inference algorithm known as Truncated Marginal Neural Ratio Estimation (TMNRE)~\cite{Miller:2022shs}, implemented within the \texttt{swyft} software package~\cite{Miller:2021hys}. There are two key features of the method that we look to take advantage of for the highly simulation efficient analysis of gravitational wave signals. For reference, the overall structure of the algorithm is described in Fig.~\ref{fig:summary}.
\begin{itemize}[leftmargin=*]
    \item \emph{The ``T".} TMNRE is a sequential method in the sense that the analysis is performed in rounds (typically around 5 to 10). At the end of each round, estimates of the posterior are generated for a \emph{specific}, targeted observation $x_0$. The priors are then truncated before generating training data for the next round, leading to a significant (empirically, of the order of 50\% to 90\%) reduction in simulation budget~\cite{Cole:2021gwr}. This, of course, has the consequence that we are generally not developing an amortised method, but rather one that can be applied to individual observations.
    \item \emph{The ``M".} The marginal aspect of TMNRE relates to the fact that the algorithm estimates \emph{marginal posteriors}. In other words, instead of estimating the full ($N$-dimensional) joint posterior $p(\boldsymbol{\theta} | x)$ for all parameters $\boldsymbol{\theta} \equiv (\theta_1, \ldots, \theta_N)$, and then marginalising, we directly estimate the marginal posterior $p(\theta_i, \theta_j,\cdots | x)$ for some parameter (or set of parameters) of interest $\left\{\theta_i, \theta_j, \ldots \right\}$. This again has the advantage that it typically reduces the simulation budget by a large fraction and results in much simpler training procedures because the data-parameter manifolds for the low dimensional marginals are much less complex.
\end{itemize}
More technically, TMNRE sets up the estimation of the ratio $r(x; \vartheta = (\theta_i, \theta_j, \ldots)) = p(x | \vartheta) / p(x)$, where $\vartheta$ represents a single parameter or some subset of parameters, as a binary classification problem through the following observation about Bayes' theorem,
\begin{equation}\label{eq:sbi_ratio}
    r(x; \vartheta) = \frac{p(x | \vartheta)}{p(x)} = \frac{p(\vartheta | x)}{p(\vartheta)} = \frac{p(x, \vartheta)}{p(x)p(\vartheta)}.
\end{equation}
In other words, we can view this likelihood-to-evidence ratio in multiple ways, \emph{(i)} as the posterior-to-prior ratio, which will ultimately allow us to generate posterior samples by generating samples from the prior $p(\vartheta)$ and weighting them by the ratio $r(x; \vartheta)$, and \emph{(ii)} as the ratio between the joint probability density $p(x, \vartheta)$ and marginal density $p(x)p(\vartheta)$. 

It is the last form that allows us to carry out a binary classification task. In particular, suppose we have $N$ samples from our simulator (or joint distribution $p(x, \vartheta) = p(x | \vartheta) p(\vartheta)$) $\left\{(\vartheta = (\theta_i, \theta_j, \ldots), x)_i\right\}_{k = 1...N}$, then we can also construct samples from the marginal distribution $p(x)p(\vartheta)$ by randomly shuffling the pair components. The method then attempts to optimise a classifier $d_\phi(x, \vartheta)$ with some trainable parameters $\phi$ to output a class $d_\phi(x, \vartheta) = 0$ (say) when $(x, \vartheta)$ is drawn marginally, and $d_\phi(x, \vartheta) = 1$ when $(x, \vartheta)$ is drawn jointly.

\begin{figure*}[t]
    \centering
    \includegraphics[width=\linewidth,trim={0.1cm 0.1cm 0.1cm 0.1cm},clip]{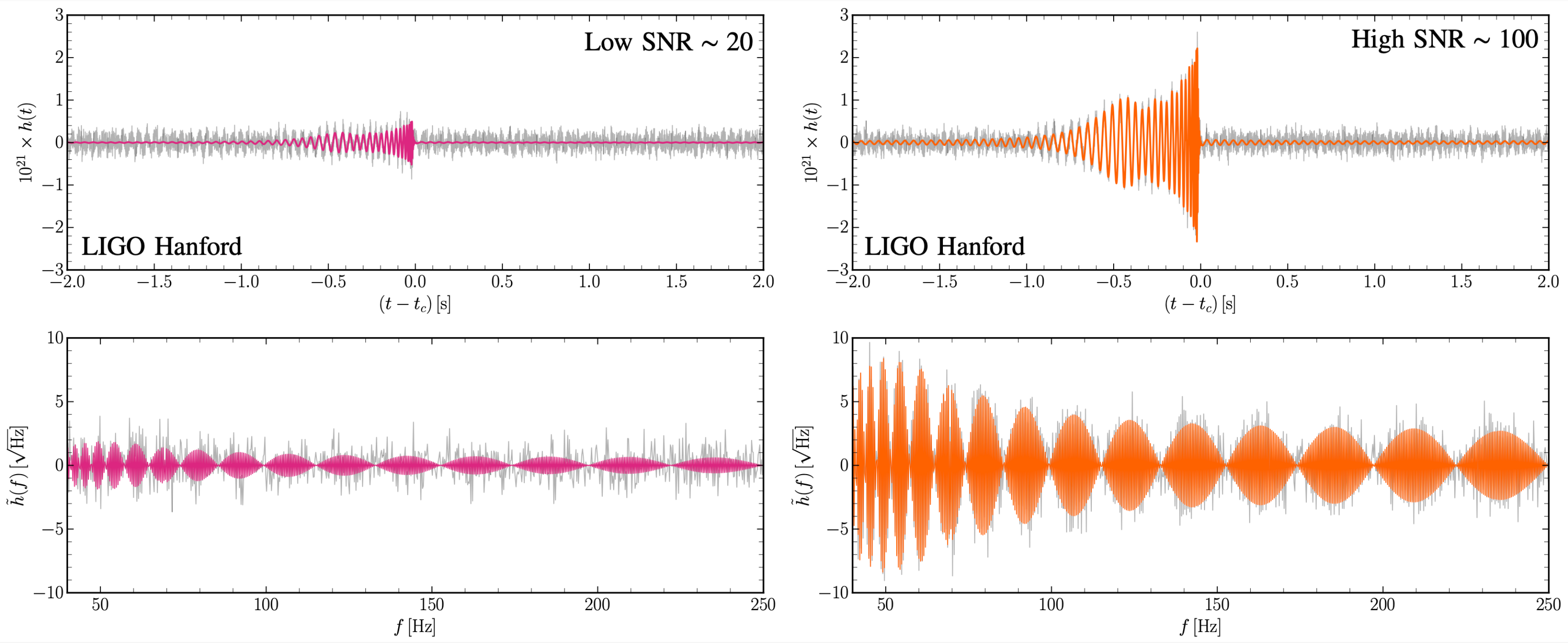}
    \caption{Two example observations that will be used in the case studies below. The parameter values for the observations can be found in the inference results below (see Figs.~\ref{fig:results_lowsn} and~\ref{fig:results_highsn}) \emph{Left:} A (typical) low signal-to-noise ratio (SNR) signal as seen in the LIGO Hanford detector. The signal is shown in pink, while the full detector response including noise is shown in grey. \emph{Right:} A similar signal (shown in orange), but with a very high SNR, again in the Hanford detector. \emph{Upper and lower panels:} The upper panels illustrate the time-domain strain $h(t)$, whilst the lower panels show the corresponding frequency domain response $\tilde{h}(f)$, both of which are used in our analysis.}
    \label{fig:observation}
\end{figure*}

In terms of TMNRE, this is technically achieved by minimising the (binary cross-entropy) loss function~\cite{Miller:2021hys,Miller:2022haf,Delaunoy:2022aaa},
\begin{multline}\label{eq:sbi_loss}
\mathcal{L}[f_\phi] = -\int{\mathrm{d}x\mathrm{d}\vartheta}\, [p(x, \vartheta) \ln \left(\sigma(f_\phi(x, \vartheta))\right) \\ + p(x)p(\vartheta) \ln \left(1 - \sigma(f_\phi(x, \vartheta))\right)],
\end{multline}
where $\sigma(x) = [1 + \exp(-x)]^{-1}$ is the sigmoid function and $d_\phi(x, \vartheta) = \sigma(f_\phi(x, \vartheta))$ in the notation above. Crucially, one of the key motivations for TMNRE is that one can actually minimise this loss analytically to obtain the optimal classifier $f_\phi^\star(x, \vartheta) \equiv \ln r(x; \vartheta)$~\cite{Miller:2021hys}. This means that by successfully optimising this loss function, \emph{we can directly obtain full information about the likelihood-to-evidence ratio or posterior-to-prior ratio.}

In practice, this is where the ``neural" aspect of TMNRE comes into play. Since the data $x$ (and parameters $\vartheta$) can in principle be extremely high-dimensional, we require both very flexible classifiers $d_\phi(x, \vartheta)$ that can process complex data structures such as images or long time series, and also an efficient way to optimise the classifier parameters $\phi$ (here neural network weights). Recent advances in machine learning architectures, training, and hardware capabilities have now made this process extremely scalable, and we make use of them here, as implemented in the \texttt{swyft} package~\cite{Miller:2021hys,Miller:2022shs}.

It is important to understand how we can achieve marginal posterior estimates with this algorithm, since it will be important for solving future sampling challenges such as those associated to the analysis of multiple waveforms~\cite{Pizzati:2021apa,Antonelli:2021vwg,Janquart:2022nyz,Langendorff:2022fzq,Samajdar:2021egv}. In TMNRE, we implicitly fully marginalise over every parameter $\boldsymbol{\theta} = (\theta_1, \theta_2,\ldots)$ by varying them all in each simulation. When we train the classifier(s), however, we only show the network the small subset of parameters we are interested in, \emph{i.e.} $(x,\vartheta = (\theta_i, \theta_j,\ldots))$. When one analytically minimises the loss, one will find that you automatically obtain the properly marginalised posterior $p(\vartheta | x)$. In practice, in this work, we will only be interested in obtaining 1- and 2-dimensional marginal posteriors, \emph{i.e.} $\vartheta \equiv \theta_k$ for some $k$ or $\vartheta \equiv (\theta_i, \theta_j)$ for some pair $(i, j)$.

We can finally turn to the inference and truncation part of TMNRE. With a fully trained ratio estimator $\hat{r}(x; \vartheta) \simeq p(\vartheta | x) / p(\vartheta)$, we can now target a specific observation $x_0$ (coming from \emph{e.g.}~an LVKC data stream, or some mock test observation). By generating samples from the prior $p(\vartheta)$, we can obtain marginal posterior inference results by weighting these samples with the ratio estimator evaluated on the target observation $\hat{r}(x_0; \vartheta) \simeq p(\vartheta | x_0) / p(\vartheta)$. Inevitably, this will lead to regions of parameter space where the posterior density is high (covering the true values if the algorithm is run successfully) and regions where the density is vanishingly low. We use these regions of low density, specifically where the posterior estimate drops below some threshold $\hat{p}(\vartheta | x_0)/\underset{\vartheta}{\mathrm{max}}\,\hat{p}(\vartheta | x_0) < \epsilon$ to truncate our prior\footnote{This introduces a certain error into the marginal posterior estimation proportional to the integrated quantity $\int_{-\Gamma} \mathrm{d}\bar{\vartheta} \, p(\boldsymbol{\theta} | x_0)$, where $-\Gamma$ is the area outside the new proposal and $\bar{\vartheta}$ is the set of parameters $\boldsymbol{\theta}$ not including the parameter(s) being inferred. But, it is specifically in this region that the posterior density is vanishingly low, so the error induced is small and controlled by $\epsilon$.} before generating samples for the next round of training and inference\footnote{We typically take $\epsilon \sim 10^{-5}$ which approximately corresponds to a $\sim4.8\sigma$ exclusion if the posterior was a pure Gaussian. We checked that the inference results are not sensitive to this choice provided $\epsilon$ is small enough, only affecting the rate at which the truncation occurs}~\cite{Miller:2022shs}. After a number of rounds, once the sensitivity is at the level of statistical uncertainty, this truncation procedure will converge, and the algorithm terminates. This process is highlighted below for the specific case study presented in this work in Fig.~\ref{fig:truncation}.

To summarise, the TMNRE algorithm splits into four steps: \begin{enumerate}
    \item \textbf{Step 1:} We generate a batch of simulations $(\vartheta, x)_{i = 1...N}$ according to our current proposal prior $p(\vartheta)$ using our forward simulator.\footnote{Note that this step can be fully parallelised (and indeed is in our implementation), and therefore made arbitrarily fast given hardware access.}
    \item \textbf{Step 2:} We train classifier(s)\footnote{In this work, we simultaneously train one classifier for each gravitational wave parameter $\theta_i$, or pair $(\theta_i, \theta_j)$, so we end up with a set of classifiers.} $f_\phi(x, \vartheta)$ to tell the difference between joint and marginal samples for the parameter(s) of interest $\vartheta$. When optimised, this gives a direct estimate of the ratio 
    $r(x;\vartheta) = p(\vartheta | x) / p(\vartheta)$ for the parameters of interest $\vartheta = (\theta_i, \theta_j, \ldots)$.
    \item \textbf{Step 3:} We use the trained ratio estimator to perform inference on a specific target observation $x_0$ and obtain marginal ratio estimates $\hat{r}(x_0; \vartheta)$.
    \item \textbf{Step 4:} By evaluating the ratios on a set of prior samples, we can truncate the initial proposal distribution to exclude regions of vanishingly low posterior density such that the variance in the training data for the next round is significantly lower.
    \item We then go back to \textbf{Step 1} and repeat until the truncation regions and posterior estimates stabilise, after which we can perform the final inference on the observation $x_0$.
\end{enumerate}
This whole process is illustrated diagrammatically in Fig.~\ref{fig:summary}.

\subsection{Design Choices for Gravitational Waves}\label{sbi:gw}

\noindent There are a number of concrete design choices and implementation steps that we must make in order to apply the TMNRE algorithm to the analysis of gravitational waves. Specifically we must \emph{(i)} choose/build the forward simulator that generates the data $x$ given parameters $\theta$ and some noise model, \emph{(ii)} implement a neural network that can parse and process the output of the simulator for training the classifier, \emph{(iii)} choose the relevant prior distributions for our parameters, and \emph{(iv)} make choices for the relevant parameters in the TMNRE algorithm.

\vspace*{8pt}
\noindent \emph{Forward Simulator.} In terms of the generative model, in this work we choose to study gravitational wave signals from compact binary black hole mergers observed in LIGO and Virgo detectors. We analyse the signal as it appears in all three detectors simultaneously\footnote{Specifically, the Livingston (L1), Hanford (H1) and Virgo (V1) detectors, although any combination or subset can be analysed also.}, across both the time and frequency domains. To model the waveforms from the merger, we use the functionality in the open-source code bilby~\cite{Ashton:2018jfp,Romero-Shaw:2020owr,Ashton:2021anp}. This allows us to inject a gravitational wave signal $h(\boldsymbol{\theta}_\mathrm{GW})$ -- typically split into the $+$ and $\times$-polarisation strains and then projected -- into a noisy detector with coloured noise strain $n_\mathrm{IFO}$ generated from a power spectral density (PSD) $S_n(f)$. Here, $\boldsymbol{\theta}_\mathrm{GW} \equiv(q, \mathcal{M}, \ldots)$ is some set of parameters sampled from the prior $p(\boldsymbol{\theta}_\mathrm{GW})$ that describe the intrinsic properties of the source under consideration such as the mass ratio $q$, (redshifted) chirp mass of the system $\mathcal{M}$, as well as extrinsic properties including the angular position on the sky $(\alpha, \delta)$ and the luminosity distance $d_L$. All of these parameters are provided as input into the waveform approximant of choice\footnote{Here we use the \texttt{SEOBNRv4PHM} waveform approximant~\cite{Ossokine:2020kjp} so as to provide case studies on the most complex spinning and precessing BBH merger systems. Importantly, the analysis pipeline remains identical regardless of this choice, although the specific parameters in $\boldsymbol{\theta}_\mathrm{GW}$ may change.}. We provide a full list of physical parameters in Tab.~\ref{tab:sbi_params}, including their prior choices. The full forward model for a single detector is then defined by:
\begin{multline}
p(x, \boldsymbol{\theta}_\mathrm{GW}) = p(x = h + n_\mathrm{IFO} | h_{+,\times}, \boldsymbol{\theta}_\mathrm{GW}) \\ \times p(h_{+,\times} | \boldsymbol{\theta}_\mathrm{GW})  p(\boldsymbol{\theta}_\mathrm{GW}),
\end{multline}
where $h$ here is the projection of the $h_{+,\times}$ strains onto the detector frame according to the extrinsic parameters in $\boldsymbol{\theta}_\mathrm{GW}$. The full implementation of the forward model can be found in the \texttt{Simulator} class of the \texttt{peregrine} package. Two example observations $x_0 = h(\boldsymbol{\theta}_\mathrm{GW}) + n_\mathrm{IFO}$ which are the direct output of the simulator and will form the basis of our case studies below are shown in Fig.~\ref{fig:observation}.

\begin{table}[t]
\centering
\resizebox{\linewidth}{!}{
\begin{tabular}{@{}lll@{}}
\hline
\textbf{Parameter}                                      & \textbf{Prior Choice}         & \textbf{Injection} \\ \hline
Mass ratio, $q$                                         & $\mathrm{U}(0.125, 1)$        & 0.8858 \\
Chirp mass $\mathcal{M}\,\mathrm{[M}_\odot\mathrm{]}$           & $\mathrm{U}(25, 100)$         & 32.14 \\
Inclination angle $\theta_{jn}\,\mathrm{[rad]}$         & $\mathrm{sine}(0, \pi)^\dagger$ & 0.4432 \\
Polarisation angle $\psi\,\mathrm{[rad]}$               & $\mathrm{U}(0, \pi)$          & 1.100 \\
Phase $\phi_c\,\mathrm{[rad]}$                          & $\mathrm{U}(0, 2\pi)$         & 5.089 \\
Tilt angles $\theta_{1}, \theta_2 \, \mathrm{[rad]}$    & $\mathrm{sine}(0, \pi)^\dagger$ & 1.497, 1.102 \\
Dimensionless spins $a_1, a_2$                          & $\mathrm{U}(0.05, 1)$         & 0.9702, 0.8118 \\
Spin angles $\phi_{12}, \phi_{jl}\,\mathrm{[rad]}$      & $\mathrm{U}(0, 2\pi)$         & 6.220, 1.885 \\
Right ascension $\alpha\,\mathrm{[rad]}$                & $\mathrm{U}(0, 2\pi)$         & 5.556 \\
Declination $\delta\,\mathrm{[rad]}$                    & $\mathrm{cosine}(-\pi/2, \pi/2)^\dagger$ & 0.071 \\
Merger time $t_c\,\mathrm{[GPS\ s]}$                    & $\mathrm{U}(-0.1, 0.1)$       & 0.000 \\
Luminosity Distance $d_\mathrm{L}\,\mathrm{[Mpc]}$      & $\mathrm{U}_{\mathrm{vol.}}(100, 2000)^\star$ & 900.0 (\textbf{C1}), 200.0 (\textbf{C2}) \\ \hline
\end{tabular}}
\caption{Definitions and prior choices for all relevant gravitational wave parameters $\boldsymbol{\theta}_\mathrm{GW}$ in this work.\\
{\footnotesize ${}^\dagger$ Note that these are the default priors used in $\mathrm{bilby}$ analyses for BBH systems (subject to calibration ranges of the waveform approximants). \\
${}^\star$ Specifically, the luminosity distance prior is taken to be uniform in comoving volume in the \emph{source frame}.}}
\label{tab:sbi_params}
\end{table}

\vspace*{8pt}
\noindent \emph{Inference Network.} To carry out the ratio estimation step in the TMNRE algorithm (\textbf{Step 2} above), we need to construct a flexible enough parameterisation of the classifier $f_\phi(x, \vartheta)$. In principle (at least in some infinite training data limit), any sufficiently general parameterisation should be able to minimise the loss in Eq.~\eqref{eq:sbi_loss}. In practice, making sensible design choices about network architectures that are well matched to the data format (particularly the signal structure) leads to huge increases in performance, robustness and general applicability of the method.

In the context of gravitational waves, the signal is a complex time series or frequency domain strain -- \emph{e.g.} at a sampling frequency of $2048 \, \mathrm{Hz}$ a $4\,\mathrm{s}$ signal will consist of over $8000$ data points. Most of the information about the parameters $\boldsymbol{\theta}_\mathrm{GW}$ is encoded in various non-linear ways. As such, there is no good analytic method for fully processing noisy data streams, or deriving an optimal architecture. On the other hand, in the context of TMNRE, it is empirically the case that the precise network architecture is not so important provided it is \emph{(a)} sensibly adjusted to the data format -- \emph{e.g.} it would be non-optimal to first flatten an image, totally removing all spatial structure -- and \emph{(b)} deep enough (i.e., with enough network parameters or structure) to be able to flexibly fit the parameter and data manifold. This relative simplicity comes about as a result of the specific targeting both of single observations and marginal posteriors, significantly reducing the complexity of the fit we are implicitly trying to perform. An interesting comparison is with the NPE implementation of gravitational wave inference where a highly optimised and bespoke network architecture is required to perform fully amortised inference on the high dimensional joint posterior~\cite{Dax:2021tsq}.

Practically, we choose a network architecture that acknowledges a number of things about the physics behind gravitational wave signals:
\begin{itemize}[leftmargin=*]
    \item \emph{Signal segments.} Broadly, compact binary coalescence gravitational wave signals can be split into 3 components (both in the time and frequency domains) corresponding to different phases of their evolution: the inspiral, merger and ringdown phases. The various parameters in $\boldsymbol{\theta}_\mathrm{GW}$ affect these components differently, for example, the coalescence time $t_c$ defines the time of merger, typically associated to the peak of the amplitude in the time domain signal $h(t)$. We want to choose a network that does not treat each part of the waveform identically -- such as a convolutional structure that simply applies the same kernel to each part. Instead we construct a 1d-analogue of the well-known \texttt{unet} architecture~\cite{Ronneberger:2015aaa}, which is excellent for image (or signal here) segmentation and subsequent analysis on the various segments.
    \item \emph{Domain choice.} Information about gravitational wave parameters is encoded differently in the time and frequency domains, although they are related directly by a Fourier transform. At the end of the day, we are interested in extracting some optimal set of features from the signal that are correlated with various physical parameters. Naturally, the effect of some parameters is much more direct in one domain {\it vs.} the other. For example, it is very simple to see the effect of varying the time of coalescence on the time domain strain $h(t)$ -- it is simply a horizontal shift and is easy to pick up with any simple network. On the other hand, in the frequency domain, the waveform gets modified as $\tilde{h}(f; t_c) = \tilde{h}(f; 0) \exp(2 \pi i f t_c)$, which affects the full structure of the frequency domain strain and is much more complicated to extract. In practice we take advantage of the full flexibility of simulation-based methods and present \emph{both} the time and frequency domain strains to the network, analysing both of them independently to extract features before combining these summary statistics to perform ratio estimation.
    \item \emph{Multiple detectors.} The current generation of detectors is set up to provide complementary information about the same gravitational wave event at each of the relevant facilities (\emph{e.g.} LIGO Hanford and Virgo). We model the detector response in all (or a subset) of these locations and stack the time (and corresponding frequency) domain strains such that the same geocentric times (or corresponding frequency bins) are aligned. Again, this is taking advantage of the freedom to use any data representation that fully encodes our knowledge about the relevant timing delays in the detectors.
\end{itemize}
The above describes the rationale for our choice of inference network used to process our data. Its general structure consists of two parallel \texttt{unet} networks which act on the time and frequency domain strains independently, before compressing down to a smaller set of summary features in each case. These features, which importantly are \emph{automatically} optimised by the classifier during training are then concatenated and passed to a simple multi-layer perceptron that is the default implemented in \texttt{swyft} which performs the ratio estimation. More generally, one can understand the network as first performing some optimised data compression into a set of summary statistics\footnote{There are some clear analogues with approximate Bayesian computation here~\cite{Sunnaker:2013aaa}, although the summary statistics are trained to be optimal rather than hand crafted.} on which the ratio estimation is then performed with (a subset of) the physical parameters $\vartheta \subset \boldsymbol{\theta}_\mathrm{GW}$. The full details of the network architecture can be found both in Appendix~\ref{app:network} as well as in the \texttt{peregrine} implementation. In terms of generalisation, we expect this network to perform well on any compact merger signal, including \emph{e.g.}, overlapping waveforms, although in principle it can also be modified arbitrarily to search for other classes of gravitational waves without any other change in the data analysis pipeline.

\vspace*{8pt}
\noindent \emph{Prior Choices.} The aim of this work is to establish and develop the TMNRE algorithm in the context of gravitational wave analysis. As such, we choose priors $p(\boldsymbol{\theta}_\mathrm{GW})$ on the physical parameters $\boldsymbol{\theta}_\mathrm{GW} = (q, \mathcal{M}, \cdots)$ that represent the standard choices in the literature for analysing this type of BBH merger signal~\cite{Thrane:2019aaa,Veitch:2014wba,Ashton:2018jfp,Biwer:2018osg}. This allows for a direct comparison to inference results from traditional algorithms that are the focus of the next section. For reference, these can be found in \emph{e.g.} the bilby documentation~\cite{Ashton:2018jfp,Romero-Shaw:2020owr,Ashton:2021anp,Handley:2015aaa,Speagle:2020aaa}. All the parameter definitions and prior distributions relevant to this work can be found in Tab.~\ref{tab:sbi_params}.

\begin{table}[t]
\centering
\resizebox{\linewidth}{!}{
\begin{tabular}{@{}ll@{}}
\hline
\textbf{TMNRE Setting}          & \textbf{Value} \\ \hline
Number of rounds                & 7${}^\star$ \\
Simulation schedule             & 30k, 60k, 90k, 120k, 120k, \\
                                & 150k, 150k \\
Bounds threshold $\epsilon$     & $10^{-5}$ \\ 
Noise shuffling                 & \texttt{True} \\ 
Min./Max. training epochs       & 30/200 \\ 
Early stopping                  & 7 \\ 
Initial learning rate           & $5 \times 10^{-4}$ \\
Training/Validation batch size  & 256/256 \\ 
Train : Validation ratio        & 0.9 : 0.1 \\ \hline
\end{tabular}}
\caption{Choices for the hyperparameters and settings for the TMNRE algorithm in this work.\\
{\footnotesize ${}^\star$ This is the minimum number of rounds, if the algorithm has not converged, we continue rounds of inference until the truncation procedure terminates.}}
\label{tab:sbi_tmnre}
\end{table}

\vspace*{8pt}
\noindent \emph{TMNRE Setup.} Finally, there are a number of choices that need to be made when running the TMNRE algorithm. These choices are fully detailed in Tab.~\ref{tab:sbi_tmnre}, or in the configuration files supplied with \texttt{peregrine}. The settings broadly split into two categories: those that control the training of the inference network, and choices for the TMNRE algorithm itself. In the former case, we can define how many epochs to train the network for (min./max. training epochs), how many epochs to wait for the validation loss to decrease\footnote{Specifically, the training keeps track of the loss on the training and validation sets, then tests whether the validation loss has increased relative to the previous round. This can be a sign of overfitting, which should of course be avoided, but we are also training via stochastic gradient descent so we instead look for a number of epochs (early stopping) where no decrease in the validation loss has been seen before stopping. We then re-initialise the network to the state with the lowest validation loss.} (early stopping), the initial learning rate (initial learning rate), the batch sizes shown to the network (training/validation batch size), and finally the split in the data between training and validation (train : validation ratio). In addition, we have the TMNRE settings such as the (minimum) number of rounds (number of rounds), the number of simulations to use in each round\footnote{In later rounds, it is typically the case that a larger simulation budget is required to obtain the necessary precision in the parameter inference.} (simulation schedule), and the bounds threshold for performing the truncation ($\epsilon$). The last setting to discuss is the noise shuffling switch. This is closely related to the question of overfitting to training data, where for small simulation batches it can be the case that the network essentially ``remembers" the training data including the noise realisations. This leads to a very low value of the loss on the training data set, but poor performance on validation data. We found that one extremely effective way of overcoming this is to shuffle the noise realisations in each batch. This effectively shows the network brand new examples every epoch, but with the same signal component relevant for inference. This strategy should be applicable for \emph{any} additive noise model (related to gravitational waves, or not) and can reduce the simulation budget by potentially an order of magnitude.

\vspace*{8pt}
\noindent To summarise this section, we have discussed broadly the field and applicability of simulation-based inference techniques, before describing in detail the development of our new gravitational wave data analysis pipeline \texttt{peregrine}. In doing so, we explained the technicalities of our particular implementation of simulation-based inference, TMNRE, that is applicable to targeted parameter inference on a specific observation of interest. In the next section, we will demonstrate the application of our analysis method to two case studies of highly spinning, precessing BBH mergers, which represent a current state-of-the-art analysis and modelling challenge.

\section{Results: Two Case Studies}\label{sec:results}

\begin{figure*}[t]
    \centering
    \includegraphics[width=\linewidth,trim={0.1cm 0.1cm 0.1cm 0.1cm},clip]{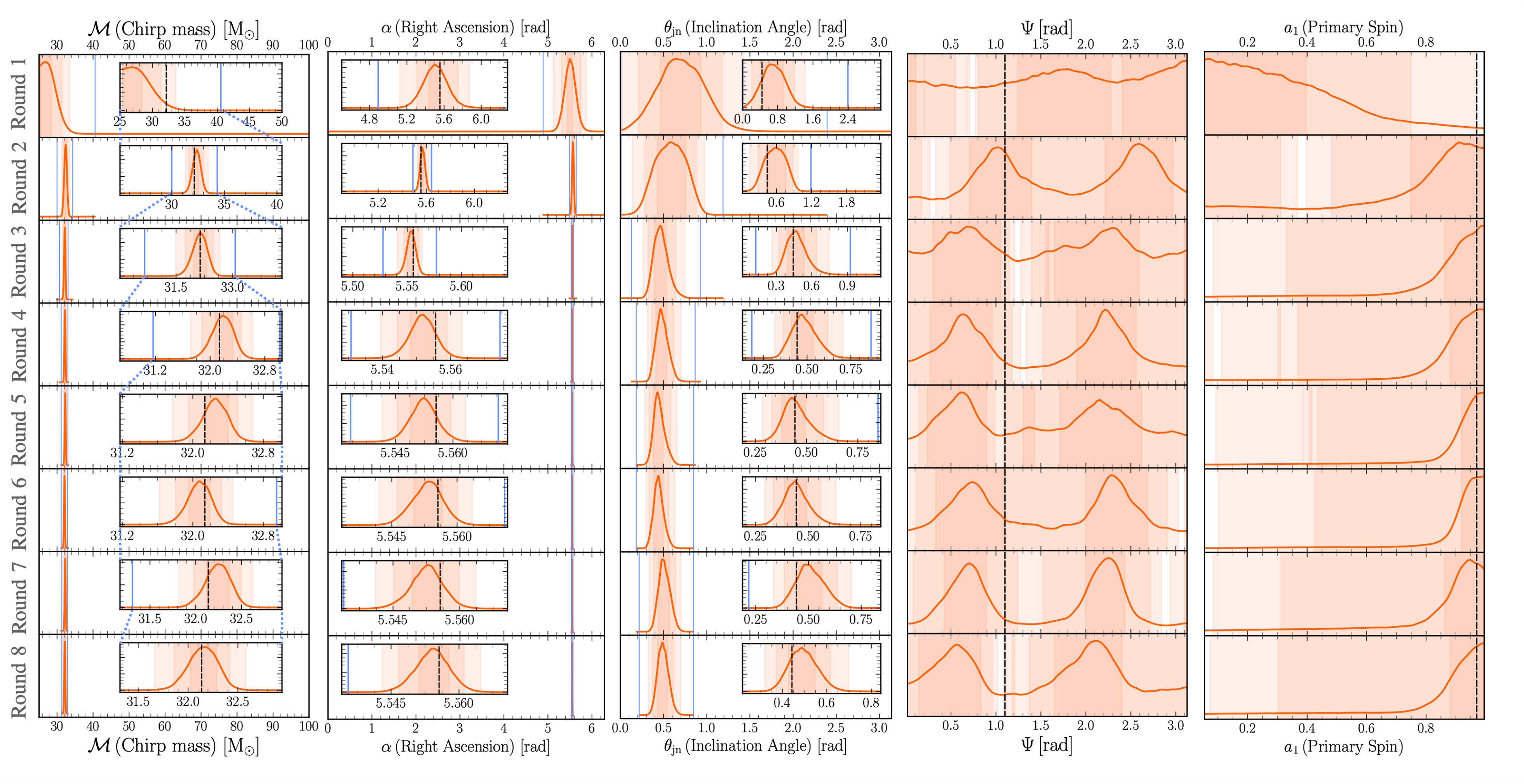}
    \caption{Illustration of the truncation procedure as applied to the high SNR case study (\textbf{C2}). Each row corresponds to one round of inference for five separate parameters (chirp mass $\mathcal{M}$, right ascension $\alpha$, inclination angle $\theta_\mathrm{jn}$, polarisation angle $\psi$, and primary spin $a_1$, from left to right). The insets are zoomed in versions of the constrained region to illustrate the coverage of the true injected values (shown by black dashed lines). The solid blue lines show the prior truncation bounds derived during the inference step.}\vspace*{-10pt}
    \label{fig:truncation}
\end{figure*}

\noindent We now turn our attention to the application of the method to two concrete case studies that represent the state-of-the-art for parameter estimation in current generation gravitational wave detectors. Specifically, in this section we consider two noisy mock observations with waveforms generated by the \texttt{SEOBNRv4PHM} approximant~\cite{Ossokine:2020kjp}. 
This time-domain waveform approximant is tuned to describe BBH merger systems that are both spinning and precessing. It employs more accurate aligned-spin as well as precessing-spin two-body dynamics than its predecessors.\footnote{In particular, it includes the $(2,\pm2), (2,\pm1), (3,\pm3),
(4,\pm4)$ and $(5,\pm5)$ modes in the co-precessing frame of the binary instead of the commonly used $(2,\pm2)$ and $(2,\pm1)$ modes, and displays improved accuracy in calibration fits with numerical relativity simulations across a larger parameter space. When generating aligned-spin waveforms, \texttt{SEOBNRv4PHM} reduces to \texttt{SEOBNRv4HM}.}
It takes as input 15 relevant physical parameters, which are detailed in Tab.~\ref{tab:sbi_params}. We do note, however, that our pipeline makes no assumption about the choice of waveform approximant, so any other model could be easily integrated.

We generate the noise according to the LIGO-O4 power spectral density for each detector (H1, L1, V1)~\cite{Ashton:2021anp}. We assume the detector noise to be gaussian-distributed, stationary and coloured by the estimated PSD of the upcoming LIGO-Virgo observing run.\footnote{In full technicality, the detector noise originates from a number of different continuous and transient sources which lead it to be non-stationary and coloured. However, it has been shown that away from these transient noise artefacts a gaussian distribution coloured by an online estimate of the PSD is an excellent approximation~\cite{LIGOScientific:2019hgc}.} In practice, particularly because we are looking to analyse individual observations, the noise model could be modified to include \emph{e.g.}, any online estimate of the PSD at the time of an event~\cite{LIGOScientific:2016aoc,LIGOScientific:2018mvr}. Whilst of course, the precise posteriors obtained would be different as a result of the different noise realisation, the data analysis pipeline would remain identical and we expect our approach to generalise to any noise model.

In the rest of the section we will describe the case studies in more detail before presenting our results when analysing the signals using the TMNRE algorithm implemented in \texttt{peregrine}. Importantly, we show that we can achieve excellent agreement compared to traditional established methods across all parameters in the 1-dimensional and 2-dimensional marginals. To obtain this comparison, we run the nested sampling code \texttt{dynesty}~\cite{Speagle:2020aaa} that comes as default with the bilby package\footnote{The results for the posteriors obtained from \texttt{dynesty}~\cite{Handley:2015aaa,Speagle:2020aaa} can be highly sensitive to the sampling settings. In this work, we follow other analyses in the literature~\cite{Romero-Shaw:2020owr} taking \texttt{nlive} $= 2000$ points as well as \texttt{n\_act} $ = 10$, and perform analytic phase marginalisation which is crucial for stable results. We do not analytically marginalise over luminosity distance or merger time, although this should not affect the results. An additional technical point that is relevant here is that for waveform models with higher modes, it is known that this phase pre-marginalisation step may break down if the $\ell > 2$ overtones give a significant contribution to the signal~\cite{Thrane:2018aaa}. We have checked explicitly that running dynesty again without phase pre-marginalisation (but with a larger \texttt{n\_act} $ = 50$) leads to identical posteriors on the well-measured parameters such as the chirp mass or luminosity distance, although with some instabilities in less well-constrained parameters such as the phase and the two tilts. Importantly, we note that \texttt{peregrine} itself performs no such phase pre-marginalisation and therefore the observed agreement is further indication that for the sources in question phase pre-marginalisation is performing well here. Regardless, this highlights the fact that for comparisons, this check should be done on a case-by-case basis when working with higher-order mode waveform approximants.}~\cite{Ashton:2018jfp,Romero-Shaw:2020owr,Ashton:2021anp}. We also discuss additional consistency checks that we can perform on simulation-based inference methods that are independent of -- and typically also infeasible to -- classical methods.

\subsection{Case Study Description}\label{sec:results_casestudy}

\noindent We test our data analysis methodology on two separate case studies, both within the context of the \texttt{SEOBNRv4PHM} waveform approximant~\cite{Ossokine:2020kjp}, which is currently one of the most detailed and computationally intensive models available (although, again, we could use any model, see e.g.~\cite{Khan:2015jqa,Pratten:2020fqn}). In particular we choose an injected signal with the true values given in Tab.~\ref{tab:sbi_params}, which are chosen because they are one set of best fit parameters for the interesting source recently discussed in Refs.~\cite{Hannam:2021pit,Payne:2022spz,Husa:2016aa}. In particular, the source comprises two objects that appear to be both precessing and highly spinning (with dimensionless spins $a_i > 0.8$)\footnote{In general, this interesting source highlights the need for development in waveform modelling, as well as an investigation surrounding the implications of noise artifacts on inference~\cite{Romero-Shaw:2022fbf,Xu:2022spd}. This is additional motivation to develop flexible, targeted inference methods so that the impact of individual components of the setup (waveform approximant, noise model etc.) on the final inference can be reliably tested. This is very challenging to achieve with fully amortised methods.}. We place the systems at two different luminosity distances to set up our two case studies:
\begin{itemize}[leftmargin=*]
    \item \textbf{C1.} $d_L = 900 \, \mathrm{Mpc}$ -- we study a quieter source (with an SNR of $\sim$20) to ensure that our method can deal reliably with inference on a data stream where there is significant amounts of noise. In other words, we test the ability of our method to distinguish between signal and noise in a gravitational wave event.
    \item \textbf{C2.} $d_L = 200 \, \mathrm{Mpc}$ -- conversely, we analyse a very loud source (with an SNR of $\sim$100) to determine our ability to achieve high precision inference. Whilst this source is arguably slightly unphysical according to current merger rates (in terms of how loud it is), the aim of this test is to confirm that we can achieve extremely high precision inference through the application of the TMNRE algorithm.
\end{itemize}
The detector response for both of these target observations are shown in Fig.~\ref{fig:observation} above for the LIGO Hanford facility.

\subsection{Parameter Estimation with TMNRE}\label{sec:results_pe}

\begin{figure*}[t]
    \centering
    \includegraphics[width=0.95 \linewidth,trim={0.1cm 0.1cm 0.1cm 0.1cm},clip]{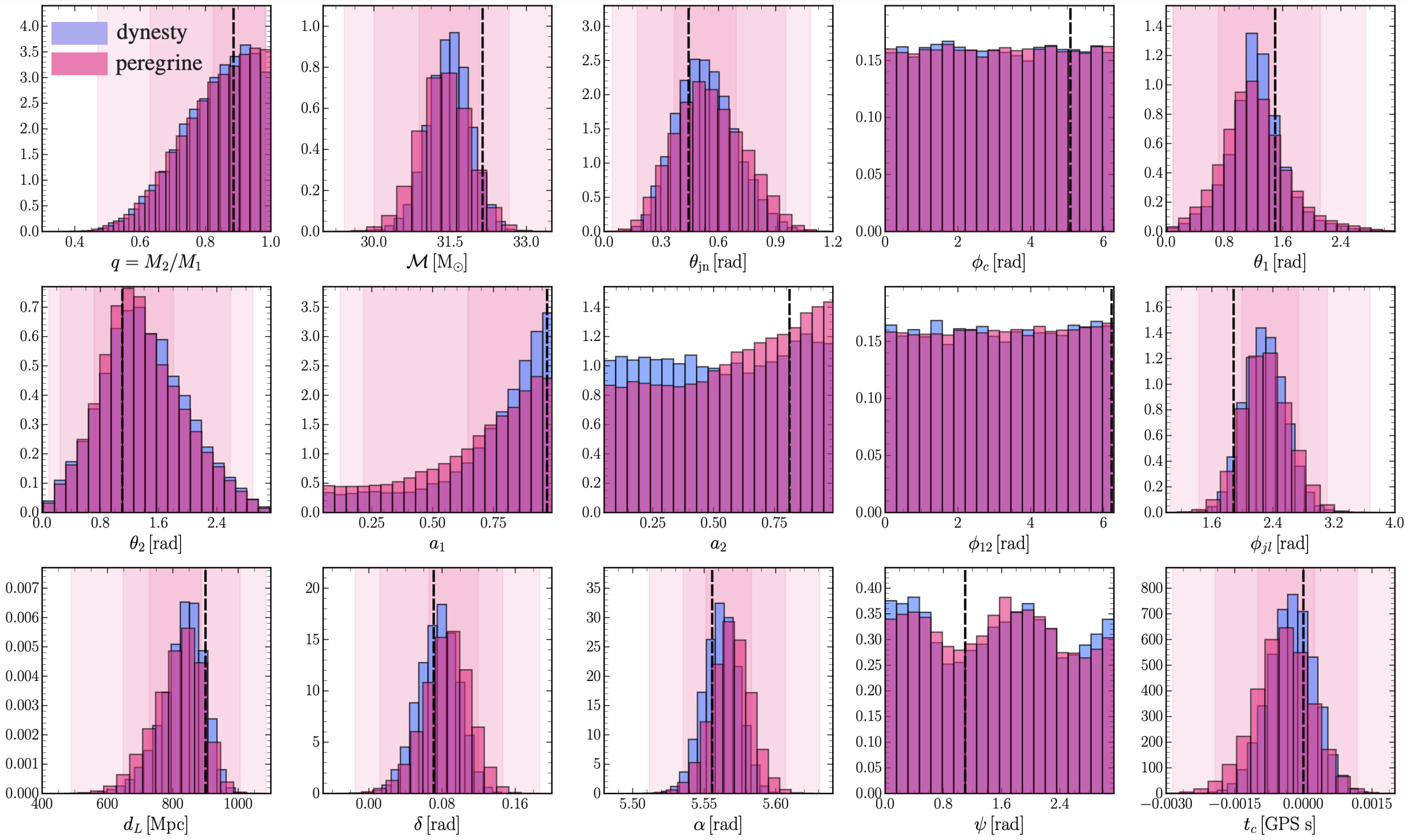}
    \caption{Full set of 1d marginal posteriors for all parameters in the low SNR case study (\textbf{C1}). The inference results from the TMNRE algorithm are shown in pink, with the $1\sigma$, $2\sigma$, and $3\sigma$ contours shown underlaid behind. The purple histograms are the \texttt{dynesty} results for the analysis performed on the same observation (see Fig.~\ref{fig:observation}). The black dashed lines indicate the true injected values.}
    \label{fig:results_lowsn}
\end{figure*}

\begin{figure*}[t]
    \centering
    \includegraphics[width=0.95 \linewidth,trim={0.1cm 0.1cm 0.1cm 0.1cm},clip]{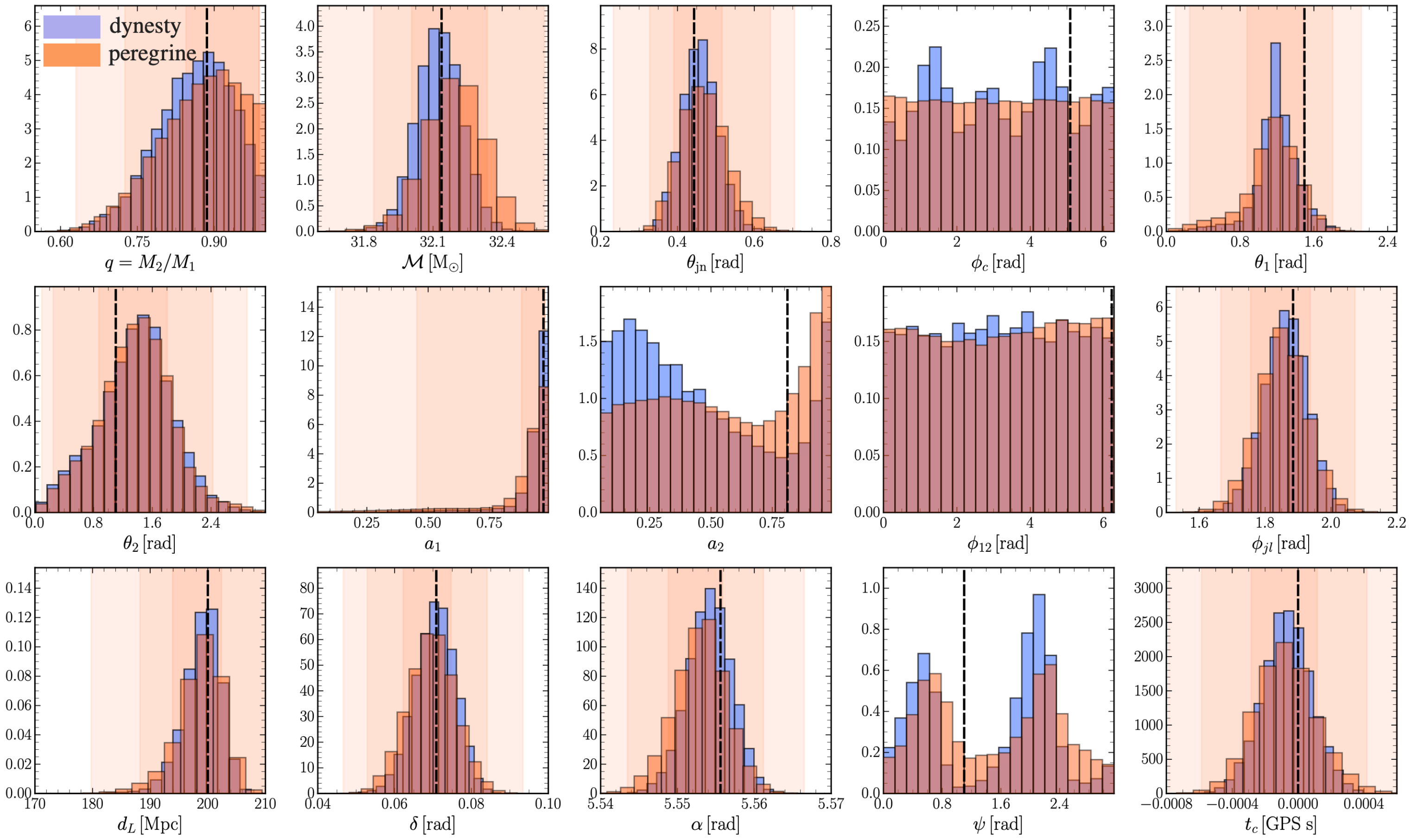}
    \caption{Full set of 1d marginal posteriors for all parameters in the high SNR case study (\textbf{C2}). The inference results from the TMNRE algorithm are shown in orange, with the $1\sigma$, $2\sigma$, and $3\sigma$ contours shown underlaid behind. The purple histograms are the \texttt{dynesty} results for the analysis performed on the same observation (see Fig.~\ref{fig:observation}). The black dashed lines indicate the true injected values.}
    \label{fig:results_highsn}
\end{figure*}

\noindent We perform parameter inference on both of the case studies using the implementation of the TMNRE algorithm in \texttt{peregrine}. In particular, we fix the algorithm settings to the ones given in Tab.~\ref{tab:sbi_tmnre}. The key results of this section are shown in Figs.~\ref{fig:results_lowsn} and~~\ref{fig:results_highsn}, while the truncation procedure is illustrated for a number of parameters in Fig.~\ref{fig:truncation}. A full set of 2d-posterior distributions can be found in the Appendix in Figs.~\ref{fig:2d_lowsn} and~\ref{fig:2d_highsn}. 

\vspace*{8pt}
\noindent \emph{Discussion.} There are a number of levels to discuss these results. Firstly we consider the TMNRE algorithm and its applicability to gravitational wave data analysis. We can see clearly from the results in Figs.~\ref{fig:results_lowsn} and~\ref{fig:results_highsn} that across all parameters we have excellent reconstruction of the injected parameters, including \emph{e.g.}, the high primary spin which can be typically challenging to analyse~\cite{Romero-Shaw:2022fbf,Xu:2022spd}. More importantly, we see that we achieve very close agreement with the traditional nested sampling method \texttt{dynesty}~\cite{Speagle:2020aaa}, both in terms of accuracy of parameter reconstruction, as well as the precision of the posteriors. To be quantitative, we compute the Jensen-Shannon divergence (JSD) between the two sets of posterior distributions. In Appendix~\ref{app:JSD}, we provide a full set of these values, averaged over all physical parameters across a number of different samplers including \texttt{dynesty}~\cite{Speagle:2020aaa}, but also \texttt{cpnest}~\cite{Veitch:2021aaa} and \texttt{ptemcee}~\cite{Foreman-Mackey:2012any,Vousden:2016aaa}, as well as a complete description of our methodology for computing the JSD\footnote{We use \texttt{nlive} $=2000$, \texttt{n\_{act}} $=10$ for \texttt{dynesty}, \texttt{nlive} $=2000$ for \texttt{cpnest} and \texttt{nsamples} $=2000$ for \texttt{ptemcee}.}. Given these results, there are two key points to highlight here. Firstly, we see that the variation between samplers as quantified by the JSD is an observation specific quantity and that in this case the variance is significantly larger for the high SNR case study (on average around $(10-30) \times 10^{-3} \, \mathrm{nat}$) compared to the low SNR one (on average around $(3-9) \times 10^{-3} \, \mathrm{nat}$). This trend follows at the level of variation between posterior distributions obtained from the same sampler (e.g. \texttt{dynesty}) run twice with different random seeds. Secondly, and more relevantly in terms of the \texttt{peregrine} posteriors, we see that for both case studies, our results are fully consistent with the variations between individual samplers (although of course we will happen to show better agreement with one compared to another). In addition, it is worth noting that we also obtain significantly better coverage statistics than \emph{e.g.}, bilby-MCMC and dynesty (see Refs.~\cite{Ashton:2018jfp,Ashton:2021anp} and the discussion below). As a final point, we can see that the main discrepancies are on parameters that are nonetheless poorly constrained, and certainly not measured to any precision -- for example the secondary dimensionless spin $a_2$, or the phase $\phi_c$ which is pre-marginalised in the sampling approach anyway.

A crucial element to achieving this level of agreement was to ensure that the same underlying noise realisation, and corresponding observation was analysed in both the TMNRE and nested sampling analysis. Performing the parameter inference on the same signal with a different noise realisation will (correctly) lead to shifts in the posterior estimates of the parameter that only \emph{on average} will be centred on the true value. This is something we actually take advantage of in the next section to perform additional quality checks of our inference. These are known broadly as coverage tests and are becoming common practice in the simulation-based inference literature to validate inference results~\cite{Hermans:2021aaa,Lemos:2023aaa}.

Two of the important features of the TMNRE algorithm that we highlighted in Sec.~\ref{sec:sbi} were the truncation and marginal elements. In Fig.~\ref{fig:truncation}, we highlight the former of these and illustrate the impact of truncation on various classes of parameters. For example, the first three columns of panels illustrate the convergence of the posterior estimates over the rounds of truncation and re-simulation for parameters that are (extremely) well measured, such as the chirp mass or inclination angle. In the final two panels, we see another class of parameters (here $\psi$ and $a_1$ as examples) which are not well constrained but are nonetheless reconstructed. We see for this second class of parameter that in the initial rounds, the marginal posterior estimate of \emph{e.g.} $p(\psi | x_0)$ is rather poor -- and indeed, one should not interpret this as the true posterior, because the TMNRE algorithm has not converged. Once the other well measured parameters are tightly constrained, however, the quality of the inference of $\psi$ and $a_1$ increases significantly, eventually converging to agree with the sampling result. This is a general trend when running TMNRE algorithms that convergence in other parameters over the course of the rounds ultimately leads to later improvements in posteriors for different parameters, see \emph{e.g.}~\cite{Karchev:2022xyn}. It is also worth briefly commenting on multi-modality here: whilst the case studies we have chosen have well-constrained parameters such as the chirp mass that are single modes, multi-modal inference in gravitational waves is a common occurrence. In the context of \texttt{peregrine}, the same level of efficiency could be realised in these scenarios by modifying how we perform the truncation of the prior proposal. Specifically, instead of bounding in some rectangular region, we could instead impose a threshold on the ratio and only sample in (potentially disconnected) regions of the prior that exceed this. This is currently under active development through \emph{e.g.}, nested sampling applied to the prior distributions, and we emphasise it is just technical hurdle, rather than any limitation of the method. The second feature we highlighted was the benefit of directly estimating marginal posteriors $p(\vartheta | x_0)$ as opposed to first generating the full ($N$-dimensional) joint distribution. This plays a crucial role in simulation efficiency at no cost to statistical accuracy -- indeed the comparison with nested sampling highlights that we obtain the correct marginal distributions even though we do not marginalise from a joint distribution.\footnote{Of course, as discussed in Sec.~\ref{sec:sbi}, we may not explicitly marginalise, but we do so implicitly by varying every parameter across all simulations} 

We can explicitly highlight this simulation efficiency by comparing the number of waveform generation steps that were required in the nested sampling case versus in the TMNRE algorithm. In the low SNR case study, we required around 720k waveform evaluations, whilst the corresponding \texttt{dynesty} run required over 44 million. This represents an \textbf{over 98\% reduction} in simulation budget for identical marginal posterior reconstruction. The difference is actually larger at the level of simulation time, since in each round of TMNRE, the waveform calls can be fully parallelised, whilst evaluations in nested sampling or MCMC broadly need to be performed sequentially. This parallelisation is implemented as default in our \texttt{peregrine} code. This is not an entirely fair comparison, since we are also required to train the classifiers at each stage. This should be compared to \emph{e.g.}, the total likelihood evaluation time for a meaningful runtime comparison. Again, taking the low SNR case study as an example, the \texttt{dynesty} run took around 43 hours versus around 12 hours for the TMNRE algorithm on the same architecture (including network training).\footnote{In future code versions, there is room for this time to be substantially improved through \emph{e.g.} accelerated waveform evaluations, reduction in network size, or a more efficient scheme for sampling from the truncated prior proposal distributions.} For scenarios where the cost per simulation is much higher, however, this trade-off and benefit from simulation efficiency will become even more pronounced.

As a final note, whilst at the level of parameter estimation the 1d marginal posteriors are an informative set of results, it is often interesting to explore the parameter degeneracies. To do so, we take advantage of the flexibility of TMNRE to estimate only the marginal posteriors we are interested in testing and construct the full set of 2d posteriors. We make a comparison to the nested sampling contours in both case studies and find similarly good agreement. To achieve this, we required an additional 500k simulations in both the low and high SNR cases on which to train the set of marginal posterior classifiers. These are shown in the Appendix in Figs.~\ref{fig:2d_lowsn} and~\ref{fig:2d_highsn}. 

\subsection{Additional Consistency Checks}\label{sec:results_consistency}

\noindent The results and comparisons to traditional methods discussed above are a helpful and meaningful way to test the data analysis pipeline that we have developed. On the other hand, the key motivation for developing \texttt{peregrine} is to apply it to scenarios where the application of traditional methods is either impossible or extremely costly. This is slightly different motivation to current research efforts to perform \emph{e.g.}, online parameter inference for LVKC data streams~\cite{Dax:2021tsq,Wong:2023lgb}. Given this, it is important to have additional ways to cross check and validate our inference results that do not rely on comparisons to sampling. This is an active area of ongoing research in simulation-based inference~\cite{Hermans:2021aaa,Lueckmann:2021aaa,Cannon:2022aaa}, however, a number of established techniques now exist. 

The most common are known as coverage tests, which implement the idea discussed above that in the context of Bayesian parameter estimation, repeated inference over different noise/statistical realisations of the same signal should result in posteriors that shift relative to the true value. The rationale behind this sort of expected coverage test is that -- simply as a result of statistical fluctuations -- the $x$\% credible interval for the posterior should contain the simulation-truth value $x$\% of the time. For a more general discussion of this in the context of simulation-based inference, see \emph{e.g.}~\cite{Hermans:2021aaa}. In practice, we can take a large sample\footnote{In the tests shown below, we generated 2000 extra simulations from the final round truncated prior region.} of observations generated from the truncated prior (where we have converged posterior estimates), and perform inference on each observation. In each case we can note how often the injected value was contained inside the $x$\% confidence interval and construct a cumulative distribution. A well calibrated posterior distribution will be a totally diagonal line when the expected coverage is plotted against the empirical findings. Due to the fact that full inference must be done for each observation, this type of test is typically infeasible to do for sequential sampling methods such as MCMC. It should be noted, however, that this test is diagnostic in the sense that a failure implies a poorly calibrated posterior, but success is not a definitive guarantee.

The coverage test results for all 15 parameters in both the high and low SNR case studies are provided in the Appendix (see Figs.~\ref{fig:pp_lowsn},~\ref{fig:zz_lowsn},~\ref{fig:pp_highsn} and~\ref{fig:zz_highsn}). We see that across every parameter, we achieve extremely good posterior coverage, something that adds additional validation to our agreement with \texttt{dynesty}.

\section{Conclusions and Outlook}\label{sec:conclusion}

\noindent In this final section, we present the key conclusions of this work, before giving some outlook as to the possible use cases for our data analysis strategy. Ultimately the motivation for our method development comes from the analysis challenges facing the field of gravitational waves as we move towards the next generation of detectors~\cite{Maggiore:2019uih,Reitze:2019iox,LISA:2017aaa}. Specifically, we will need methods that are highly scalable, simulation efficient, and flexible to take full advantage of future data. In this context, the key research contributions from this work are:
\begin{itemize}[leftmargin=*]
    \item \emph{Simulation-based inference pipeline.} The key development put forward in this work is a sequential simulation-based inference pipeline for analysing gravitational waves (see Sec.~\ref{sec:sbi} for a discussion). Specifically, we implemented an algorithm known as Truncated Marginal Neural Ratio Estimation (TMNRE)~\cite{Miller:2022shs}. The motivation and advantages associated to this choice for gravitational waves are broadly centred around simulation efficiency and its applicability to individual observations. We showed for example that in a standard analysis for a binary black hole merger, we reduced the simulation budget (number of waveform evaluations) \textbf{by over 98\%} compared to likelihood-based methods. This scaling is achieved primarily through the ability of the TMNRE algorithm to directly estimate marginal posteriors, and is the key argument for the use of our method in currently intractable sampling problems such as the analysis of overlapping sources~\cite{Pizzati:2021apa,Antonelli:2021vwg,Samajdar:2021egv,Janquart:2022nyz,Langendorff:2022fzq}.
    \item \emph{Case studies.} We studied two cases where we analysed highly spinning precessing binary black hole mergers, one with a high SNR of around 100, and another with a relatively low SNR of approximately 20. In each case, we performed full parameter inference using our method on observations generated using the state-of-the-art \texttt{SEOBNRv4PHM} waveform approximant~\cite{Ossokine:2020kjp} (see Fig.~\ref{fig:observation}), although the method can be applied to any waveform model. We compared our results for the 1d- and 2d-marginal posteriors to posterior samples obtained from nested sampling and demonstrated the excellent agreement across all parameters (see Figs.~\ref{fig:results_lowsn} and~\ref{fig:results_highsn}). In addition, we carried out expected coverage tests to validate the behaviour of our posterior estimates (see \emph{e.g.} Figs.~\ref{fig:pp_highsn} and~\ref{fig:zz_highsn} for the high SNR case). 
    \item \emph{Robust and flexible method.} Our implementation of TMNRE is not the first method to apply modern machine learning techniques to the analysis of gravitational waves~\cite{Gabbard:2019rde,Green:2020dnx,Chua:2019wwt,Alvares:2020bjg,Dax:2021tsq,Delaunoy:2020zcu,Dax:2022pxd}. On the other hand, these other impressive parameter estimation pipelines typically have a different goal in mind -- rapid or instantaneous parameter inference. This has great relevance when attempting to \emph{e.g.} optimise signal triggering or follow up quickly on electromagnetic counterparts. However, this comes at the cost of complicated architectures, or bespoke likelihood designs~\cite{Wong:2023lgb,Zackay:2018qdy,Leslie:2021ssu} which render these approaches less flexible for studying arbitrary classes of signals. In contrast, our motivation is to develop a method that is robust and flexible enough to be used to analyse a wide range of signal classes in specific observations. In order to study a new class of signal -- only a new forward simulation model is required, whilst the rest of the pipeline will remain identical. Importantly, in cases where traditional methods are not available, the coverage tests described in Sec.~\ref{sec:results} will still be applicable, so we can still validate our results.
    \item \emph{Code release.} Alongside this work, we have developed an implementation of our analysis method, known as \texttt{peregrine} (built on top of \texttt{swyft}~\cite{Miller:2021hys,Miller:2022shs}). This is a highly modular and scalable library that allows the user to implement their own \texttt{Simulator} class. Given the simulation efficiency and relatively small network size, it is usable on an individual basis with reasonable access to consumer hardware. Both the algorithm as well as the coverage tests are fully developed and tested, with examples available. Access to \texttt{peregrine} can be requested at \href{https://github.com/undark-lab/peregrine-public}{this link }. 
\end{itemize}

\subsection{Outlook}\label{sec:conclusion_outlook}

\noindent As discussed in the introduction to this work, we are currently in an exciting era of gravitational wave experimental development. With future detectors such as the Einstein Telescope~\cite{Maggiore:2019uih} or LISA~\cite{LISA:2017aaa} taking shape, the future gravitational wave sky could be extremely loud, varied, and informative. On the other hand, there are clear data analysis challenges associated to fully utilising this data~\cite{MockLISADataChallengeTaskForce:2009wir,Schafer:2022dxv}. Two important examples of these challenges include \emph{(i)} the analysis of two (or more) overlapping gravitational wave signals, which will arise as a result of large increases in detector sensitivity~\cite{Pizzati:2021apa,Antonelli:2021vwg,Samajdar:2021egv,Janquart:2022nyz,Langendorff:2022fzq,Buscicchio:2019aaa}, and \emph{(ii)} the identification and analysis of stochastic signals that will be relevant to space-based observatories such as LISA~\cite{LISA:2017aaa}. In the former case, the challenge is a sampling one, with current methods potentially taking months to analyse a single event. In the latter case, there is also potentially a large sampling challenge in an attempt to fit multiple signal templates or noise models, but there are also statistical challenges that could arise from correlated signal or detector noise, as well as significant simulation costs.

It is this class of data analysis challenges that we had in mind when developing \texttt{peregrine}.\footnote{Although in addition, motivated by the observed simulation efficiency, it would be interesting to explore the extent to which the inference can be fully amortised by significantly expanding the initial simulation budget, after which approximately real time inference could be performed. We will leave this aspect to future work, however, along with investigations as to how to include the effect of non-stationary noise in such an approach.} There are a number of reasons to believe that the framework put forward here can approach these problems and extract the maximum science/physics results. Firstly, the method is designed to be highly simulation efficient, something we have demonstrated explicitly in the case of binary black hole mergers where we saw a $\sim$98\% reduction in simulation budget. This is one of the keys to solving sampling based problems, where we can directly perform inference at the level of marginal distributions rather than solve the full joint problem. Assuming the same simulation efficiency holds, this could lead to a reduction in analysis time for a problem such as parameter inference for multiple overlapping waveforms from $\mathcal{O}$(months)~\cite{Janquart:2022nyz} to less than a day. Similarly, TMNRE and other simulation-based inference approaches are by construction implicit-likelihood methods. This is equivalent to the statement that one only needs a forward simulator to perform inference, rather than an explicit likelihood. In the case of \emph{e.g.}, possible correlated noise in the stochastic gravitational wave background~\cite{Caprini:2018mtu,Christensen:2018iqi}, this feature could unlock a number of physics questions that are currently intractable. It is also worth noting that non-stationary noise, which is typically challenging to deal with in a likelihood-based framework is simple in the context of simulation-based inference and \texttt{peregrine}. In particular, the analysis would remain identical if \emph{e.g.}, the PSD was time-dependent, provided this was included in the detector aspect of the simulator. We will leave these to future works to analyse these situations in detail.

To summarise, it is clear that in the era of high-sensitivity gravitational wave experiments, flexible and scalable data analysis strategies will be crucial to uncovering exciting fundamental physics discoveries. In this work, we have presented \texttt{peregrine} as a community-based tool that can hopefully take some steps towards that goal.

\vspace*{-10pt}
\section*{Acknowledgements}
\noindent 
This work is part of the project CORTEX (NWA.1160.18.316) of the research programme NWA-ORC which is (partly) financed by the Dutch Research Council (NWO). Additionally, CW acknowledges funding from the European Research Council (ERC) under the European Union’s Horizon 2020 research and innovation programme (Grant agreement No. 864035). UB is supported through the CORTEX project of the NWA-ORC with project number NWA.1160.18.316 which is partly financed by the Dutch Research Council (NWO). JA is supported through the research program ``The Hidden Universe of Weakly Interacting Particles" with project number 680.92.18.03 (NWO Vrije Programma), which is partly financed by the Nederlandse Organisatie voor Wetenschappelijk Onderzoek (Dutch Research Council). SN acknowledges support from the NWO Projectruimte grant (Samaya Nissanke). BKM is funded by the University of Amsterdam Faculty of Science (FNWI), Informatics Institute (IvI), and the Institute of Physics (IoP). BKM is also affiliated with the European Laboratory for Learning and Intelligent Systems (ELLIS Society). The main analysis for this work was carried out on the Lisa and Snellius Computing Clusters at SURFsara. We are very grateful to Thomas Edwards and Sam Witte for their careful reading of the draft and useful comments. We also thank Andrew Williamson and Pablo Bosch for helpful conversations about this project.

\bibliography{biblio}

\begin{thebibliography}{103}%
\makeatletter
\providecommand \@ifxundefined [1]{%
 \@ifx{#1\undefined}
}%
\providecommand \@ifnum [1]{%
 \ifnum #1\expandafter \@firstoftwo
 \else \expandafter \@secondoftwo
 \fi
}%
\providecommand \@ifx [1]{%
 \ifx #1\expandafter \@firstoftwo
 \else \expandafter \@secondoftwo
 \fi
}%
\providecommand \natexlab [1]{#1}%
\providecommand \enquote  [1]{``#1''}%
\providecommand \bibnamefont  [1]{#1}%
\providecommand \bibfnamefont [1]{#1}%
\providecommand \citenamefont [1]{#1}%
\providecommand \href@noop [0]{\@secondoftwo}%
\providecommand \href [0]{\begingroup \@sanitize@url \@href}%
\providecommand \@href[1]{\@@startlink{#1}\@@href}%
\providecommand \@@href[1]{\endgroup#1\@@endlink}%
\providecommand \@sanitize@url [0]{\catcode `\\12\catcode `\$12\catcode
  `\&12\catcode `\#12\catcode `\^12\catcode `\_12\catcode `\%12\relax}%
\providecommand \@@startlink[1]{}%
\providecommand \@@endlink[0]{}%
\providecommand \url  [0]{\begingroup\@sanitize@url \@url }%
\providecommand \@url [1]{\endgroup\@href {#1}{\urlprefix }}%
\providecommand \urlprefix  [0]{URL }%
\providecommand \Eprint [0]{\href }%
\providecommand \doibase [0]{http://dx.doi.org/}%
\providecommand \selectlanguage [0]{\@gobble}%
\providecommand \bibinfo  [0]{\@secondoftwo}%
\providecommand \bibfield  [0]{\@secondoftwo}%
\providecommand \translation [1]{[#1]}%
\providecommand \BibitemOpen [0]{}%
\providecommand \bibitemStop [0]{}%
\providecommand \bibitemNoStop [0]{.\EOS\space}%
\providecommand \EOS [0]{\spacefactor3000\relax}%
\providecommand \BibitemShut  [1]{\csname bibitem#1\endcsname}%
\let\auto@bib@innerbib\@empty
\bibitem [{\citenamefont {Abbott}\ \emph {et~al.}(2016)\citenamefont {Abbott}
  \emph {et~al.}}]{LIGOScientific:2016aoc}%
  \BibitemOpen
  \bibfield  {author} {\bibinfo {author} {\bibfnamefont {B.~P.}\ \bibnamefont
  {Abbott}} \emph {et~al.} (\bibinfo {collaboration} {LIGO Scientific,
  Virgo}),\ }\href {\doibase 10.1103/PhysRevLett.116.061102} {\bibfield
  {journal} {\bibinfo  {journal} {Phys. Rev. Lett.}\ }\textbf {\bibinfo
  {volume} {116}},\ \bibinfo {pages} {061102} (\bibinfo {year} {2016})},\
  \Eprint {http://arxiv.org/abs/1602.03837} {arXiv:1602.03837 [gr-qc]}
  \BibitemShut {NoStop}%
\bibitem [{\citenamefont {Abbott}\ \emph
  {et~al.}(2021{\natexlab{a}})\citenamefont {Abbott} \emph
  {et~al.}}]{LIGOScientific:2020tif}%
  \BibitemOpen
  \bibfield  {author} {\bibinfo {author} {\bibfnamefont {R.}~\bibnamefont
  {Abbott}} \emph {et~al.} (\bibinfo {collaboration} {LIGO Scientific,
  Virgo}),\ }\href {\doibase 10.1103/PhysRevD.103.122002} {\bibfield  {journal}
  {\bibinfo  {journal} {Phys. Rev. D}\ }\textbf {\bibinfo {volume} {103}},\
  \bibinfo {pages} {122002} (\bibinfo {year} {2021}{\natexlab{a}})},\ \Eprint
  {http://arxiv.org/abs/2010.14529} {arXiv:2010.14529 [gr-qc]} \BibitemShut
  {NoStop}%
\bibitem [{\citenamefont {Abbott}\ \emph
  {et~al.}(2021{\natexlab{b}})\citenamefont {Abbott} \emph
  {et~al.}}]{LIGOScientific:2019zcs}%
  \BibitemOpen
  \bibfield  {author} {\bibinfo {author} {\bibfnamefont {B.~P.}\ \bibnamefont
  {Abbott}} \emph {et~al.} (\bibinfo {collaboration} {LIGO Scientific, Virgo,
  VIRGO}),\ }\href {\doibase 10.3847/1538-4357/abdcb7} {\bibfield  {journal}
  {\bibinfo  {journal} {Astrophys. J.}\ }\textbf {\bibinfo {volume} {909}},\
  \bibinfo {pages} {218} (\bibinfo {year} {2021}{\natexlab{b}})},\ \Eprint
  {http://arxiv.org/abs/1908.06060} {arXiv:1908.06060 [astro-ph.CO]}
  \BibitemShut {NoStop}%
\bibitem [{\citenamefont {Nissanke}\ \emph {et~al.}(2010)\citenamefont
  {Nissanke}, \citenamefont {Holz}, \citenamefont {Hughes}, \citenamefont
  {Dalal},\ and\ \citenamefont {Sievers}}]{Nissanke:2009kt}%
  \BibitemOpen
  \bibfield  {author} {\bibinfo {author} {\bibfnamefont {S.}~\bibnamefont
  {Nissanke}}, \bibinfo {author} {\bibfnamefont {D.~E.}\ \bibnamefont {Holz}},
  \bibinfo {author} {\bibfnamefont {S.~A.}\ \bibnamefont {Hughes}}, \bibinfo
  {author} {\bibfnamefont {N.}~\bibnamefont {Dalal}}, \ and\ \bibinfo {author}
  {\bibfnamefont {J.~L.}\ \bibnamefont {Sievers}},\ }\href {\doibase
  10.1088/0004-637X/725/1/496} {\bibfield  {journal} {\bibinfo  {journal}
  {Astrophys. J.}\ }\textbf {\bibinfo {volume} {725}},\ \bibinfo {pages} {496}
  (\bibinfo {year} {2010})},\ \Eprint {http://arxiv.org/abs/0904.1017}
  {arXiv:0904.1017 [astro-ph.CO]} \BibitemShut {NoStop}%
\bibitem [{\citenamefont {Nissanke}\ \emph {et~al.}(2011)\citenamefont
  {Nissanke}, \citenamefont {Sievers}, \citenamefont {Dalal},\ and\
  \citenamefont {Holz}}]{Nissanke:2011ax}%
  \BibitemOpen
  \bibfield  {author} {\bibinfo {author} {\bibfnamefont {S.}~\bibnamefont
  {Nissanke}}, \bibinfo {author} {\bibfnamefont {J.}~\bibnamefont {Sievers}},
  \bibinfo {author} {\bibfnamefont {N.}~\bibnamefont {Dalal}}, \ and\ \bibinfo
  {author} {\bibfnamefont {D.}~\bibnamefont {Holz}},\ }\href {\doibase
  10.1088/0004-637X/739/2/99} {\bibfield  {journal} {\bibinfo  {journal}
  {Astrophys. J.}\ }\textbf {\bibinfo {volume} {739}},\ \bibinfo {pages} {99}
  (\bibinfo {year} {2011})},\ \Eprint {http://arxiv.org/abs/1105.3184}
  {arXiv:1105.3184 [astro-ph.CO]} \BibitemShut {NoStop}%
\bibitem [{\citenamefont {Abbott}\ \emph
  {et~al.}(2021{\natexlab{c}})\citenamefont {Abbott} \emph
  {et~al.}}]{LIGOScientific:2020kqk}%
  \BibitemOpen
  \bibfield  {author} {\bibinfo {author} {\bibfnamefont {R.}~\bibnamefont
  {Abbott}} \emph {et~al.} (\bibinfo {collaboration} {LIGO Scientific,
  Virgo}),\ }\href {\doibase 10.3847/2041-8213/abe949} {\bibfield  {journal}
  {\bibinfo  {journal} {Astrophys. J. Lett.}\ }\textbf {\bibinfo {volume}
  {913}},\ \bibinfo {pages} {L7} (\bibinfo {year} {2021}{\natexlab{c}})},\
  \Eprint {http://arxiv.org/abs/2010.14533} {arXiv:2010.14533 [astro-ph.HE]}
  \BibitemShut {NoStop}%
\bibitem [{\citenamefont {Abbott}\ \emph
  {et~al.}(2018{\natexlab{a}})\citenamefont {Abbott} \emph
  {et~al.}}]{LIGOScientific:2018cki}%
  \BibitemOpen
  \bibfield  {author} {\bibinfo {author} {\bibfnamefont {B.~P.}\ \bibnamefont
  {Abbott}} \emph {et~al.} (\bibinfo {collaboration} {LIGO Scientific,
  Virgo}),\ }\href {\doibase 10.1103/PhysRevLett.121.161101} {\bibfield
  {journal} {\bibinfo  {journal} {Phys. Rev. Lett.}\ }\textbf {\bibinfo
  {volume} {121}},\ \bibinfo {pages} {161101} (\bibinfo {year}
  {2018}{\natexlab{a}})},\ \Eprint {http://arxiv.org/abs/1805.11581}
  {arXiv:1805.11581 [gr-qc]} \BibitemShut {NoStop}%
\bibitem [{\citenamefont {Abbott}\ \emph
  {et~al.}(2021{\natexlab{d}})\citenamefont {Abbott} \emph
  {et~al.}}]{LIGOScientific:2021usb}%
  \BibitemOpen
  \bibfield  {author} {\bibinfo {author} {\bibfnamefont {R.}~\bibnamefont
  {Abbott}} \emph {et~al.} (\bibinfo {collaboration} {LIGO Scientific,
  VIRGO}),\ }\href@noop {} {\  (\bibinfo {year} {2021}{\natexlab{d}})},\
  \Eprint {http://arxiv.org/abs/2108.01045} {arXiv:2108.01045 [gr-qc]}
  \BibitemShut {NoStop}%
\bibitem [{\citenamefont {Nitz}\ \emph {et~al.}(2021)\citenamefont {Nitz},
  \citenamefont {Capano}, \citenamefont {Kumar}, \citenamefont {Wang},
  \citenamefont {Kastha}, \citenamefont {Sch\"afer}, \citenamefont
  {Dhurkunde},\ and\ \citenamefont {Cabero}}]{Nitz:2021uxj}%
  \BibitemOpen
  \bibfield  {author} {\bibinfo {author} {\bibfnamefont {A.~H.}\ \bibnamefont
  {Nitz}}, \bibinfo {author} {\bibfnamefont {C.~D.}\ \bibnamefont {Capano}},
  \bibinfo {author} {\bibfnamefont {S.}~\bibnamefont {Kumar}}, \bibinfo
  {author} {\bibfnamefont {Y.-F.}\ \bibnamefont {Wang}}, \bibinfo {author}
  {\bibfnamefont {S.}~\bibnamefont {Kastha}}, \bibinfo {author} {\bibfnamefont
  {M.}~\bibnamefont {Sch\"afer}}, \bibinfo {author} {\bibfnamefont
  {R.}~\bibnamefont {Dhurkunde}}, \ and\ \bibinfo {author} {\bibfnamefont
  {M.}~\bibnamefont {Cabero}},\ }\href {\doibase 10.3847/1538-4357/ac1c03}
  {\bibfield  {journal} {\bibinfo  {journal} {Astrophys. J.}\ }\textbf
  {\bibinfo {volume} {922}},\ \bibinfo {pages} {76} (\bibinfo {year} {2021})},\
  \Eprint {http://arxiv.org/abs/2105.09151} {arXiv:2105.09151 [astro-ph.HE]}
  \BibitemShut {NoStop}%
\bibitem [{\citenamefont {Venumadhav}\ \emph
  {et~al.}(2019{\natexlab{a}})\citenamefont {Venumadhav}, \citenamefont
  {Zackay}, \citenamefont {Roulet}, \citenamefont {Dai},\ and\ \citenamefont
  {Zaldarriaga}}]{Venumadhav:2019aa}%
  \BibitemOpen
  \bibfield  {author} {\bibinfo {author} {\bibfnamefont {T.}~\bibnamefont
  {Venumadhav}}, \bibinfo {author} {\bibfnamefont {B.}~\bibnamefont {Zackay}},
  \bibinfo {author} {\bibfnamefont {J.}~\bibnamefont {Roulet}}, \bibinfo
  {author} {\bibfnamefont {L.}~\bibnamefont {Dai}}, \ and\ \bibinfo {author}
  {\bibfnamefont {M.}~\bibnamefont {Zaldarriaga}},\ }\href {\doibase
  10.1103/PhysRevD.100.023011} {\bibfield  {journal} {\bibinfo  {journal}
  {Phys. Rev. D}\ }\textbf {\bibinfo {volume} {100}},\ \bibinfo {pages}
  {023011} (\bibinfo {year} {2019}{\natexlab{a}})}\BibitemShut {NoStop}%
\bibitem [{\citenamefont {Venumadhav}\ \emph {et~al.}(2020)\citenamefont
  {Venumadhav}, \citenamefont {Zackay}, \citenamefont {Roulet}, \citenamefont
  {Dai},\ and\ \citenamefont {Zaldarriaga}}]{Venumadhav:2020aa}%
  \BibitemOpen
  \bibfield  {author} {\bibinfo {author} {\bibfnamefont {T.}~\bibnamefont
  {Venumadhav}}, \bibinfo {author} {\bibfnamefont {B.}~\bibnamefont {Zackay}},
  \bibinfo {author} {\bibfnamefont {J.}~\bibnamefont {Roulet}}, \bibinfo
  {author} {\bibfnamefont {L.}~\bibnamefont {Dai}}, \ and\ \bibinfo {author}
  {\bibfnamefont {M.}~\bibnamefont {Zaldarriaga}},\ }\href {\doibase
  10.1103/PhysRevD.101.083030} {\bibfield  {journal} {\bibinfo  {journal}
  {Phys. Rev. D}\ }\textbf {\bibinfo {volume} {101}},\ \bibinfo {pages}
  {083030} (\bibinfo {year} {2020})}\BibitemShut {NoStop}%
\bibitem [{\citenamefont {Abbott}\ \emph
  {et~al.}(2021{\natexlab{e}})\citenamefont {Abbott} \emph
  {et~al.}}]{LIGOScientific:2020ibl}%
  \BibitemOpen
  \bibfield  {author} {\bibinfo {author} {\bibfnamefont {R.}~\bibnamefont
  {Abbott}} \emph {et~al.} (\bibinfo {collaboration} {LIGO Scientific,
  Virgo}),\ }\href {\doibase 10.1103/PhysRevX.11.021053} {\bibfield  {journal}
  {\bibinfo  {journal} {Phys. Rev. X}\ }\textbf {\bibinfo {volume} {11}},\
  \bibinfo {pages} {021053} (\bibinfo {year} {2021}{\natexlab{e}})},\ \Eprint
  {http://arxiv.org/abs/2010.14527} {arXiv:2010.14527 [gr-qc]} \BibitemShut
  {NoStop}%
\bibitem [{\citenamefont {Abbott}\ \emph
  {et~al.}(2021{\natexlab{f}})\citenamefont {Abbott} \emph
  {et~al.}}]{LIGOScientific:2021djp}%
  \BibitemOpen
  \bibfield  {author} {\bibinfo {author} {\bibfnamefont {R.}~\bibnamefont
  {Abbott}} \emph {et~al.} (\bibinfo {collaboration} {LIGO Scientific, VIRGO,
  KAGRA}),\ }\href@noop {} {\  (\bibinfo {year} {2021}{\natexlab{f}})},\
  \Eprint {http://arxiv.org/abs/2111.03606} {arXiv:2111.03606 [gr-qc]}
  \BibitemShut {NoStop}%
\bibitem [{\citenamefont {Allen}\ \emph {et~al.}(2012)\citenamefont {Allen},
  \citenamefont {Anderson}, \citenamefont {Brady}, \citenamefont {Brown},\ and\
  \citenamefont {Creighton}}]{Allen:2005fk}%
  \BibitemOpen
  \bibfield  {author} {\bibinfo {author} {\bibfnamefont {B.}~\bibnamefont
  {Allen}}, \bibinfo {author} {\bibfnamefont {W.~G.}\ \bibnamefont {Anderson}},
  \bibinfo {author} {\bibfnamefont {P.~R.}\ \bibnamefont {Brady}}, \bibinfo
  {author} {\bibfnamefont {D.~A.}\ \bibnamefont {Brown}}, \ and\ \bibinfo
  {author} {\bibfnamefont {J.~D.~E.}\ \bibnamefont {Creighton}},\ }\href
  {\doibase 10.1103/PhysRevD.85.122006} {\bibfield  {journal} {\bibinfo
  {journal} {Phys. Rev. D}\ }\textbf {\bibinfo {volume} {85}},\ \bibinfo
  {pages} {122006} (\bibinfo {year} {2012})},\ \Eprint
  {http://arxiv.org/abs/gr-qc/0509116} {arXiv:gr-qc/0509116} \BibitemShut
  {NoStop}%
\bibitem [{\citenamefont {Isi}\ \emph {et~al.}(2018)\citenamefont {Isi},
  \citenamefont {Smith}, \citenamefont {Vitale}, \citenamefont {Massinger},
  \citenamefont {Kanner},\ and\ \citenamefont {Vajpeyi}}]{Isi:2018vst}%
  \BibitemOpen
  \bibfield  {author} {\bibinfo {author} {\bibfnamefont {M.}~\bibnamefont
  {Isi}}, \bibinfo {author} {\bibfnamefont {R.}~\bibnamefont {Smith}}, \bibinfo
  {author} {\bibfnamefont {S.}~\bibnamefont {Vitale}}, \bibinfo {author}
  {\bibfnamefont {T.~J.}\ \bibnamefont {Massinger}}, \bibinfo {author}
  {\bibfnamefont {J.}~\bibnamefont {Kanner}}, \ and\ \bibinfo {author}
  {\bibfnamefont {A.}~\bibnamefont {Vajpeyi}},\ }\href {\doibase
  10.1103/PhysRevD.98.042007} {\bibfield  {journal} {\bibinfo  {journal} {Phys.
  Rev. D}\ }\textbf {\bibinfo {volume} {98}},\ \bibinfo {pages} {042007}
  (\bibinfo {year} {2018})},\ \Eprint {http://arxiv.org/abs/1803.09783}
  {arXiv:1803.09783 [gr-qc]} \BibitemShut {NoStop}%
\bibitem [{\citenamefont {Venumadhav}\ \emph
  {et~al.}(2019{\natexlab{b}})\citenamefont {Venumadhav}, \citenamefont
  {Zackay}, \citenamefont {Roulet}, \citenamefont {Dai},\ and\ \citenamefont
  {Zaldarriaga}}]{Venumadhav:2019tad}%
  \BibitemOpen
  \bibfield  {author} {\bibinfo {author} {\bibfnamefont {T.}~\bibnamefont
  {Venumadhav}}, \bibinfo {author} {\bibfnamefont {B.}~\bibnamefont {Zackay}},
  \bibinfo {author} {\bibfnamefont {J.}~\bibnamefont {Roulet}}, \bibinfo
  {author} {\bibfnamefont {L.}~\bibnamefont {Dai}}, \ and\ \bibinfo {author}
  {\bibfnamefont {M.}~\bibnamefont {Zaldarriaga}},\ }\href {\doibase
  10.1103/PhysRevD.100.023011} {\bibfield  {journal} {\bibinfo  {journal}
  {Phys. Rev. D}\ }\textbf {\bibinfo {volume} {100}},\ \bibinfo {pages}
  {023011} (\bibinfo {year} {2019}{\natexlab{b}})},\ \Eprint
  {http://arxiv.org/abs/1902.10341} {arXiv:1902.10341 [astro-ph.IM]}
  \BibitemShut {NoStop}%
\bibitem [{\citenamefont {Abbott}\ \emph
  {et~al.}(2018{\natexlab{b}})\citenamefont {Abbott} \emph
  {et~al.}}]{KAGRA:2013rdx}%
  \BibitemOpen
  \bibfield  {author} {\bibinfo {author} {\bibfnamefont {B.~P.}\ \bibnamefont
  {Abbott}} \emph {et~al.} (\bibinfo {collaboration} {KAGRA, LIGO Scientific,
  Virgo, VIRGO}),\ }\href {\doibase 10.1007/s41114-020-00026-9} {\bibfield
  {journal} {\bibinfo  {journal} {Living Rev. Rel.}\ }\textbf {\bibinfo
  {volume} {21}},\ \bibinfo {pages} {3} (\bibinfo {year}
  {2018}{\natexlab{b}})},\ \Eprint {http://arxiv.org/abs/1304.0670}
  {arXiv:1304.0670 [gr-qc]} \BibitemShut {NoStop}%
\bibitem [{\citenamefont {Petrov}\ \emph {et~al.}(2022)\citenamefont {Petrov},
  \citenamefont {Singer}, \citenamefont {Coughlin}, \citenamefont {Kumar},
  \citenamefont {Almualla}, \citenamefont {Anand}, \citenamefont {Bulla},
  \citenamefont {Dietrich}, \citenamefont {Foucart},\ and\ \citenamefont
  {Guessoum}}]{Petrov:2021bqm}%
  \BibitemOpen
  \bibfield  {author} {\bibinfo {author} {\bibfnamefont {P.}~\bibnamefont
  {Petrov}}, \bibinfo {author} {\bibfnamefont {L.~P.}\ \bibnamefont {Singer}},
  \bibinfo {author} {\bibfnamefont {M.~W.}\ \bibnamefont {Coughlin}}, \bibinfo
  {author} {\bibfnamefont {V.}~\bibnamefont {Kumar}}, \bibinfo {author}
  {\bibfnamefont {M.}~\bibnamefont {Almualla}}, \bibinfo {author}
  {\bibfnamefont {S.}~\bibnamefont {Anand}}, \bibinfo {author} {\bibfnamefont
  {M.}~\bibnamefont {Bulla}}, \bibinfo {author} {\bibfnamefont
  {T.}~\bibnamefont {Dietrich}}, \bibinfo {author} {\bibfnamefont
  {F.}~\bibnamefont {Foucart}}, \ and\ \bibinfo {author} {\bibfnamefont
  {N.}~\bibnamefont {Guessoum}},\ }\href {\doibase 10.3847/1538-4357/ac366d}
  {\bibfield  {journal} {\bibinfo  {journal} {Astrophys. J.}\ }\textbf
  {\bibinfo {volume} {924}},\ \bibinfo {pages} {54} (\bibinfo {year} {2022})},\
  \Eprint {http://arxiv.org/abs/2108.07277} {arXiv:2108.07277 [astro-ph.HE]}
  \BibitemShut {NoStop}%
\bibitem [{\citenamefont {Cutler}\ and\ \citenamefont
  {Flanagan}(1994)}]{Cutler:1994ys}%
  \BibitemOpen
  \bibfield  {author} {\bibinfo {author} {\bibfnamefont {C.}~\bibnamefont
  {Cutler}}\ and\ \bibinfo {author} {\bibfnamefont {E.~E.}\ \bibnamefont
  {Flanagan}},\ }\href {\doibase 10.1103/PhysRevD.49.2658} {\bibfield
  {journal} {\bibinfo  {journal} {Phys. Rev. D}\ }\textbf {\bibinfo {volume}
  {49}},\ \bibinfo {pages} {2658} (\bibinfo {year} {1994})},\ \Eprint
  {http://arxiv.org/abs/gr-qc/9402014} {arXiv:gr-qc/9402014} \BibitemShut
  {NoStop}%
\bibitem [{\citenamefont {{Thrane}}\ and\ \citenamefont
  {{Talbot}}(2019{\natexlab{a}})}]{Thrane:2019aaa}%
  \BibitemOpen
  \bibfield  {author} {\bibinfo {author} {\bibfnamefont {E.}~\bibnamefont
  {{Thrane}}}\ and\ \bibinfo {author} {\bibfnamefont {C.}~\bibnamefont
  {{Talbot}}},\ }\href {\doibase 10.1017/pasa.2019.2} {\bibfield  {journal}
  {\bibinfo  {journal} {PASA}\ }\textbf {\bibinfo {volume} {36}},\ \bibinfo
  {eid} {e010} (\bibinfo {year} {2019}{\natexlab{a}})},\ \Eprint
  {http://arxiv.org/abs/1809.02293} {arXiv:1809.02293 [astro-ph.IM]}
  \BibitemShut {NoStop}%
\bibitem [{\citenamefont {Ossokine}\ \emph {et~al.}(2020)\citenamefont
  {Ossokine} \emph {et~al.}}]{Ossokine:2020kjp}%
  \BibitemOpen
  \bibfield  {author} {\bibinfo {author} {\bibfnamefont {S.}~\bibnamefont
  {Ossokine}} \emph {et~al.},\ }\href {\doibase 10.1103/PhysRevD.102.044055}
  {\bibfield  {journal} {\bibinfo  {journal} {Phys. Rev. D}\ }\textbf {\bibinfo
  {volume} {102}},\ \bibinfo {pages} {044055} (\bibinfo {year} {2020})},\
  \Eprint {http://arxiv.org/abs/2004.09442} {arXiv:2004.09442 [gr-qc]}
  \BibitemShut {NoStop}%
\bibitem [{\citenamefont {Khan}\ \emph {et~al.}(2016)\citenamefont {Khan},
  \citenamefont {Husa}, \citenamefont {Hannam}, \citenamefont {Ohme},
  \citenamefont {P\"urrer}, \citenamefont {Jim\'enez~Forteza},\ and\
  \citenamefont {Boh\'e}}]{Khan:2015jqa}%
  \BibitemOpen
  \bibfield  {author} {\bibinfo {author} {\bibfnamefont {S.}~\bibnamefont
  {Khan}}, \bibinfo {author} {\bibfnamefont {S.}~\bibnamefont {Husa}}, \bibinfo
  {author} {\bibfnamefont {M.}~\bibnamefont {Hannam}}, \bibinfo {author}
  {\bibfnamefont {F.}~\bibnamefont {Ohme}}, \bibinfo {author} {\bibfnamefont
  {M.}~\bibnamefont {P\"urrer}}, \bibinfo {author} {\bibfnamefont
  {X.}~\bibnamefont {Jim\'enez~Forteza}}, \ and\ \bibinfo {author}
  {\bibfnamefont {A.}~\bibnamefont {Boh\'e}},\ }\href {\doibase
  10.1103/PhysRevD.93.044007} {\bibfield  {journal} {\bibinfo  {journal} {Phys.
  Rev. D}\ }\textbf {\bibinfo {volume} {93}},\ \bibinfo {pages} {044007}
  (\bibinfo {year} {2016})},\ \Eprint {http://arxiv.org/abs/1508.07253}
  {arXiv:1508.07253 [gr-qc]} \BibitemShut {NoStop}%
\bibitem [{\citenamefont {Pratten}\ \emph {et~al.}(2020)\citenamefont
  {Pratten}, \citenamefont {Husa}, \citenamefont {Garcia-Quiros}, \citenamefont
  {Colleoni}, \citenamefont {Ramos-Buades}, \citenamefont {Estelles},\ and\
  \citenamefont {Jaume}}]{Pratten:2020fqn}%
  \BibitemOpen
  \bibfield  {author} {\bibinfo {author} {\bibfnamefont {G.}~\bibnamefont
  {Pratten}}, \bibinfo {author} {\bibfnamefont {S.}~\bibnamefont {Husa}},
  \bibinfo {author} {\bibfnamefont {C.}~\bibnamefont {Garcia-Quiros}}, \bibinfo
  {author} {\bibfnamefont {M.}~\bibnamefont {Colleoni}}, \bibinfo {author}
  {\bibfnamefont {A.}~\bibnamefont {Ramos-Buades}}, \bibinfo {author}
  {\bibfnamefont {H.}~\bibnamefont {Estelles}}, \ and\ \bibinfo {author}
  {\bibfnamefont {R.}~\bibnamefont {Jaume}},\ }\href {\doibase
  10.1103/PhysRevD.102.064001} {\bibfield  {journal} {\bibinfo  {journal}
  {Phys. Rev. D}\ }\textbf {\bibinfo {volume} {102}},\ \bibinfo {pages}
  {064001} (\bibinfo {year} {2020})},\ \Eprint
  {http://arxiv.org/abs/2001.11412} {arXiv:2001.11412 [gr-qc]} \BibitemShut
  {NoStop}%
\bibitem [{\citenamefont {Edwards}\ \emph {et~al.}(2023)\citenamefont
  {Edwards}, \citenamefont {Wong}, \citenamefont {Lam}, \citenamefont {Coogan},
  \citenamefont {Foreman-Mackey}, \citenamefont {Isi},\ and\ \citenamefont
  {Zimmerman}}]{Edwards:2023sak}%
  \BibitemOpen
  \bibfield  {author} {\bibinfo {author} {\bibfnamefont {T.~D.~P.}\
  \bibnamefont {Edwards}}, \bibinfo {author} {\bibfnamefont {K.~W.~K.}\
  \bibnamefont {Wong}}, \bibinfo {author} {\bibfnamefont {K.~K.~H.}\
  \bibnamefont {Lam}}, \bibinfo {author} {\bibfnamefont {A.}~\bibnamefont
  {Coogan}}, \bibinfo {author} {\bibfnamefont {D.}~\bibnamefont
  {Foreman-Mackey}}, \bibinfo {author} {\bibfnamefont {M.}~\bibnamefont {Isi}},
  \ and\ \bibinfo {author} {\bibfnamefont {A.}~\bibnamefont {Zimmerman}},\
  }\href@noop {} {\  (\bibinfo {year} {2023})},\ \Eprint
  {http://arxiv.org/abs/2302.05329} {arXiv:2302.05329 [astro-ph.IM]}
  \BibitemShut {NoStop}%
\bibitem [{\citenamefont {Mackay}(2003)}]{Mackay:2003aaa}%
  \BibitemOpen
  \bibfield  {author} {\bibinfo {author} {\bibfnamefont {D.}~\bibnamefont
  {Mackay}},\ }\href@noop {} {\emph {\bibinfo {title} {Informaiton Theory,
  Inference, and Learning Algorithms}}}\ (\bibinfo  {publisher} {Cambridge
  University Press},\ \bibinfo {address} {Cambridge},\ \bibinfo {year}
  {2003})\BibitemShut {NoStop}%
\bibitem [{\citenamefont {Foreman-Mackey}\ \emph {et~al.}(2013)\citenamefont
  {Foreman-Mackey}, \citenamefont {Hogg}, \citenamefont {Lang},\ and\
  \citenamefont {Goodman}}]{Foreman-Mackey:2012any}%
  \BibitemOpen
  \bibfield  {author} {\bibinfo {author} {\bibfnamefont {D.}~\bibnamefont
  {Foreman-Mackey}}, \bibinfo {author} {\bibfnamefont {D.~W.}\ \bibnamefont
  {Hogg}}, \bibinfo {author} {\bibfnamefont {D.}~\bibnamefont {Lang}}, \ and\
  \bibinfo {author} {\bibfnamefont {J.}~\bibnamefont {Goodman}},\ }\href
  {\doibase 10.1086/670067} {\bibfield  {journal} {\bibinfo  {journal} {Publ.
  Astron. Soc. Pac.}\ }\textbf {\bibinfo {volume} {125}},\ \bibinfo {pages}
  {306} (\bibinfo {year} {2013})},\ \Eprint {http://arxiv.org/abs/1202.3665}
  {arXiv:1202.3665 [astro-ph.IM]} \BibitemShut {NoStop}%
\bibitem [{\citenamefont {Skilling}(2006)}]{Skilling:2006gxv}%
  \BibitemOpen
  \bibfield  {author} {\bibinfo {author} {\bibfnamefont {J.}~\bibnamefont
  {Skilling}},\ }\href {\doibase 10.1214/06-BA127} {\bibfield  {journal}
  {\bibinfo  {journal} {Bayesian Analysis}\ }\textbf {\bibinfo {volume} {1}},\
  \bibinfo {pages} {833} (\bibinfo {year} {2006})}\BibitemShut {NoStop}%
\bibitem [{\citenamefont {{Handley}}\ \emph {et~al.}(2015)\citenamefont
  {{Handley}}, \citenamefont {{Hobson}},\ and\ \citenamefont
  {{Lasenby}}}]{Handley:2015aaa}%
  \BibitemOpen
  \bibfield  {author} {\bibinfo {author} {\bibfnamefont {W.~J.}\ \bibnamefont
  {{Handley}}}, \bibinfo {author} {\bibfnamefont {M.~P.}\ \bibnamefont
  {{Hobson}}}, \ and\ \bibinfo {author} {\bibfnamefont {A.~N.}\ \bibnamefont
  {{Lasenby}}},\ }\href {\doibase 10.1093/mnras/stv1911} {\bibfield  {journal}
  {\bibinfo  {journal} {Mon. Not. Roy. Astron. Soc.}\ }\textbf {\bibinfo
  {volume} {453}},\ \bibinfo {pages} {4384} (\bibinfo {year} {2015})},\ \Eprint
  {http://arxiv.org/abs/1506.00171} {arXiv:1506.00171 [astro-ph.IM]}
  \BibitemShut {NoStop}%
\bibitem [{\citenamefont {Ashton}\ \emph {et~al.}(2022)\citenamefont {Ashton}
  \emph {et~al.}}]{Ashton:2022grj}%
  \BibitemOpen
  \bibfield  {author} {\bibinfo {author} {\bibfnamefont {G.}~\bibnamefont
  {Ashton}} \emph {et~al.},\ }\href {\doibase 10.1038/s43586-022-00121-x}
  {\bibfield  {journal} {\bibinfo  {journal} {Nature}\ }\textbf {\bibinfo
  {volume} {2}} (\bibinfo {year} {2022}),\ 10.1038/s43586-022-00121-x},\
  \Eprint {http://arxiv.org/abs/2205.15570} {arXiv:2205.15570 [stat.CO]}
  \BibitemShut {NoStop}%
\bibitem [{\citenamefont {{Speagle}}(2020)}]{Speagle:2020aaa}%
  \BibitemOpen
  \bibfield  {author} {\bibinfo {author} {\bibfnamefont {J.~S.}\ \bibnamefont
  {{Speagle}}},\ }\href {\doibase 10.1093/mnras/staa278} {\bibfield  {journal}
  {\bibinfo  {journal} {Mon. Not. Roy. Astron. Soc.}\ }\textbf {\bibinfo
  {volume} {493}},\ \bibinfo {pages} {3132} (\bibinfo {year} {2020})},\ \Eprint
  {http://arxiv.org/abs/1904.02180} {arXiv:1904.02180 [astro-ph.IM]}
  \BibitemShut {NoStop}%
\bibitem [{\citenamefont {Veitch}\ \emph {et~al.}(2015)\citenamefont {Veitch}
  \emph {et~al.}}]{Veitch:2014wba}%
  \BibitemOpen
  \bibfield  {author} {\bibinfo {author} {\bibfnamefont {J.}~\bibnamefont
  {Veitch}} \emph {et~al.},\ }\href {\doibase 10.1103/PhysRevD.91.042003}
  {\bibfield  {journal} {\bibinfo  {journal} {Phys. Rev. D}\ }\textbf {\bibinfo
  {volume} {91}},\ \bibinfo {pages} {042003} (\bibinfo {year} {2015})},\
  \Eprint {http://arxiv.org/abs/1409.7215} {arXiv:1409.7215 [gr-qc]}
  \BibitemShut {NoStop}%
\bibitem [{\citenamefont {Biwer}\ \emph {et~al.}(2019)\citenamefont {Biwer},
  \citenamefont {Capano}, \citenamefont {De}, \citenamefont {Cabero},
  \citenamefont {Brown}, \citenamefont {Nitz},\ and\ \citenamefont
  {Raymond}}]{Biwer:2018osg}%
  \BibitemOpen
  \bibfield  {author} {\bibinfo {author} {\bibfnamefont {C.~M.}\ \bibnamefont
  {Biwer}}, \bibinfo {author} {\bibfnamefont {C.~D.}\ \bibnamefont {Capano}},
  \bibinfo {author} {\bibfnamefont {S.}~\bibnamefont {De}}, \bibinfo {author}
  {\bibfnamefont {M.}~\bibnamefont {Cabero}}, \bibinfo {author} {\bibfnamefont
  {D.~A.}\ \bibnamefont {Brown}}, \bibinfo {author} {\bibfnamefont {A.~H.}\
  \bibnamefont {Nitz}}, \ and\ \bibinfo {author} {\bibfnamefont
  {V.}~\bibnamefont {Raymond}},\ }\href {\doibase 10.1088/1538-3873/aaef0b}
  {\bibfield  {journal} {\bibinfo  {journal} {Publ. Astron. Soc. Pac.}\
  }\textbf {\bibinfo {volume} {131}},\ \bibinfo {pages} {024503} (\bibinfo
  {year} {2019})},\ \Eprint {http://arxiv.org/abs/1807.10312} {arXiv:1807.10312
  [astro-ph.IM]} \BibitemShut {NoStop}%
\bibitem [{\citenamefont {Ashton}\ \emph {et~al.}(2019)\citenamefont {Ashton}
  \emph {et~al.}}]{Ashton:2018jfp}%
  \BibitemOpen
  \bibfield  {author} {\bibinfo {author} {\bibfnamefont {G.}~\bibnamefont
  {Ashton}} \emph {et~al.},\ }\href {\doibase 10.3847/1538-4365/ab06fc}
  {\bibfield  {journal} {\bibinfo  {journal} {Astrophys. J. Suppl.}\ }\textbf
  {\bibinfo {volume} {241}},\ \bibinfo {pages} {27} (\bibinfo {year} {2019})},\
  \Eprint {http://arxiv.org/abs/1811.02042} {arXiv:1811.02042 [astro-ph.IM]}
  \BibitemShut {NoStop}%
\bibitem [{\citenamefont {Romero-Shaw}\ \emph {et~al.}(2020)\citenamefont
  {Romero-Shaw} \emph {et~al.}}]{Romero-Shaw:2020owr}%
  \BibitemOpen
  \bibfield  {author} {\bibinfo {author} {\bibfnamefont {I.~M.}\ \bibnamefont
  {Romero-Shaw}} \emph {et~al.},\ }\href {\doibase 10.1093/mnras/staa2850}
  {\bibfield  {journal} {\bibinfo  {journal} {Mon. Not. Roy. Astron. Soc.}\
  }\textbf {\bibinfo {volume} {499}},\ \bibinfo {pages} {3295} (\bibinfo {year}
  {2020})},\ \Eprint {http://arxiv.org/abs/2006.00714} {arXiv:2006.00714
  [astro-ph.IM]} \BibitemShut {NoStop}%
\bibitem [{\citenamefont {Ashton}\ and\ \citenamefont
  {Talbot}(2021)}]{Ashton:2021anp}%
  \BibitemOpen
  \bibfield  {author} {\bibinfo {author} {\bibfnamefont {G.}~\bibnamefont
  {Ashton}}\ and\ \bibinfo {author} {\bibfnamefont {C.}~\bibnamefont
  {Talbot}},\ }\href {\doibase 10.1093/mnras/stab2236} {\bibfield  {journal}
  {\bibinfo  {journal} {Mon. Not. Roy. Astron. Soc.}\ }\textbf {\bibinfo
  {volume} {507}},\ \bibinfo {pages} {2037} (\bibinfo {year} {2021})},\ \Eprint
  {http://arxiv.org/abs/2106.08730} {arXiv:2106.08730 [gr-qc]} \BibitemShut
  {NoStop}%
\bibitem [{\citenamefont {Williams}\ \emph {et~al.}(2023)\citenamefont
  {Williams}, \citenamefont {Veitch},\ and\ \citenamefont
  {Messenger}}]{Williams:2023ppp}%
  \BibitemOpen
  \bibfield  {author} {\bibinfo {author} {\bibfnamefont {M.~J.}\ \bibnamefont
  {Williams}}, \bibinfo {author} {\bibfnamefont {J.}~\bibnamefont {Veitch}}, \
  and\ \bibinfo {author} {\bibfnamefont {C.}~\bibnamefont {Messenger}},\
  }\href@noop {} {\  (\bibinfo {year} {2023})},\ \Eprint
  {http://arxiv.org/abs/2302.08526} {arXiv:2302.08526 [astro-ph.IM]}
  \BibitemShut {NoStop}%
\bibitem [{\citenamefont {Iacovelli}\ \emph {et~al.}(2022)\citenamefont
  {Iacovelli}, \citenamefont {Mancarella}, \citenamefont {Foffa},\ and\
  \citenamefont {Maggiore}}]{Iacovelli:2022bbs}%
  \BibitemOpen
  \bibfield  {author} {\bibinfo {author} {\bibfnamefont {F.}~\bibnamefont
  {Iacovelli}}, \bibinfo {author} {\bibfnamefont {M.}~\bibnamefont
  {Mancarella}}, \bibinfo {author} {\bibfnamefont {S.}~\bibnamefont {Foffa}}, \
  and\ \bibinfo {author} {\bibfnamefont {M.}~\bibnamefont {Maggiore}},\ }\href
  {\doibase 10.3847/1538-4357/ac9cd4} {\bibfield  {journal} {\bibinfo
  {journal} {Astrophys. J.}\ }\textbf {\bibinfo {volume} {941}},\ \bibinfo
  {pages} {208} (\bibinfo {year} {2022})},\ \Eprint
  {http://arxiv.org/abs/2207.02771} {arXiv:2207.02771 [gr-qc]} \BibitemShut
  {NoStop}%
\bibitem [{\citenamefont {Pizzati}\ \emph {et~al.}(2022)\citenamefont
  {Pizzati}, \citenamefont {Sachdev}, \citenamefont {Gupta},\ and\
  \citenamefont {Sathyaprakash}}]{Pizzati:2021apa}%
  \BibitemOpen
  \bibfield  {author} {\bibinfo {author} {\bibfnamefont {E.}~\bibnamefont
  {Pizzati}}, \bibinfo {author} {\bibfnamefont {S.}~\bibnamefont {Sachdev}},
  \bibinfo {author} {\bibfnamefont {A.}~\bibnamefont {Gupta}}, \ and\ \bibinfo
  {author} {\bibfnamefont {B.}~\bibnamefont {Sathyaprakash}},\ }\href {\doibase
  10.1103/PhysRevD.105.104016} {\bibfield  {journal} {\bibinfo  {journal}
  {Phys. Rev. D}\ }\textbf {\bibinfo {volume} {105}},\ \bibinfo {pages}
  {104016} (\bibinfo {year} {2022})},\ \Eprint
  {http://arxiv.org/abs/2102.07692} {arXiv:2102.07692 [gr-qc]} \BibitemShut
  {NoStop}%
\bibitem [{\citenamefont {Antonelli}\ \emph {et~al.}(2021)\citenamefont
  {Antonelli}, \citenamefont {Burke},\ and\ \citenamefont
  {Gair}}]{Antonelli:2021vwg}%
  \BibitemOpen
  \bibfield  {author} {\bibinfo {author} {\bibfnamefont {A.}~\bibnamefont
  {Antonelli}}, \bibinfo {author} {\bibfnamefont {O.}~\bibnamefont {Burke}}, \
  and\ \bibinfo {author} {\bibfnamefont {J.~R.}\ \bibnamefont {Gair}},\ }\href
  {\doibase 10.1093/mnras/stab2358} {\bibfield  {journal} {\bibinfo  {journal}
  {Mon. Not. Roy. Astron. Soc.}\ }\textbf {\bibinfo {volume} {507}},\ \bibinfo
  {pages} {5069} (\bibinfo {year} {2021})},\ \Eprint
  {http://arxiv.org/abs/2104.01897} {arXiv:2104.01897 [gr-qc]} \BibitemShut
  {NoStop}%
\bibitem [{\citenamefont {Samajdar}\ \emph {et~al.}(2021)\citenamefont
  {Samajdar}, \citenamefont {Janquart}, \citenamefont {Van Den~Broeck},\ and\
  \citenamefont {Dietrich}}]{Samajdar:2021egv}%
  \BibitemOpen
  \bibfield  {author} {\bibinfo {author} {\bibfnamefont {A.}~\bibnamefont
  {Samajdar}}, \bibinfo {author} {\bibfnamefont {J.}~\bibnamefont {Janquart}},
  \bibinfo {author} {\bibfnamefont {C.}~\bibnamefont {Van Den~Broeck}}, \ and\
  \bibinfo {author} {\bibfnamefont {T.}~\bibnamefont {Dietrich}},\ }\href
  {\doibase 10.1103/PhysRevD.104.044003} {\bibfield  {journal} {\bibinfo
  {journal} {Phys. Rev. D}\ }\textbf {\bibinfo {volume} {104}},\ \bibinfo
  {pages} {044003} (\bibinfo {year} {2021})},\ \Eprint
  {http://arxiv.org/abs/2102.07544} {arXiv:2102.07544 [gr-qc]} \BibitemShut
  {NoStop}%
\bibitem [{\citenamefont {Janquart}\ \emph {et~al.}(2022)\citenamefont
  {Janquart}, \citenamefont {Baka}, \citenamefont {Samajdar}, \citenamefont
  {Dietrich},\ and\ \citenamefont {Van Den~Broeck}}]{Janquart:2022nyz}%
  \BibitemOpen
  \bibfield  {author} {\bibinfo {author} {\bibfnamefont {J.}~\bibnamefont
  {Janquart}}, \bibinfo {author} {\bibfnamefont {T.}~\bibnamefont {Baka}},
  \bibinfo {author} {\bibfnamefont {A.}~\bibnamefont {Samajdar}}, \bibinfo
  {author} {\bibfnamefont {T.}~\bibnamefont {Dietrich}}, \ and\ \bibinfo
  {author} {\bibfnamefont {C.}~\bibnamefont {Van Den~Broeck}},\ }\href@noop {}
  {\  (\bibinfo {year} {2022})},\ \Eprint {http://arxiv.org/abs/2211.01304}
  {arXiv:2211.01304 [gr-qc]} \BibitemShut {NoStop}%
\bibitem [{\citenamefont {Langendorff}\ \emph {et~al.}(2022)\citenamefont
  {Langendorff}, \citenamefont {Kolmus}, \citenamefont {Janquart},\ and\
  \citenamefont {Van Den~Broeck}}]{Langendorff:2022fzq}%
  \BibitemOpen
  \bibfield  {author} {\bibinfo {author} {\bibfnamefont {J.}~\bibnamefont
  {Langendorff}}, \bibinfo {author} {\bibfnamefont {A.}~\bibnamefont {Kolmus}},
  \bibinfo {author} {\bibfnamefont {J.}~\bibnamefont {Janquart}}, \ and\
  \bibinfo {author} {\bibfnamefont {C.}~\bibnamefont {Van Den~Broeck}},\
  }\href@noop {} {\  (\bibinfo {year} {2022})},\ \Eprint
  {http://arxiv.org/abs/2211.15097} {arXiv:2211.15097 [gr-qc]} \BibitemShut
  {NoStop}%
\bibitem [{\citenamefont {Caprini}\ and\ \citenamefont
  {Figueroa}(2018)}]{Caprini:2018mtu}%
  \BibitemOpen
  \bibfield  {author} {\bibinfo {author} {\bibfnamefont {C.}~\bibnamefont
  {Caprini}}\ and\ \bibinfo {author} {\bibfnamefont {D.~G.}\ \bibnamefont
  {Figueroa}},\ }\href {\doibase 10.1088/1361-6382/aac608} {\bibfield
  {journal} {\bibinfo  {journal} {Class. Quant. Grav.}\ }\textbf {\bibinfo
  {volume} {35}},\ \bibinfo {pages} {163001} (\bibinfo {year} {2018})},\
  \Eprint {http://arxiv.org/abs/1801.04268} {arXiv:1801.04268 [astro-ph.CO]}
  \BibitemShut {NoStop}%
\bibitem [{\citenamefont {Christensen}(2019)}]{Christensen:2018iqi}%
  \BibitemOpen
  \bibfield  {author} {\bibinfo {author} {\bibfnamefont {N.}~\bibnamefont
  {Christensen}},\ }\href {\doibase 10.1088/1361-6633/aae6b5} {\bibfield
  {journal} {\bibinfo  {journal} {Rept. Prog. Phys.}\ }\textbf {\bibinfo
  {volume} {82}},\ \bibinfo {pages} {016903} (\bibinfo {year} {2019})},\
  \Eprint {http://arxiv.org/abs/1811.08797} {arXiv:1811.08797 [gr-qc]}
  \BibitemShut {NoStop}%
\bibitem [{\citenamefont {Abbott}\ \emph
  {et~al.}(2019{\natexlab{a}})\citenamefont {Abbott} \emph
  {et~al.}}]{LIGOScientific:2019vic}%
  \BibitemOpen
  \bibfield  {author} {\bibinfo {author} {\bibfnamefont {B.~P.}\ \bibnamefont
  {Abbott}} \emph {et~al.} (\bibinfo {collaboration} {LIGO Scientific,
  Virgo}),\ }\href {\doibase 10.1103/PhysRevD.100.061101} {\bibfield  {journal}
  {\bibinfo  {journal} {Phys. Rev. D}\ }\textbf {\bibinfo {volume} {100}},\
  \bibinfo {pages} {061101} (\bibinfo {year} {2019}{\natexlab{a}})},\ \Eprint
  {http://arxiv.org/abs/1903.02886} {arXiv:1903.02886 [gr-qc]} \BibitemShut
  {NoStop}%
\bibitem [{\citenamefont {Cranmer}\ \emph {et~al.}(2020)\citenamefont
  {Cranmer}, \citenamefont {Brehmer},\ and\ \citenamefont
  {Louppe}}]{Cranmer:2019eaq}%
  \BibitemOpen
  \bibfield  {author} {\bibinfo {author} {\bibfnamefont {K.}~\bibnamefont
  {Cranmer}}, \bibinfo {author} {\bibfnamefont {J.}~\bibnamefont {Brehmer}}, \
  and\ \bibinfo {author} {\bibfnamefont {G.}~\bibnamefont {Louppe}},\ }\href
  {\doibase 10.1073/pnas.1912789117} {\bibfield  {journal} {\bibinfo  {journal}
  {Proc. Nat. Acad. Sci.}\ }\textbf {\bibinfo {volume} {117}},\ \bibinfo
  {pages} {30055} (\bibinfo {year} {2020})},\ \Eprint
  {http://arxiv.org/abs/1911.01429} {arXiv:1911.01429 [stat.ML]} \BibitemShut
  {NoStop}%
\bibitem [{\citenamefont {Brehmer}\ and\ \citenamefont
  {Cranmer}(2020)}]{Brehmer:2020cvb}%
  \BibitemOpen
  \bibfield  {author} {\bibinfo {author} {\bibfnamefont {J.}~\bibnamefont
  {Brehmer}}\ and\ \bibinfo {author} {\bibfnamefont {K.}~\bibnamefont
  {Cranmer}},\ }\href@noop {} {\  (\bibinfo {year} {2020})},\ \Eprint
  {http://arxiv.org/abs/2010.06439} {arXiv:2010.06439 [hep-ph]} \BibitemShut
  {NoStop}%
\bibitem [{\citenamefont {{Lueckmann}}\ \emph {et~al.}(2021)\citenamefont
  {{Lueckmann}}, \citenamefont {{Boelts}}, \citenamefont {{Greenberg}},
  \citenamefont {{Gon{\c{c}}alves}},\ and\ \citenamefont
  {{Macke}}}]{Lueckmann:2021aaa}%
  \BibitemOpen
  \bibfield  {author} {\bibinfo {author} {\bibfnamefont {J.-M.}\ \bibnamefont
  {{Lueckmann}}}, \bibinfo {author} {\bibfnamefont {J.}~\bibnamefont
  {{Boelts}}}, \bibinfo {author} {\bibfnamefont {D.~S.}\ \bibnamefont
  {{Greenberg}}}, \bibinfo {author} {\bibfnamefont {P.~J.}\ \bibnamefont
  {{Gon{\c{c}}alves}}}, \ and\ \bibinfo {author} {\bibfnamefont {J.~H.}\
  \bibnamefont {{Macke}}},\ }\href@noop {} {\  (\bibinfo {year} {2021})},\
  \Eprint {http://arxiv.org/abs/2101.04653} {arXiv:2101.04653 [stat.ML]}
  \BibitemShut {NoStop}%
\bibitem [{\citenamefont {Miller}\ \emph {et~al.}(2021)\citenamefont {Miller},
  \citenamefont {Cole}, \citenamefont {Forr\'e}, \citenamefont {Louppe},\ and\
  \citenamefont {Weniger}}]{Miller:2021hys}%
  \BibitemOpen
  \bibfield  {author} {\bibinfo {author} {\bibfnamefont {B.~K.}\ \bibnamefont
  {Miller}}, \bibinfo {author} {\bibfnamefont {A.}~\bibnamefont {Cole}},
  \bibinfo {author} {\bibfnamefont {P.}~\bibnamefont {Forr\'e}}, \bibinfo
  {author} {\bibfnamefont {G.}~\bibnamefont {Louppe}}, \ and\ \bibinfo {author}
  {\bibfnamefont {C.}~\bibnamefont {Weniger}},\ }in\ \href {\doibase
  10.5281/zenodo.5043706} {\emph {\bibinfo {booktitle} {{35th Conference on
  Neural Information Processing Systems}}}}\ (\bibinfo {year} {2021})\ \Eprint
  {http://arxiv.org/abs/2107.01214} {arXiv:2107.01214 [stat.ML]} \BibitemShut
  {NoStop}%
\bibitem [{\citenamefont {Miller}\ \emph
  {et~al.}(2022{\natexlab{a}})\citenamefont {Miller}, \citenamefont {Cole},
  \citenamefont {Weniger}, \citenamefont {Nattino}, \citenamefont {Ku},\ and\
  \citenamefont {Grootes}}]{Miller:2022shs}%
  \BibitemOpen
  \bibfield  {author} {\bibinfo {author} {\bibfnamefont {B.~K.}\ \bibnamefont
  {Miller}}, \bibinfo {author} {\bibfnamefont {A.}~\bibnamefont {Cole}},
  \bibinfo {author} {\bibfnamefont {C.}~\bibnamefont {Weniger}}, \bibinfo
  {author} {\bibfnamefont {F.}~\bibnamefont {Nattino}}, \bibinfo {author}
  {\bibfnamefont {O.}~\bibnamefont {Ku}}, \ and\ \bibinfo {author}
  {\bibfnamefont {M.~W.}\ \bibnamefont {Grootes}},\ }\href {\doibase
  10.21105/joss.04205} {\bibfield  {journal} {\bibinfo  {journal} {J. Open
  Source Softw.}\ }\textbf {\bibinfo {volume} {7}},\ \bibinfo {pages} {4205}
  (\bibinfo {year} {2022}{\natexlab{a}})}\BibitemShut {NoStop}%
\bibitem [{\citenamefont {Tejero-Cantero}\ \emph {et~al.}(2020)\citenamefont
  {Tejero-Cantero}, \citenamefont {Boelts}, \citenamefont {Deistler},
  \citenamefont {Lueckmann}, \citenamefont {Durkan}, \citenamefont
  {Gonçalves}, \citenamefont {Greenberg},\ and\ \citenamefont
  {Macke}}]{Tejero-cantero:2020aaa}%
  \BibitemOpen
  \bibfield  {author} {\bibinfo {author} {\bibfnamefont {A.}~\bibnamefont
  {Tejero-Cantero}}, \bibinfo {author} {\bibfnamefont {J.}~\bibnamefont
  {Boelts}}, \bibinfo {author} {\bibfnamefont {M.}~\bibnamefont {Deistler}},
  \bibinfo {author} {\bibfnamefont {J.-M.}\ \bibnamefont {Lueckmann}}, \bibinfo
  {author} {\bibfnamefont {C.}~\bibnamefont {Durkan}}, \bibinfo {author}
  {\bibfnamefont {P.~J.}\ \bibnamefont {Gonçalves}}, \bibinfo {author}
  {\bibfnamefont {D.~S.}\ \bibnamefont {Greenberg}}, \ and\ \bibinfo {author}
  {\bibfnamefont {J.~H.}\ \bibnamefont {Macke}},\ }\href {\doibase
  10.21105/joss.02505} {\bibfield  {journal} {\bibinfo  {journal} {Journal of
  Open Source Software}\ }\textbf {\bibinfo {volume} {5}},\ \bibinfo {pages}
  {2505} (\bibinfo {year} {2020})}\BibitemShut {NoStop}%
\bibitem [{\citenamefont {Alsing}\ \emph {et~al.}(2019)\citenamefont {Alsing},
  \citenamefont {Charnock}, \citenamefont {Feeney},\ and\ \citenamefont
  {Wandelt}}]{Alsing:2019xrx}%
  \BibitemOpen
  \bibfield  {author} {\bibinfo {author} {\bibfnamefont {J.}~\bibnamefont
  {Alsing}}, \bibinfo {author} {\bibfnamefont {T.}~\bibnamefont {Charnock}},
  \bibinfo {author} {\bibfnamefont {S.}~\bibnamefont {Feeney}}, \ and\ \bibinfo
  {author} {\bibfnamefont {B.}~\bibnamefont {Wandelt}},\ }\href {\doibase
  10.1093/mnras/stz1960} {\bibfield  {journal} {\bibinfo  {journal} {Mon. Not.
  Roy. Astron. Soc.}\ }\textbf {\bibinfo {volume} {488}},\ \bibinfo {pages}
  {4440} (\bibinfo {year} {2019})},\ \Eprint {http://arxiv.org/abs/1903.00007}
  {arXiv:1903.00007 [astro-ph.CO]} \BibitemShut {NoStop}%
\bibitem [{\citenamefont {Cole}\ \emph {et~al.}(2022)\citenamefont {Cole},
  \citenamefont {Miller}, \citenamefont {Witte}, \citenamefont {Cai},
  \citenamefont {Grootes}, \citenamefont {Nattino},\ and\ \citenamefont
  {Weniger}}]{Cole:2021gwr}%
  \BibitemOpen
  \bibfield  {author} {\bibinfo {author} {\bibfnamefont {A.}~\bibnamefont
  {Cole}}, \bibinfo {author} {\bibfnamefont {B.~K.}\ \bibnamefont {Miller}},
  \bibinfo {author} {\bibfnamefont {S.~J.}\ \bibnamefont {Witte}}, \bibinfo
  {author} {\bibfnamefont {M.~X.}\ \bibnamefont {Cai}}, \bibinfo {author}
  {\bibfnamefont {M.~W.}\ \bibnamefont {Grootes}}, \bibinfo {author}
  {\bibfnamefont {F.}~\bibnamefont {Nattino}}, \ and\ \bibinfo {author}
  {\bibfnamefont {C.}~\bibnamefont {Weniger}},\ }\href {\doibase
  10.1088/1475-7516/2022/09/004} {\bibfield  {journal} {\bibinfo  {journal}
  {JCAP}\ }\textbf {\bibinfo {volume} {09}},\ \bibinfo {pages} {004} (\bibinfo
  {year} {2022})},\ \Eprint {http://arxiv.org/abs/2111.08030} {arXiv:2111.08030
  [astro-ph.CO]} \BibitemShut {NoStop}%
\bibitem [{\citenamefont {Montel}\ \emph {et~al.}(2022)\citenamefont {Montel},
  \citenamefont {Coogan}, \citenamefont {Correa}, \citenamefont {Karchev},\
  and\ \citenamefont {Weniger}}]{Montel:2022fhv}%
  \BibitemOpen
  \bibfield  {author} {\bibinfo {author} {\bibfnamefont {N.~A.}\ \bibnamefont
  {Montel}}, \bibinfo {author} {\bibfnamefont {A.}~\bibnamefont {Coogan}},
  \bibinfo {author} {\bibfnamefont {C.}~\bibnamefont {Correa}}, \bibinfo
  {author} {\bibfnamefont {K.}~\bibnamefont {Karchev}}, \ and\ \bibinfo
  {author} {\bibfnamefont {C.}~\bibnamefont {Weniger}},\ }\href {\doibase
  10.1093/mnras/stac3215} {\bibfield  {journal} {\bibinfo  {journal} {Mon. Not.
  Roy. Astron. Soc.}\ }\textbf {\bibinfo {volume} {518}},\ \bibinfo {pages}
  {2746} (\bibinfo {year} {2022})},\ \Eprint {http://arxiv.org/abs/2205.09126}
  {arXiv:2205.09126 [astro-ph.CO]} \BibitemShut {NoStop}%
\bibitem [{\citenamefont {Anau~Montel}\ and\ \citenamefont
  {Weniger}(2022)}]{AnauMontel:2022ppb}%
  \BibitemOpen
  \bibfield  {author} {\bibinfo {author} {\bibfnamefont {N.}~\bibnamefont
  {Anau~Montel}}\ and\ \bibinfo {author} {\bibfnamefont {C.}~\bibnamefont
  {Weniger}},\ }in\ \href@noop {} {\emph {\bibinfo {booktitle} {{36th
  Conference on Neural Information Processing Systems}}}}\ (\bibinfo {year}
  {2022})\ \Eprint {http://arxiv.org/abs/2211.04291} {arXiv:2211.04291
  [astro-ph.IM]} \BibitemShut {NoStop}%
\bibitem [{\citenamefont {Makinen}\ \emph {et~al.}(2021)\citenamefont
  {Makinen}, \citenamefont {Charnock}, \citenamefont {Alsing},\ and\
  \citenamefont {Wandelt}}]{Makinen:2021nly}%
  \BibitemOpen
  \bibfield  {author} {\bibinfo {author} {\bibfnamefont {T.~L.}\ \bibnamefont
  {Makinen}}, \bibinfo {author} {\bibfnamefont {T.}~\bibnamefont {Charnock}},
  \bibinfo {author} {\bibfnamefont {J.}~\bibnamefont {Alsing}}, \ and\ \bibinfo
  {author} {\bibfnamefont {B.~D.}\ \bibnamefont {Wandelt}},\ }\href {\doibase
  10.1088/1475-7516/2021/11/049} {\bibfield  {journal} {\bibinfo  {journal}
  {JCAP}\ }\textbf {\bibinfo {volume} {11}},\ \bibinfo {pages} {049} (\bibinfo
  {year} {2021})},\ \Eprint {http://arxiv.org/abs/2107.07405} {arXiv:2107.07405
  [astro-ph.CO]} \BibitemShut {NoStop}%
\bibitem [{\citenamefont {Dimitriou}\ \emph {et~al.}(2022)\citenamefont
  {Dimitriou}, \citenamefont {Weniger},\ and\ \citenamefont
  {Correa}}]{Dimitriou:2022cvc}%
  \BibitemOpen
  \bibfield  {author} {\bibinfo {author} {\bibfnamefont {A.}~\bibnamefont
  {Dimitriou}}, \bibinfo {author} {\bibfnamefont {C.}~\bibnamefont {Weniger}},
  \ and\ \bibinfo {author} {\bibfnamefont {C.~A.}\ \bibnamefont {Correa}},\
  }\href@noop {} {\  (\bibinfo {year} {2022})},\ \Eprint
  {http://arxiv.org/abs/2206.11312} {arXiv:2206.11312 [astro-ph.CO]}
  \BibitemShut {NoStop}%
\bibitem [{\citenamefont {Gagnon-Hartman}\ \emph {et~al.}(2023)\citenamefont
  {Gagnon-Hartman}, \citenamefont {Ruan},\ and\ \citenamefont
  {Haggard}}]{Gagnon-Hartman:2023soa}%
  \BibitemOpen
  \bibfield  {author} {\bibinfo {author} {\bibfnamefont {S.}~\bibnamefont
  {Gagnon-Hartman}}, \bibinfo {author} {\bibfnamefont {J.}~\bibnamefont
  {Ruan}}, \ and\ \bibinfo {author} {\bibfnamefont {D.}~\bibnamefont
  {Haggard}},\ }\href {\doibase 10.1093/mnras/stad069} {\bibfield  {journal}
  {\bibinfo  {journal} {Mon. Not. Roy. Astron. Soc.}\ }\textbf {\bibinfo
  {volume} {520}},\ \bibinfo {pages} {1} (\bibinfo {year} {2023})},\ \Eprint
  {http://arxiv.org/abs/2301.05241} {arXiv:2301.05241 [astro-ph.CO]}
  \BibitemShut {NoStop}%
\bibitem [{\citenamefont {Delaunoy}\ \emph {et~al.}(2020)\citenamefont
  {Delaunoy}, \citenamefont {Wehenkel}, \citenamefont {Hinderer}, \citenamefont
  {Nissanke}, \citenamefont {Weniger}, \citenamefont {Williamson},\ and\
  \citenamefont {Louppe}}]{Delaunoy:2020zcu}%
  \BibitemOpen
  \bibfield  {author} {\bibinfo {author} {\bibfnamefont {A.}~\bibnamefont
  {Delaunoy}}, \bibinfo {author} {\bibfnamefont {A.}~\bibnamefont {Wehenkel}},
  \bibinfo {author} {\bibfnamefont {T.}~\bibnamefont {Hinderer}}, \bibinfo
  {author} {\bibfnamefont {S.}~\bibnamefont {Nissanke}}, \bibinfo {author}
  {\bibfnamefont {C.}~\bibnamefont {Weniger}}, \bibinfo {author} {\bibfnamefont
  {A.~R.}\ \bibnamefont {Williamson}}, \ and\ \bibinfo {author} {\bibfnamefont
  {G.}~\bibnamefont {Louppe}},\ }\href@noop {} {\  (\bibinfo {year} {2020})},\
  \Eprint {http://arxiv.org/abs/2010.12931} {arXiv:2010.12931 [astro-ph.IM]}
  \BibitemShut {NoStop}%
\bibitem [{\citenamefont {Karchev}\ \emph {et~al.}(2022)\citenamefont
  {Karchev}, \citenamefont {Trotta},\ and\ \citenamefont
  {Weniger}}]{Karchev:2022xyn}%
  \BibitemOpen
  \bibfield  {author} {\bibinfo {author} {\bibfnamefont {K.}~\bibnamefont
  {Karchev}}, \bibinfo {author} {\bibfnamefont {R.}~\bibnamefont {Trotta}}, \
  and\ \bibinfo {author} {\bibfnamefont {C.}~\bibnamefont {Weniger}},\ }\href
  {\doibase 10.1093/mnras/stac3785} {\  (\bibinfo {year} {2022}),\
  10.1093/mnras/stac3785},\ \Eprint {http://arxiv.org/abs/2209.06733}
  {arXiv:2209.06733 [astro-ph.CO]} \BibitemShut {NoStop}%
\bibitem [{\citenamefont {Lin}\ \emph {et~al.}(2022)\citenamefont {Lin},
  \citenamefont {von Wietersheim-Kramsta}, \citenamefont {Joachimi},\ and\
  \citenamefont {Feeney}}]{Lin:2022ayr}%
  \BibitemOpen
  \bibfield  {author} {\bibinfo {author} {\bibfnamefont {K.}~\bibnamefont
  {Lin}}, \bibinfo {author} {\bibfnamefont {M.}~\bibnamefont {von
  Wietersheim-Kramsta}}, \bibinfo {author} {\bibfnamefont {B.}~\bibnamefont
  {Joachimi}}, \ and\ \bibinfo {author} {\bibfnamefont {S.}~\bibnamefont
  {Feeney}},\ }\href@noop {} {\  (\bibinfo {year} {2022})},\ \Eprint
  {http://arxiv.org/abs/2212.04521} {arXiv:2212.04521 [astro-ph.CO]}
  \BibitemShut {NoStop}%
\bibitem [{\citenamefont {Dax}\ \emph {et~al.}(2021{\natexlab{a}})\citenamefont
  {Dax}, \citenamefont {Green}, \citenamefont {Gair}, \citenamefont {Deistler},
  \citenamefont {Sch\"olkopf},\ and\ \citenamefont {Macke}}]{Dax:2021myb}%
  \BibitemOpen
  \bibfield  {author} {\bibinfo {author} {\bibfnamefont {M.}~\bibnamefont
  {Dax}}, \bibinfo {author} {\bibfnamefont {S.~R.}\ \bibnamefont {Green}},
  \bibinfo {author} {\bibfnamefont {J.}~\bibnamefont {Gair}}, \bibinfo {author}
  {\bibfnamefont {M.}~\bibnamefont {Deistler}}, \bibinfo {author}
  {\bibfnamefont {B.}~\bibnamefont {Sch\"olkopf}}, \ and\ \bibinfo {author}
  {\bibfnamefont {J.~H.}\ \bibnamefont {Macke}},\ }\href@noop {} {\  (\bibinfo
  {year} {2021}{\natexlab{a}})},\ \Eprint {http://arxiv.org/abs/2111.13139}
  {arXiv:2111.13139 [cs.LG]} \BibitemShut {NoStop}%
\bibitem [{\citenamefont {Wildberger}\ \emph {et~al.}(2022)\citenamefont
  {Wildberger}, \citenamefont {Dax}, \citenamefont {Green}, \citenamefont
  {Gair}, \citenamefont {P\"urrer}, \citenamefont {Macke}, \citenamefont
  {Buonanno},\ and\ \citenamefont {Sch\"olkopf}}]{Wildberger:2022agw}%
  \BibitemOpen
  \bibfield  {author} {\bibinfo {author} {\bibfnamefont {J.}~\bibnamefont
  {Wildberger}}, \bibinfo {author} {\bibfnamefont {M.}~\bibnamefont {Dax}},
  \bibinfo {author} {\bibfnamefont {S.~R.}\ \bibnamefont {Green}}, \bibinfo
  {author} {\bibfnamefont {J.}~\bibnamefont {Gair}}, \bibinfo {author}
  {\bibfnamefont {M.}~\bibnamefont {P\"urrer}}, \bibinfo {author}
  {\bibfnamefont {J.~H.}\ \bibnamefont {Macke}}, \bibinfo {author}
  {\bibfnamefont {A.}~\bibnamefont {Buonanno}}, \ and\ \bibinfo {author}
  {\bibfnamefont {B.}~\bibnamefont {Sch\"olkopf}},\ }\href@noop {} {\
  (\bibinfo {year} {2022})},\ \Eprint {http://arxiv.org/abs/2211.08801}
  {arXiv:2211.08801 [gr-qc]} \BibitemShut {NoStop}%
\bibitem [{\citenamefont {Dax}\ \emph {et~al.}(2021{\natexlab{b}})\citenamefont
  {Dax}, \citenamefont {Green}, \citenamefont {Gair}, \citenamefont {Macke},
  \citenamefont {Buonanno},\ and\ \citenamefont {Sch\"olkopf}}]{Dax:2021tsq}%
  \BibitemOpen
  \bibfield  {author} {\bibinfo {author} {\bibfnamefont {M.}~\bibnamefont
  {Dax}}, \bibinfo {author} {\bibfnamefont {S.~R.}\ \bibnamefont {Green}},
  \bibinfo {author} {\bibfnamefont {J.}~\bibnamefont {Gair}}, \bibinfo {author}
  {\bibfnamefont {J.~H.}\ \bibnamefont {Macke}}, \bibinfo {author}
  {\bibfnamefont {A.}~\bibnamefont {Buonanno}}, \ and\ \bibinfo {author}
  {\bibfnamefont {B.}~\bibnamefont {Sch\"olkopf}},\ }\href {\doibase
  10.1103/PhysRevLett.127.241103} {\bibfield  {journal} {\bibinfo  {journal}
  {Phys. Rev. Lett.}\ }\textbf {\bibinfo {volume} {127}},\ \bibinfo {pages}
  {241103} (\bibinfo {year} {2021}{\natexlab{b}})},\ \Eprint
  {http://arxiv.org/abs/2106.12594} {arXiv:2106.12594 [gr-qc]} \BibitemShut
  {NoStop}%
\bibitem [{\citenamefont {Pratten}\ \emph {et~al.}(2021)\citenamefont {Pratten}
  \emph {et~al.}}]{Pratten:2020ceb}%
  \BibitemOpen
  \bibfield  {author} {\bibinfo {author} {\bibfnamefont {G.}~\bibnamefont
  {Pratten}} \emph {et~al.},\ }\href {\doibase 10.1103/PhysRevD.103.104056}
  {\bibfield  {journal} {\bibinfo  {journal} {Phys. Rev. D}\ }\textbf {\bibinfo
  {volume} {103}},\ \bibinfo {pages} {104056} (\bibinfo {year} {2021})},\
  \Eprint {http://arxiv.org/abs/2004.06503} {arXiv:2004.06503 [gr-qc]}
  \BibitemShut {NoStop}%
\bibitem [{\citenamefont {Hannam}\ \emph {et~al.}(2014)\citenamefont {Hannam},
  \citenamefont {Schmidt}, \citenamefont {Boh\'e}, \citenamefont {Haegel},
  \citenamefont {Husa}, \citenamefont {Ohme}, \citenamefont {Pratten},\ and\
  \citenamefont {P\"urrer}}]{Hannam:2013oca}%
  \BibitemOpen
  \bibfield  {author} {\bibinfo {author} {\bibfnamefont {M.}~\bibnamefont
  {Hannam}}, \bibinfo {author} {\bibfnamefont {P.}~\bibnamefont {Schmidt}},
  \bibinfo {author} {\bibfnamefont {A.}~\bibnamefont {Boh\'e}}, \bibinfo
  {author} {\bibfnamefont {L.}~\bibnamefont {Haegel}}, \bibinfo {author}
  {\bibfnamefont {S.}~\bibnamefont {Husa}}, \bibinfo {author} {\bibfnamefont
  {F.}~\bibnamefont {Ohme}}, \bibinfo {author} {\bibfnamefont {G.}~\bibnamefont
  {Pratten}}, \ and\ \bibinfo {author} {\bibfnamefont {M.}~\bibnamefont
  {P\"urrer}},\ }\href {\doibase 10.1103/PhysRevLett.113.151101} {\bibfield
  {journal} {\bibinfo  {journal} {Phys. Rev. Lett.}\ }\textbf {\bibinfo
  {volume} {113}},\ \bibinfo {pages} {151101} (\bibinfo {year} {2014})},\
  \Eprint {http://arxiv.org/abs/1308.3271} {arXiv:1308.3271 [gr-qc]}
  \BibitemShut {NoStop}%
\bibitem [{\citenamefont {Babak}\ \emph {et~al.}(2010)\citenamefont {Babak}
  \emph {et~al.}}]{MockLISADataChallengeTaskForce:2009wir}%
  \BibitemOpen
  \bibfield  {author} {\bibinfo {author} {\bibfnamefont {S.}~\bibnamefont
  {Babak}} \emph {et~al.} (\bibinfo {collaboration} {Mock LISA Data Challenge
  Task Force}),\ }\href {\doibase 10.1088/0264-9381/27/8/084009} {\bibfield
  {journal} {\bibinfo  {journal} {Class. Quant. Grav.}\ }\textbf {\bibinfo
  {volume} {27}},\ \bibinfo {pages} {084009} (\bibinfo {year} {2010})},\
  \Eprint {http://arxiv.org/abs/0912.0548} {arXiv:0912.0548 [gr-qc]}
  \BibitemShut {NoStop}%
\bibitem [{\citenamefont {Sch\"afer}\ \emph {et~al.}(2023)\citenamefont
  {Sch\"afer} \emph {et~al.}}]{Schafer:2022dxv}%
  \BibitemOpen
  \bibfield  {author} {\bibinfo {author} {\bibfnamefont {M.~B.}\ \bibnamefont
  {Sch\"afer}} \emph {et~al.},\ }\href {\doibase 10.1103/PhysRevD.107.023021}
  {\bibfield  {journal} {\bibinfo  {journal} {Phys. Rev. D}\ }\textbf {\bibinfo
  {volume} {107}},\ \bibinfo {pages} {023021} (\bibinfo {year} {2023})},\
  \Eprint {http://arxiv.org/abs/2209.11146} {arXiv:2209.11146 [astro-ph.IM]}
  \BibitemShut {NoStop}%
\bibitem [{\citenamefont {{Papamakarios}}\ and\ \citenamefont
  {{Murray}}(2016)}]{Papamakarios:2016aaa}%
  \BibitemOpen
  \bibfield  {author} {\bibinfo {author} {\bibfnamefont {G.}~\bibnamefont
  {{Papamakarios}}}\ and\ \bibinfo {author} {\bibfnamefont {I.}~\bibnamefont
  {{Murray}}},\ }\href@noop {} {\  (\bibinfo {year} {2016})},\ \Eprint
  {http://arxiv.org/abs/1605.06376} {arXiv:1605.06376 [stat.ML]} \BibitemShut
  {NoStop}%
\bibitem [{\citenamefont {{Zeghal}}\ \emph {et~al.}(2022)\citenamefont
  {{Zeghal}}, \citenamefont {{Lanusse}}, \citenamefont {{Boucaud}},
  \citenamefont {{Remy}},\ and\ \citenamefont {{Aubourg}}}]{Zeghal:2022aaa}%
  \BibitemOpen
  \bibfield  {author} {\bibinfo {author} {\bibfnamefont {J.}~\bibnamefont
  {{Zeghal}}}, \bibinfo {author} {\bibfnamefont {F.}~\bibnamefont {{Lanusse}}},
  \bibinfo {author} {\bibfnamefont {A.}~\bibnamefont {{Boucaud}}}, \bibinfo
  {author} {\bibfnamefont {B.}~\bibnamefont {{Remy}}}, \ and\ \bibinfo {author}
  {\bibfnamefont {E.}~\bibnamefont {{Aubourg}}},\ }\href@noop {} {\  (\bibinfo
  {year} {2022})},\ \Eprint {http://arxiv.org/abs/2207.05636} {arXiv:2207.05636
  [astro-ph.IM]} \BibitemShut {NoStop}%
\bibitem [{\citenamefont {Dax}\ \emph {et~al.}(2022)\citenamefont {Dax},
  \citenamefont {Green}, \citenamefont {Gair}, \citenamefont {P\"urrer},
  \citenamefont {Wildberger}, \citenamefont {Macke}, \citenamefont {Buonanno},\
  and\ \citenamefont {Sch\"olkopf}}]{Dax:2022pxd}%
  \BibitemOpen
  \bibfield  {author} {\bibinfo {author} {\bibfnamefont {M.}~\bibnamefont
  {Dax}}, \bibinfo {author} {\bibfnamefont {S.~R.}\ \bibnamefont {Green}},
  \bibinfo {author} {\bibfnamefont {J.}~\bibnamefont {Gair}}, \bibinfo {author}
  {\bibfnamefont {M.}~\bibnamefont {P\"urrer}}, \bibinfo {author}
  {\bibfnamefont {J.}~\bibnamefont {Wildberger}}, \bibinfo {author}
  {\bibfnamefont {J.~H.}\ \bibnamefont {Macke}}, \bibinfo {author}
  {\bibfnamefont {A.}~\bibnamefont {Buonanno}}, \ and\ \bibinfo {author}
  {\bibfnamefont {B.}~\bibnamefont {Sch\"olkopf}},\ }\href@noop {} {\
  (\bibinfo {year} {2022})},\ \Eprint {http://arxiv.org/abs/2210.05686}
  {arXiv:2210.05686 [gr-qc]} \BibitemShut {NoStop}%
\bibitem [{\citenamefont {{Rozet}}\ and\ \citenamefont
  {{Louppe}}(2021)}]{Rozet:2022aaa}%
  \BibitemOpen
  \bibfield  {author} {\bibinfo {author} {\bibfnamefont {F.}~\bibnamefont
  {{Rozet}}}\ and\ \bibinfo {author} {\bibfnamefont {G.}~\bibnamefont
  {{Louppe}}},\ }\href@noop {} {\  (\bibinfo {year} {2021})},\ \Eprint
  {http://arxiv.org/abs/2110.00449} {arXiv:2110.00449 [cs.LG]} \BibitemShut
  {NoStop}%
\bibitem [{\citenamefont {{Delaunoy}}\ \emph {et~al.}(2022)\citenamefont
  {{Delaunoy}}, \citenamefont {{Hermans}}, \citenamefont {{Rozet}},
  \citenamefont {{Wehenkel}},\ and\ \citenamefont
  {{Louppe}}}]{Delaunoy:2022aaa}%
  \BibitemOpen
  \bibfield  {author} {\bibinfo {author} {\bibfnamefont {A.}~\bibnamefont
  {{Delaunoy}}}, \bibinfo {author} {\bibfnamefont {J.}~\bibnamefont
  {{Hermans}}}, \bibinfo {author} {\bibfnamefont {F.}~\bibnamefont {{Rozet}}},
  \bibinfo {author} {\bibfnamefont {A.}~\bibnamefont {{Wehenkel}}}, \ and\
  \bibinfo {author} {\bibfnamefont {G.}~\bibnamefont {{Louppe}}},\ }\href@noop
  {} {\  (\bibinfo {year} {2022})},\ \Eprint {http://arxiv.org/abs/2208.13624}
  {arXiv:2208.13624 [stat.ML]} \BibitemShut {NoStop}%
\bibitem [{\citenamefont {Miller}\ \emph
  {et~al.}(2022{\natexlab{b}})\citenamefont {Miller}, \citenamefont {Weniger},\
  and\ \citenamefont {Forr\'e}}]{Miller:2022haf}%
  \BibitemOpen
  \bibfield  {author} {\bibinfo {author} {\bibfnamefont {B.~K.}\ \bibnamefont
  {Miller}}, \bibinfo {author} {\bibfnamefont {C.}~\bibnamefont {Weniger}}, \
  and\ \bibinfo {author} {\bibfnamefont {P.}~\bibnamefont {Forr\'e}},\
  }\href@noop {} {\  (\bibinfo {year} {2022}{\natexlab{b}})},\ \Eprint
  {http://arxiv.org/abs/2210.06170} {arXiv:2210.06170 [stat.ML]} \BibitemShut
  {NoStop}%
\bibitem [{\citenamefont {{Hermans}}\ \emph {et~al.}(2019)\citenamefont
  {{Hermans}}, \citenamefont {{Begy}},\ and\ \citenamefont
  {{Louppe}}}]{Hermans:2019aaa}%
  \BibitemOpen
  \bibfield  {author} {\bibinfo {author} {\bibfnamefont {J.}~\bibnamefont
  {{Hermans}}}, \bibinfo {author} {\bibfnamefont {V.}~\bibnamefont {{Begy}}}, \
  and\ \bibinfo {author} {\bibfnamefont {G.}~\bibnamefont {{Louppe}}},\
  }\href@noop {} {\  (\bibinfo {year} {2019})},\ \Eprint
  {http://arxiv.org/abs/1903.04057} {arXiv:1903.04057 [stat.ML]} \BibitemShut
  {NoStop}%
\bibitem [{\citenamefont {{Durkan}}\ \emph {et~al.}(2020)\citenamefont
  {{Durkan}}, \citenamefont {{Murray}},\ and\ \citenamefont
  {{Papamakarios}}}]{Durkan:2020aaa}%
  \BibitemOpen
  \bibfield  {author} {\bibinfo {author} {\bibfnamefont {C.}~\bibnamefont
  {{Durkan}}}, \bibinfo {author} {\bibfnamefont {I.}~\bibnamefont {{Murray}}},
  \ and\ \bibinfo {author} {\bibfnamefont {G.}~\bibnamefont {{Papamakarios}}},\
  }\href@noop {} {\  (\bibinfo {year} {2020})},\ \Eprint
  {http://arxiv.org/abs/2002.03712} {arXiv:2002.03712 [stat.ML]} \BibitemShut
  {NoStop}%
\bibitem [{\citenamefont {{Ronneberger}}\ \emph {et~al.}(2015)\citenamefont
  {{Ronneberger}}, \citenamefont {{Fischer}},\ and\ \citenamefont
  {{Brox}}}]{Ronneberger:2015aaa}%
  \BibitemOpen
  \bibfield  {author} {\bibinfo {author} {\bibfnamefont {O.}~\bibnamefont
  {{Ronneberger}}}, \bibinfo {author} {\bibfnamefont {P.}~\bibnamefont
  {{Fischer}}}, \ and\ \bibinfo {author} {\bibfnamefont {T.}~\bibnamefont
  {{Brox}}},\ }\href@noop {} {\  (\bibinfo {year} {2015})},\ \Eprint
  {http://arxiv.org/abs/1505.04597} {arXiv:1505.04597 [cs.CV]} \BibitemShut
  {NoStop}%
\bibitem [{\citenamefont {Sunnaker}\ \emph {et~al.}(2013)\citenamefont
  {Sunnaker}, \citenamefont {Busetto}, \citenamefont {Numminen}, \citenamefont
  {Corander}, \citenamefont {Foll},\ and\ \citenamefont
  {Dessimoz}}]{Sunnaker:2013aaa}%
  \BibitemOpen
  \bibfield  {author} {\bibinfo {author} {\bibfnamefont {M.}~\bibnamefont
  {Sunnaker}}, \bibinfo {author} {\bibfnamefont {A.}~\bibnamefont {Busetto}},
  \bibinfo {author} {\bibfnamefont {E.}~\bibnamefont {Numminen}}, \bibinfo
  {author} {\bibfnamefont {J.}~\bibnamefont {Corander}}, \bibinfo {author}
  {\bibfnamefont {M.}~\bibnamefont {Foll}}, \ and\ \bibinfo {author}
  {\bibfnamefont {C.}~\bibnamefont {Dessimoz}},\ }\href {\doibase
  10.1371/journal.pcbi.1002803} {\bibfield  {journal} {\bibinfo  {journal}
  {PLoS Comput Biol}\ }\textbf {\bibinfo {volume} {9}},\ \bibinfo {pages}
  {e1002803} (\bibinfo {year} {2013})}\BibitemShut {NoStop}%
\bibitem [{\citenamefont {Abbott}\ \emph {et~al.}(2020)\citenamefont {Abbott}
  \emph {et~al.}}]{LIGOScientific:2019hgc}%
  \BibitemOpen
  \bibfield  {author} {\bibinfo {author} {\bibfnamefont {B.~P.}\ \bibnamefont
  {Abbott}} \emph {et~al.} (\bibinfo {collaboration} {LIGO Scientific,
  Virgo}),\ }\href {\doibase 10.1088/1361-6382/ab685e} {\bibfield  {journal}
  {\bibinfo  {journal} {Class. Quant. Grav.}\ }\textbf {\bibinfo {volume}
  {37}},\ \bibinfo {pages} {055002} (\bibinfo {year} {2020})},\ \Eprint
  {http://arxiv.org/abs/1908.11170} {arXiv:1908.11170 [gr-qc]} \BibitemShut
  {NoStop}%
\bibitem [{\citenamefont {Abbott}\ \emph
  {et~al.}(2019{\natexlab{b}})\citenamefont {Abbott} \emph
  {et~al.}}]{LIGOScientific:2018mvr}%
  \BibitemOpen
  \bibfield  {author} {\bibinfo {author} {\bibfnamefont {B.~P.}\ \bibnamefont
  {Abbott}} \emph {et~al.} (\bibinfo {collaboration} {LIGO Scientific,
  Virgo}),\ }\href {\doibase 10.1103/PhysRevX.9.031040} {\bibfield  {journal}
  {\bibinfo  {journal} {Phys. Rev. X}\ }\textbf {\bibinfo {volume} {9}},\
  \bibinfo {pages} {031040} (\bibinfo {year} {2019}{\natexlab{b}})},\ \Eprint
  {http://arxiv.org/abs/1811.12907} {arXiv:1811.12907 [astro-ph.HE]}
  \BibitemShut {NoStop}%
\bibitem [{\citenamefont {{Thrane}}\ and\ \citenamefont
  {{Talbot}}(2019{\natexlab{b}})}]{Thrane:2018aaa}%
  \BibitemOpen
  \bibfield  {author} {\bibinfo {author} {\bibfnamefont {E.}~\bibnamefont
  {{Thrane}}}\ and\ \bibinfo {author} {\bibfnamefont {C.}~\bibnamefont
  {{Talbot}}},\ }\href {\doibase 10.1017/pasa.2019.2} {\bibfield  {journal}
  {\bibinfo  {journal} {Publications of the Astronomical Society of Australia}\
  }\textbf {\bibinfo {volume} {36}},\ \bibinfo {eid} {e010} (\bibinfo {year}
  {2019}{\natexlab{b}})},\ \Eprint {http://arxiv.org/abs/1809.02293}
  {arXiv:1809.02293 [astro-ph.IM]} \BibitemShut {NoStop}%
\bibitem [{\citenamefont {Hannam}\ \emph {et~al.}(2022)\citenamefont {Hannam}
  \emph {et~al.}}]{Hannam:2021pit}%
  \BibitemOpen
  \bibfield  {author} {\bibinfo {author} {\bibfnamefont {M.}~\bibnamefont
  {Hannam}} \emph {et~al.},\ }\href {\doibase 10.1038/s41586-022-05212-z}
  {\bibfield  {journal} {\bibinfo  {journal} {Nature}\ }\textbf {\bibinfo
  {volume} {610}},\ \bibinfo {pages} {652} (\bibinfo {year} {2022})},\ \Eprint
  {http://arxiv.org/abs/2112.11300} {arXiv:2112.11300 [gr-qc]} \BibitemShut
  {NoStop}%
\bibitem [{\citenamefont {Payne}\ \emph {et~al.}(2022)\citenamefont {Payne},
  \citenamefont {Hourihane}, \citenamefont {Golomb}, \citenamefont {Udall},
  \citenamefont {Udall}, \citenamefont {Davis},\ and\ \citenamefont
  {Chatziioannou}}]{Payne:2022spz}%
  \BibitemOpen
  \bibfield  {author} {\bibinfo {author} {\bibfnamefont {E.}~\bibnamefont
  {Payne}}, \bibinfo {author} {\bibfnamefont {S.}~\bibnamefont {Hourihane}},
  \bibinfo {author} {\bibfnamefont {J.}~\bibnamefont {Golomb}}, \bibinfo
  {author} {\bibfnamefont {R.}~\bibnamefont {Udall}}, \bibinfo {author}
  {\bibfnamefont {R.}~\bibnamefont {Udall}}, \bibinfo {author} {\bibfnamefont
  {D.}~\bibnamefont {Davis}}, \ and\ \bibinfo {author} {\bibfnamefont
  {K.}~\bibnamefont {Chatziioannou}},\ }\href {\doibase
  10.1103/PhysRevD.106.104017} {\bibfield  {journal} {\bibinfo  {journal}
  {Phys. Rev. D}\ }\textbf {\bibinfo {volume} {106}},\ \bibinfo {pages}
  {104017} (\bibinfo {year} {2022})},\ \Eprint
  {http://arxiv.org/abs/2206.11932} {arXiv:2206.11932 [gr-qc]} \BibitemShut
  {NoStop}%
\bibitem [{\citenamefont {Husa}\ \emph {et~al.}(2016)\citenamefont {Husa},
  \citenamefont {Khan}, \citenamefont {Hannam}, \citenamefont {P\"urrer},
  \citenamefont {Ohme}, \citenamefont {Forteza},\ and\ \citenamefont
  {Boh\'e}}]{Husa:2016aa}%
  \BibitemOpen
  \bibfield  {author} {\bibinfo {author} {\bibfnamefont {S.}~\bibnamefont
  {Husa}}, \bibinfo {author} {\bibfnamefont {S.}~\bibnamefont {Khan}}, \bibinfo
  {author} {\bibfnamefont {M.}~\bibnamefont {Hannam}}, \bibinfo {author}
  {\bibfnamefont {M.}~\bibnamefont {P\"urrer}}, \bibinfo {author}
  {\bibfnamefont {F.}~\bibnamefont {Ohme}}, \bibinfo {author} {\bibfnamefont
  {X.~J.}\ \bibnamefont {Forteza}}, \ and\ \bibinfo {author} {\bibfnamefont
  {A.}~\bibnamefont {Boh\'e}},\ }\href {\doibase 10.1103/PhysRevD.93.044006}
  {\bibfield  {journal} {\bibinfo  {journal} {Phys. Rev. D}\ }\textbf {\bibinfo
  {volume} {93}},\ \bibinfo {pages} {044006} (\bibinfo {year}
  {2016})}\BibitemShut {NoStop}%
\bibitem [{\citenamefont {Romero-Shaw}\ \emph {et~al.}(2023)\citenamefont
  {Romero-Shaw}, \citenamefont {Gerosa},\ and\ \citenamefont
  {Loutrel}}]{Romero-Shaw:2022fbf}%
  \BibitemOpen
  \bibfield  {author} {\bibinfo {author} {\bibfnamefont {I.~M.}\ \bibnamefont
  {Romero-Shaw}}, \bibinfo {author} {\bibfnamefont {D.}~\bibnamefont {Gerosa}},
  \ and\ \bibinfo {author} {\bibfnamefont {N.}~\bibnamefont {Loutrel}},\ }\href
  {\doibase 10.1093/mnras/stad031} {\bibfield  {journal} {\bibinfo  {journal}
  {Mon. Not. Roy. Astron. Soc.}\ }\textbf {\bibinfo {volume} {519}},\ \bibinfo
  {pages} {5352} (\bibinfo {year} {2023})},\ \Eprint
  {http://arxiv.org/abs/2211.07528} {arXiv:2211.07528 [astro-ph.HE]}
  \BibitemShut {NoStop}%
\bibitem [{\citenamefont {Xu}\ and\ \citenamefont
  {Hamilton}(2022)}]{Xu:2022spd}%
  \BibitemOpen
  \bibfield  {author} {\bibinfo {author} {\bibfnamefont {Y.}~\bibnamefont
  {Xu}}\ and\ \bibinfo {author} {\bibfnamefont {E.}~\bibnamefont {Hamilton}},\
  }\href@noop {} {\  (\bibinfo {year} {2022})},\ \Eprint
  {http://arxiv.org/abs/2211.09561} {arXiv:2211.09561 [gr-qc]} \BibitemShut
  {NoStop}%
\bibitem [{\citenamefont {Veitch}\ \emph {et~al.}(2021)\citenamefont {Veitch},
  \citenamefont {Pozzo}, \citenamefont {Lyttle}, \citenamefont {Williams},
  \citenamefont {Talbot}, \citenamefont {Pitkin}, \citenamefont {Ashton},
  \citenamefont {Cody}, \citenamefont {Hübner}, \citenamefont {Nitz},
  \citenamefont {Macleod}, \citenamefont {Carullo}, \citenamefont {Davies},\
  and\ \citenamefont {Tony}}]{Veitch:2021aaa}%
  \BibitemOpen
  \bibfield  {author} {\bibinfo {author} {\bibfnamefont {J.}~\bibnamefont
  {Veitch}}, \bibinfo {author} {\bibfnamefont {W.~D.}\ \bibnamefont {Pozzo}},
  \bibinfo {author} {\bibfnamefont {A.}~\bibnamefont {Lyttle}}, \bibinfo
  {author} {\bibfnamefont {M.}~\bibnamefont {Williams}}, \bibinfo {author}
  {\bibfnamefont {C.}~\bibnamefont {Talbot}}, \bibinfo {author} {\bibfnamefont
  {M.}~\bibnamefont {Pitkin}}, \bibinfo {author} {\bibfnamefont
  {G.}~\bibnamefont {Ashton}}, \bibinfo {author} {\bibnamefont {Cody}},
  \bibinfo {author} {\bibfnamefont {M.}~\bibnamefont {Hübner}}, \bibinfo
  {author} {\bibfnamefont {A.}~\bibnamefont {Nitz}}, \bibinfo {author}
  {\bibfnamefont {D.}~\bibnamefont {Macleod}}, \bibinfo {author} {\bibfnamefont
  {G.}~\bibnamefont {Carullo}}, \bibinfo {author} {\bibfnamefont
  {G.}~\bibnamefont {Davies}}, \ and\ \bibinfo {author} {\bibnamefont {Tony}},\
  }\href {\doibase 10.5281/zenodo.4470001} {\enquote {\bibinfo {title} {cpnest:
  v0.11.3},}\ } (\bibinfo {year} {2021})\BibitemShut {NoStop}%
\bibitem [{\citenamefont {{Vousden}}\ \emph {et~al.}(2016)\citenamefont
  {{Vousden}}, \citenamefont {{Farr}},\ and\ \citenamefont
  {{Mandel}}}]{Vousden:2016aaa}%
  \BibitemOpen
  \bibfield  {author} {\bibinfo {author} {\bibfnamefont {W.~D.}\ \bibnamefont
  {{Vousden}}}, \bibinfo {author} {\bibfnamefont {W.~M.}\ \bibnamefont
  {{Farr}}}, \ and\ \bibinfo {author} {\bibfnamefont {I.}~\bibnamefont
  {{Mandel}}},\ }\href {\doibase 10.1093/mnras/stv2422} {\bibfield  {journal}
  {\bibinfo  {journal} {Mon. Not. Roy. Astron. Soc.}\ }\textbf {\bibinfo
  {volume} {455}},\ \bibinfo {pages} {1919} (\bibinfo {year} {2016})},\ \Eprint
  {http://arxiv.org/abs/1501.05823} {arXiv:1501.05823} \BibitemShut {NoStop}%
\bibitem [{\citenamefont {{Hermans}}\ \emph {et~al.}(2021)\citenamefont
  {{Hermans}}, \citenamefont {{Delaunoy}}, \citenamefont {{Rozet}},
  \citenamefont {{Wehenkel}}, \citenamefont {{Begy}},\ and\ \citenamefont
  {{Louppe}}}]{Hermans:2021aaa}%
  \BibitemOpen
  \bibfield  {author} {\bibinfo {author} {\bibfnamefont {J.}~\bibnamefont
  {{Hermans}}}, \bibinfo {author} {\bibfnamefont {A.}~\bibnamefont
  {{Delaunoy}}}, \bibinfo {author} {\bibfnamefont {F.}~\bibnamefont {{Rozet}}},
  \bibinfo {author} {\bibfnamefont {A.}~\bibnamefont {{Wehenkel}}}, \bibinfo
  {author} {\bibfnamefont {V.}~\bibnamefont {{Begy}}}, \ and\ \bibinfo {author}
  {\bibfnamefont {G.}~\bibnamefont {{Louppe}}},\ }\href@noop {} {\  (\bibinfo
  {year} {2021})},\ \Eprint {http://arxiv.org/abs/2110.06581} {arXiv:2110.06581
  [stat.ML]} \BibitemShut {NoStop}%
\bibitem [{\citenamefont {{Lemos}}\ \emph {et~al.}(2023)\citenamefont
  {{Lemos}}, \citenamefont {{Coogan}}, \citenamefont {{Hezaveh}},\ and\
  \citenamefont {{Perreault-Levasseur}}}]{Lemos:2023aaa}%
  \BibitemOpen
  \bibfield  {author} {\bibinfo {author} {\bibfnamefont {P.}~\bibnamefont
  {{Lemos}}}, \bibinfo {author} {\bibfnamefont {A.}~\bibnamefont {{Coogan}}},
  \bibinfo {author} {\bibfnamefont {Y.}~\bibnamefont {{Hezaveh}}}, \ and\
  \bibinfo {author} {\bibfnamefont {L.}~\bibnamefont {{Perreault-Levasseur}}},\
  }\href@noop {} {\  (\bibinfo {year} {2023})},\ \Eprint
  {http://arxiv.org/abs/2302.03026} {arXiv:2302.03026 [stat.ML]} \BibitemShut
  {NoStop}%
\bibitem [{\citenamefont {Wong}\ \emph {et~al.}(2023)\citenamefont {Wong},
  \citenamefont {Isi},\ and\ \citenamefont {Edwards}}]{Wong:2023lgb}%
  \BibitemOpen
  \bibfield  {author} {\bibinfo {author} {\bibfnamefont {K.~W.~K.}\
  \bibnamefont {Wong}}, \bibinfo {author} {\bibfnamefont {M.}~\bibnamefont
  {Isi}}, \ and\ \bibinfo {author} {\bibfnamefont {T.~D.~P.}\ \bibnamefont
  {Edwards}},\ }\href@noop {} {\  (\bibinfo {year} {2023})},\ \Eprint
  {http://arxiv.org/abs/2302.05333} {arXiv:2302.05333 [astro-ph.IM]}
  \BibitemShut {NoStop}%
\bibitem [{\citenamefont {{Cannon}}\ \emph {et~al.}(2022)\citenamefont
  {{Cannon}}, \citenamefont {{Ward}},\ and\ \citenamefont
  {{Schmon}}}]{Cannon:2022aaa}%
  \BibitemOpen
  \bibfield  {author} {\bibinfo {author} {\bibfnamefont {P.}~\bibnamefont
  {{Cannon}}}, \bibinfo {author} {\bibfnamefont {D.}~\bibnamefont {{Ward}}}, \
  and\ \bibinfo {author} {\bibfnamefont {S.~M.}\ \bibnamefont {{Schmon}}},\
  }\href@noop {} {\  (\bibinfo {year} {2022})},\ \Eprint
  {http://arxiv.org/abs/2209.01845} {arXiv:2209.01845 [stat.ML]} \BibitemShut
  {NoStop}%
\bibitem [{\citenamefont {Maggiore}\ \emph {et~al.}(2020)\citenamefont
  {Maggiore} \emph {et~al.}}]{Maggiore:2019uih}%
  \BibitemOpen
  \bibfield  {author} {\bibinfo {author} {\bibfnamefont {M.}~\bibnamefont
  {Maggiore}} \emph {et~al.},\ }\href {\doibase 10.1088/1475-7516/2020/03/050}
  {\bibfield  {journal} {\bibinfo  {journal} {JCAP}\ }\textbf {\bibinfo
  {volume} {03}},\ \bibinfo {pages} {050} (\bibinfo {year} {2020})},\ \Eprint
  {http://arxiv.org/abs/1912.02622} {arXiv:1912.02622 [astro-ph.CO]}
  \BibitemShut {NoStop}%
\bibitem [{\citenamefont {Reitze}\ \emph {et~al.}(2019)\citenamefont {Reitze}
  \emph {et~al.}}]{Reitze:2019iox}%
  \BibitemOpen
  \bibfield  {author} {\bibinfo {author} {\bibfnamefont {D.}~\bibnamefont
  {Reitze}} \emph {et~al.},\ }\href@noop {} {\bibfield  {journal} {\bibinfo
  {journal} {Bull. Am. Astron. Soc.}\ }\textbf {\bibinfo {volume} {51}},\
  \bibinfo {pages} {035} (\bibinfo {year} {2019})},\ \Eprint
  {http://arxiv.org/abs/1907.04833} {arXiv:1907.04833 [astro-ph.IM]}
  \BibitemShut {NoStop}%
\bibitem [{\citenamefont {Amaro-Seoane}\ \emph {et~al.}(2017)\citenamefont
  {Amaro-Seoane}, \citenamefont {Audley}, \citenamefont {Babak}, \citenamefont
  {Baker}, \citenamefont {Barausse}, \citenamefont {Bender}, \citenamefont
  {Berti} \emph {et~al.}}]{LISA:2017aaa}%
  \BibitemOpen
  \bibfield  {author} {\bibinfo {author} {\bibfnamefont {P.}~\bibnamefont
  {Amaro-Seoane}}, \bibinfo {author} {\bibfnamefont {H.}~\bibnamefont
  {Audley}}, \bibinfo {author} {\bibfnamefont {S.}~\bibnamefont {Babak}},
  \bibinfo {author} {\bibfnamefont {J.}~\bibnamefont {Baker}}, \bibinfo
  {author} {\bibfnamefont {E.}~\bibnamefont {Barausse}}, \bibinfo {author}
  {\bibfnamefont {P.}~\bibnamefont {Bender}}, \bibinfo {author} {\bibfnamefont
  {E.}~\bibnamefont {Berti}},  \emph {et~al.},\ }\href@noop {} {\  (\bibinfo
  {year} {2017})},\ \Eprint {http://arxiv.org/abs/1702.00786} {arXiv:1702.00786
  [astro-ph.IM]} \BibitemShut {NoStop}%
\bibitem [{\citenamefont {Gabbard}\ \emph {et~al.}(2022)\citenamefont
  {Gabbard}, \citenamefont {Messenger}, \citenamefont {Heng}, \citenamefont
  {Tonolini},\ and\ \citenamefont {Murray-Smith}}]{Gabbard:2019rde}%
  \BibitemOpen
  \bibfield  {author} {\bibinfo {author} {\bibfnamefont {H.}~\bibnamefont
  {Gabbard}}, \bibinfo {author} {\bibfnamefont {C.}~\bibnamefont {Messenger}},
  \bibinfo {author} {\bibfnamefont {I.~S.}\ \bibnamefont {Heng}}, \bibinfo
  {author} {\bibfnamefont {F.}~\bibnamefont {Tonolini}}, \ and\ \bibinfo
  {author} {\bibfnamefont {R.}~\bibnamefont {Murray-Smith}},\ }\href {\doibase
  10.1038/s41567-021-01425-7} {\bibfield  {journal} {\bibinfo  {journal}
  {Nature Phys.}\ }\textbf {\bibinfo {volume} {18}},\ \bibinfo {pages} {112}
  (\bibinfo {year} {2022})},\ \Eprint {http://arxiv.org/abs/1909.06296}
  {arXiv:1909.06296 [astro-ph.IM]} \BibitemShut {NoStop}%
\bibitem [{\citenamefont {Green}\ and\ \citenamefont
  {Gair}(2021)}]{Green:2020dnx}%
  \BibitemOpen
  \bibfield  {author} {\bibinfo {author} {\bibfnamefont {S.~R.}\ \bibnamefont
  {Green}}\ and\ \bibinfo {author} {\bibfnamefont {J.}~\bibnamefont {Gair}},\
  }\href {\doibase 10.1088/2632-2153/abfaed} {\bibfield  {journal} {\bibinfo
  {journal} {Mach. Learn. Sci. Tech.}\ }\textbf {\bibinfo {volume} {2}},\
  \bibinfo {pages} {03LT01} (\bibinfo {year} {2021})},\ \Eprint
  {http://arxiv.org/abs/2008.03312} {arXiv:2008.03312 [astro-ph.IM]}
  \BibitemShut {NoStop}%
\bibitem [{\citenamefont {Chua}\ and\ \citenamefont
  {Vallisneri}(2020)}]{Chua:2019wwt}%
  \BibitemOpen
  \bibfield  {author} {\bibinfo {author} {\bibfnamefont {A.~J.~K.}\
  \bibnamefont {Chua}}\ and\ \bibinfo {author} {\bibfnamefont {M.}~\bibnamefont
  {Vallisneri}},\ }\href {\doibase 10.1103/PhysRevLett.124.041102} {\bibfield
  {journal} {\bibinfo  {journal} {Phys. Rev. Lett.}\ }\textbf {\bibinfo
  {volume} {124}},\ \bibinfo {pages} {041102} (\bibinfo {year} {2020})},\
  \Eprint {http://arxiv.org/abs/1909.05966} {arXiv:1909.05966 [gr-qc]}
  \BibitemShut {NoStop}%
\bibitem [{\citenamefont {\'Alvares}\ \emph {et~al.}(2020)\citenamefont
  {\'Alvares}, \citenamefont {Font}, \citenamefont {Freitas}, \citenamefont
  {Freitas}, \citenamefont {Morais}, \citenamefont {Nunes}, \citenamefont
  {Onofre},\ and\ \citenamefont {Torres-Forn\'e}}]{Alvares:2020bjg}%
  \BibitemOpen
  \bibfield  {author} {\bibinfo {author} {\bibfnamefont {J.~a.~D.}\
  \bibnamefont {\'Alvares}}, \bibinfo {author} {\bibfnamefont {J.~A.}\
  \bibnamefont {Font}}, \bibinfo {author} {\bibfnamefont {F.~F.}\ \bibnamefont
  {Freitas}}, \bibinfo {author} {\bibfnamefont {O.~G.}\ \bibnamefont
  {Freitas}}, \bibinfo {author} {\bibfnamefont {A.~P.}\ \bibnamefont {Morais}},
  \bibinfo {author} {\bibfnamefont {S.}~\bibnamefont {Nunes}}, \bibinfo
  {author} {\bibfnamefont {A.}~\bibnamefont {Onofre}}, \ and\ \bibinfo {author}
  {\bibfnamefont {A.}~\bibnamefont {Torres-Forn\'e}}\ }(\bibinfo {year}
  {2020})\ \Eprint {http://arxiv.org/abs/2011.10425} {arXiv:2011.10425 [gr-qc]}
  \BibitemShut {NoStop}%
\bibitem [{\citenamefont {Zackay}\ \emph {et~al.}(2018)\citenamefont {Zackay},
  \citenamefont {Dai},\ and\ \citenamefont {Venumadhav}}]{Zackay:2018qdy}%
  \BibitemOpen
  \bibfield  {author} {\bibinfo {author} {\bibfnamefont {B.}~\bibnamefont
  {Zackay}}, \bibinfo {author} {\bibfnamefont {L.}~\bibnamefont {Dai}}, \ and\
  \bibinfo {author} {\bibfnamefont {T.}~\bibnamefont {Venumadhav}},\
  }\href@noop {} {\  (\bibinfo {year} {2018})},\ \Eprint
  {http://arxiv.org/abs/1806.08792} {arXiv:1806.08792 [astro-ph.IM]}
  \BibitemShut {NoStop}%
\bibitem [{\citenamefont {Leslie}\ \emph {et~al.}(2021)\citenamefont {Leslie},
  \citenamefont {Dai},\ and\ \citenamefont {Pratten}}]{Leslie:2021ssu}%
  \BibitemOpen
  \bibfield  {author} {\bibinfo {author} {\bibfnamefont {N.}~\bibnamefont
  {Leslie}}, \bibinfo {author} {\bibfnamefont {L.}~\bibnamefont {Dai}}, \ and\
  \bibinfo {author} {\bibfnamefont {G.}~\bibnamefont {Pratten}},\ }\href
  {\doibase 10.1103/PhysRevD.104.123030} {\bibfield  {journal} {\bibinfo
  {journal} {Phys. Rev. D}\ }\textbf {\bibinfo {volume} {104}},\ \bibinfo
  {pages} {123030} (\bibinfo {year} {2021})},\ \Eprint
  {http://arxiv.org/abs/2109.09872} {arXiv:2109.09872 [astro-ph.IM]}
  \BibitemShut {NoStop}%
\bibitem [{\citenamefont {{Buscicchio}}\ \emph {et~al.}(2019)\citenamefont
  {{Buscicchio}}, \citenamefont {{Roebber}}, \citenamefont {{Goldstein}},\ and\
  \citenamefont {{Moore}}}]{Buscicchio:2019aaa}%
  \BibitemOpen
  \bibfield  {author} {\bibinfo {author} {\bibfnamefont {R.}~\bibnamefont
  {{Buscicchio}}}, \bibinfo {author} {\bibfnamefont {E.}~\bibnamefont
  {{Roebber}}}, \bibinfo {author} {\bibfnamefont {J.~M.}\ \bibnamefont
  {{Goldstein}}}, \ and\ \bibinfo {author} {\bibfnamefont {C.~J.}\ \bibnamefont
  {{Moore}}},\ }\href {\doibase 10.1103/PhysRevD.100.084041} {\bibfield
  {journal} {\bibinfo  {journal} {Phys. Rev. D}\ }\textbf {\bibinfo {volume}
  {100}},\ \bibinfo {eid} {084041} (\bibinfo {year} {2019})},\ \Eprint
  {http://arxiv.org/abs/1907.11631} {arXiv:1907.11631 [astro-ph.IM]}
  \BibitemShut {NoStop}%
\bibitem [{\citenamefont {{Paszke}}\ \emph {et~al.}(2019)\citenamefont
  {{Paszke}}, \citenamefont {{Gross}}, \citenamefont {{Massa}} \emph
  {et~al.}}]{Paszke:2019aaa}%
  \BibitemOpen
  \bibfield  {author} {\bibinfo {author} {\bibfnamefont {A.}~\bibnamefont
  {{Paszke}}}, \bibinfo {author} {\bibfnamefont {S.}~\bibnamefont {{Gross}}},
  \bibinfo {author} {\bibfnamefont {F.}~\bibnamefont {{Massa}}},  \emph
  {et~al.},\ }\href@noop {} {\  (\bibinfo {year} {2019})},\ \Eprint
  {http://arxiv.org/abs/1912.01703} {arXiv:1912.01703 [cs.LG]} \BibitemShut
  {NoStop}%
\end{thebibliography}%

\appendix
\section{Network Architecture}\label{app:network}
\noindent The network architecture used for the analysis in \texttt{peregrine} is shown in the figure below. As described in the text in Sec.~\ref{sec:sbi}, we use the flexibility of the TMNRE implementation to automatically extract a small set of summary statistics from both the time and frequency domain strains after processing and compressing the input data. This summary can then be passed to the ratio estimator along with the physical parameters $\boldsymbol{\theta}_\mathrm{GW}$ to perform the inference. The full \texttt{pytorch}~\cite{Paszke:2019aaa} implementation of the network components shown below (specifically the \texttt{unet} and linear compression networks) can be found in the \texttt{InferenceNetwork} class of \texttt{peregrine}.
\begin{figure}[h]
    \centering
    \includegraphics[width=\linewidth,trim={0.1cm 0.1cm 0.1cm 0.1cm},clip]{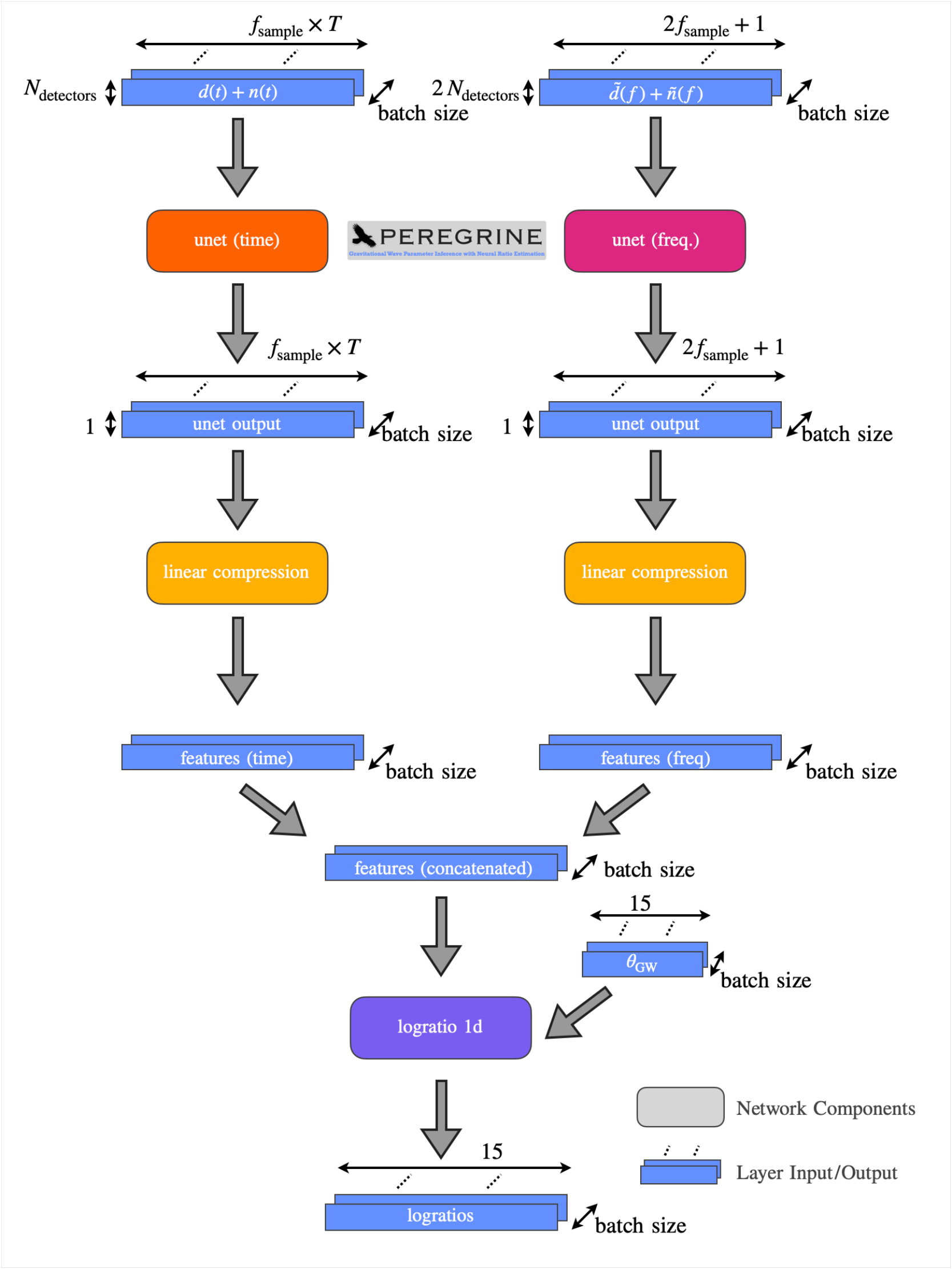}
    \label{fig:net}
\end{figure}

\newpage

\section{Comparison with Likelihood Based Samplers}\label{app:JSD}
\noindent In order to validate our results in Sec.~\ref{sec:results}, we quantitatively compare the \texttt{peregrine} posteriors against those obtained using traditional sampling methods such as nested sampling and MCMC. Specifically, we compute the Jensen-Shannon divergence amongst the 1d posteriors obtained using \texttt{peregrine}, \texttt{dynesty}, \texttt{ptemcee}, and \texttt{cpnest}. Our implementation including all the data relevant to this article are available publicly on our \texttt{Zenodo} page (\href{https://zenodo.org/records/7788596}{link}), along with files that can regenerate the JSD values (both on average and on a parameter-by-parameter basis) given below in Tab.~\ref{tab:JSD}.

\begingroup
\begin{table}[h]
\centering
\setlength{\tabcolsep}{-0.1pt}

\resizebox{\linewidth}{!}{
\begin{tabular}{l | ccccc}\hline
$\mathbf{JSD}\,\mathrm{[}10^{-3}\,\mathrm{nat]}$ & \texttt{peregrine\;\;}           & \texttt{dynesty\;\;}               & \texttt{dynesty} (re-run)\;\;      & \texttt{ptemcee\;\;}               & \texttt{cpnest  }                 \\ \hline
\texttt{peregrine}                        &         &\cellcolor[HTML]{FDAE6B}$18.0$ &\cellcolor[HTML]{FDAE6B}$13.1$ &\cellcolor[HTML]{FDAE6B}$21.0$       &\cellcolor[HTML]{FDAE6B}$20.6$ \\\noalign{\vskip-0.1pt}
\texttt{dynesty}                          &\cellcolor[HTML]{E772AB}$9.75$ &        &\cellcolor[HTML]{FDAE6B}$4.02$ &   \cellcolor[HTML]{FDAE6B}$10.0$     &\cellcolor[HTML]{FDAE6B}$35.2$ \\\noalign{\vskip-0.1pt}
\texttt{dynesty} (re-run)                 &\cellcolor[HTML]{E772AB} $10.9$ &\cellcolor[HTML]{E772AB} $0.55$ &        &  \cellcolor[HTML]{FDAE6B}$10.6$      & \cellcolor[HTML]{FDAE6B}$29.9$ \\\noalign{\vskip-0.1pt}
\texttt{ptemcee}                          &\cellcolor[HTML]{E772AB} $16.5$ &\cellcolor[HTML]{E772AB} $7.01$ &\cellcolor[HTML]{E772AB} $7.56$ &       &\cellcolor[HTML]{FDAE6B}$35.9$        \\\noalign{\vskip-0.1pt}
\texttt{cpnest}                           &\cellcolor[HTML]{E772AB} $6.82$ &\cellcolor[HTML]{E772AB} $3.52$ &\cellcolor[HTML]{E772AB} $4.46$ &\cellcolor[HTML]{E772AB} $8.60$ &  \\\noalign{\vskip-0.1pt}\hline     
\end{tabular}}
\caption{Average Jensen-Shannon divergences over all physical parameters $\boldsymbol{\theta}_\mathrm{GW} = (q, \mathcal{M}, \ldots)$ for different sampler combinations. Values below the diagonal (in pink boxes) refer to the low SNR case study (\textbf{C1}), whilst those above the diagonal (in orange boxes) are for the high SNR case (\textbf{C2}).} 
\label{tab:JSD}
\end{table}
\endgroup

\clearpage

\newpage

\begin{figure*}[t]
    \centering
    \includegraphics[width=\linewidth,trim={0.1cm 0.1cm 0.1cm 0.1cm},clip]{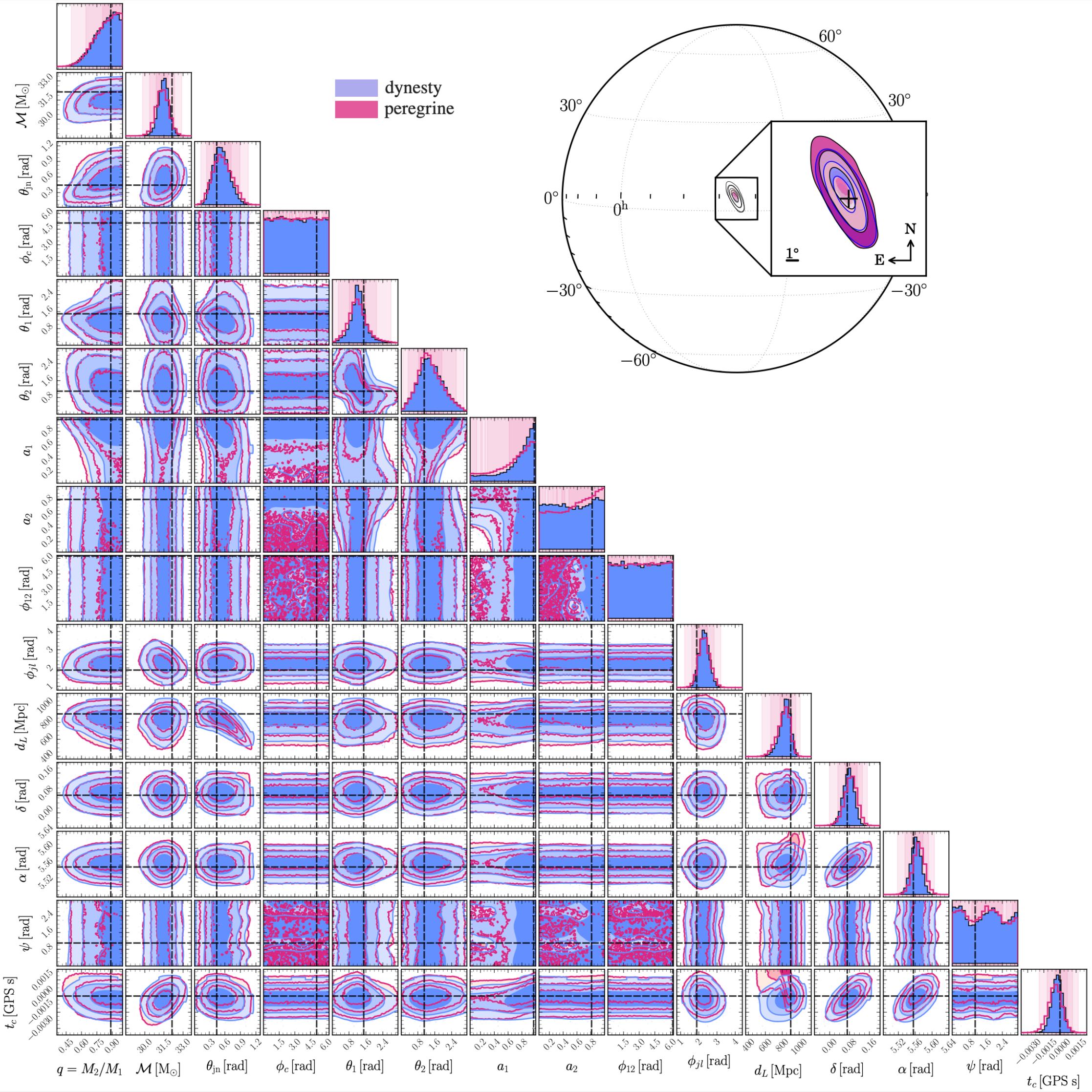}
    \caption{Full set of 2d marginal posteriors in the low SNR case study (\textbf{C1}). The results from the TMNRE analysis are shown in pink, whilst the corresponding \texttt{dynesty} analysis is shown in blue. The darker and subsequently lighter contours in the 2d marginals indicate the $1\sigma$, $2\sigma$ and $3\sigma$ confidence intervals respectively. The sky map shows the ($\alpha$,$\delta$) contours centered at the injection value.}
    \label{fig:2d_lowsn}
\end{figure*}

\begin{figure*}[t]
    \centering
    \includegraphics[width=0.92\linewidth,trim={0.1cm 0.1cm 0.1cm 0.1cm},clip]{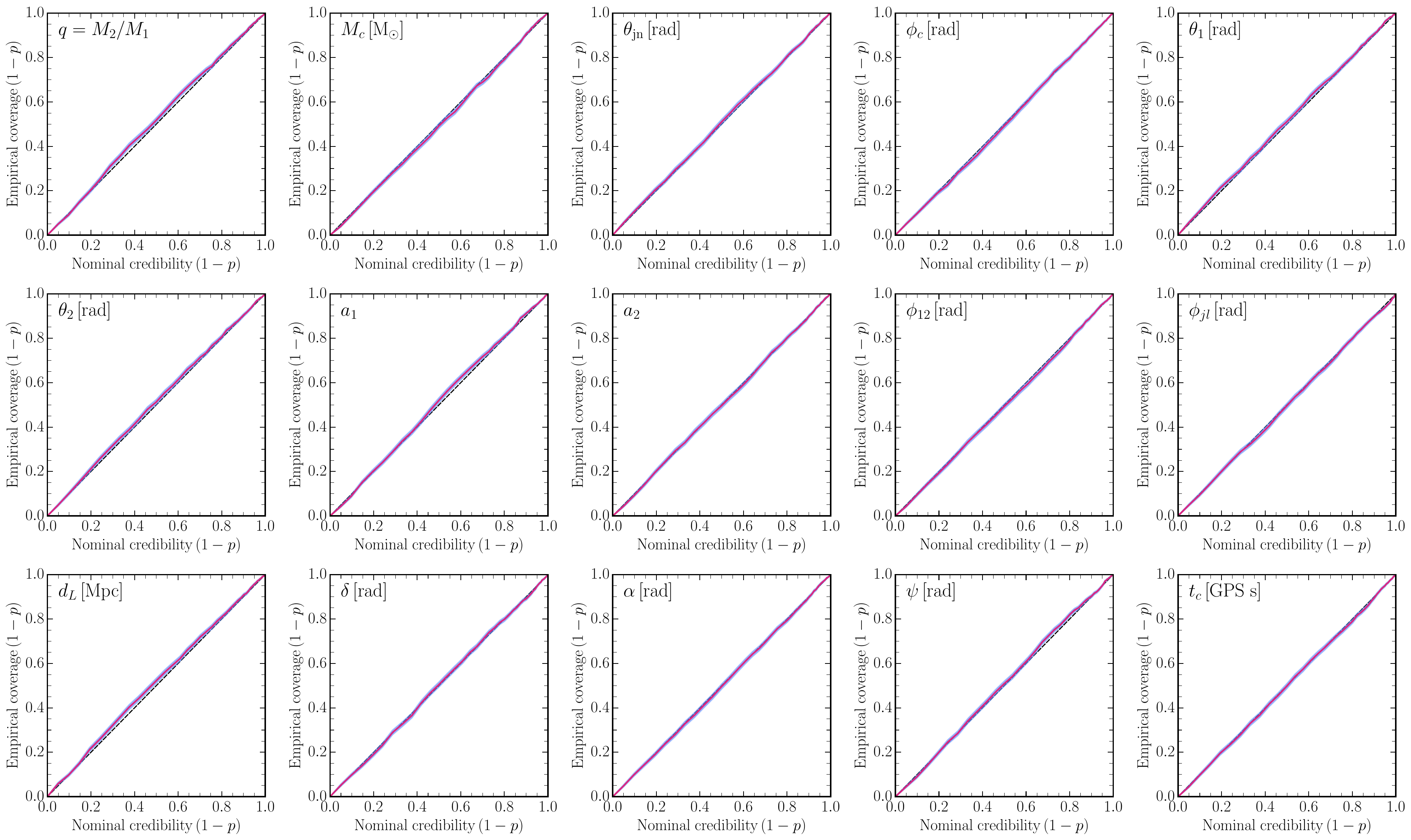}
    \caption{Coverage results for the low SNR case study (\textbf{C1}) for all parameters. This compares the expected coverage of the true value as a percentage on the $x$-axis against the actual coverage of our ratio estimator on the $y$-axis. The pink line indicates the average coverage, while the blue contour indicates the $68$\% confidence interval on the coverage.}
    \label{fig:pp_lowsn}
\end{figure*}

\begin{figure*}[t]
    \centering
    \includegraphics[width=0.92\linewidth,trim={0.1cm 0.1cm 0.1cm 0.1cm},clip]{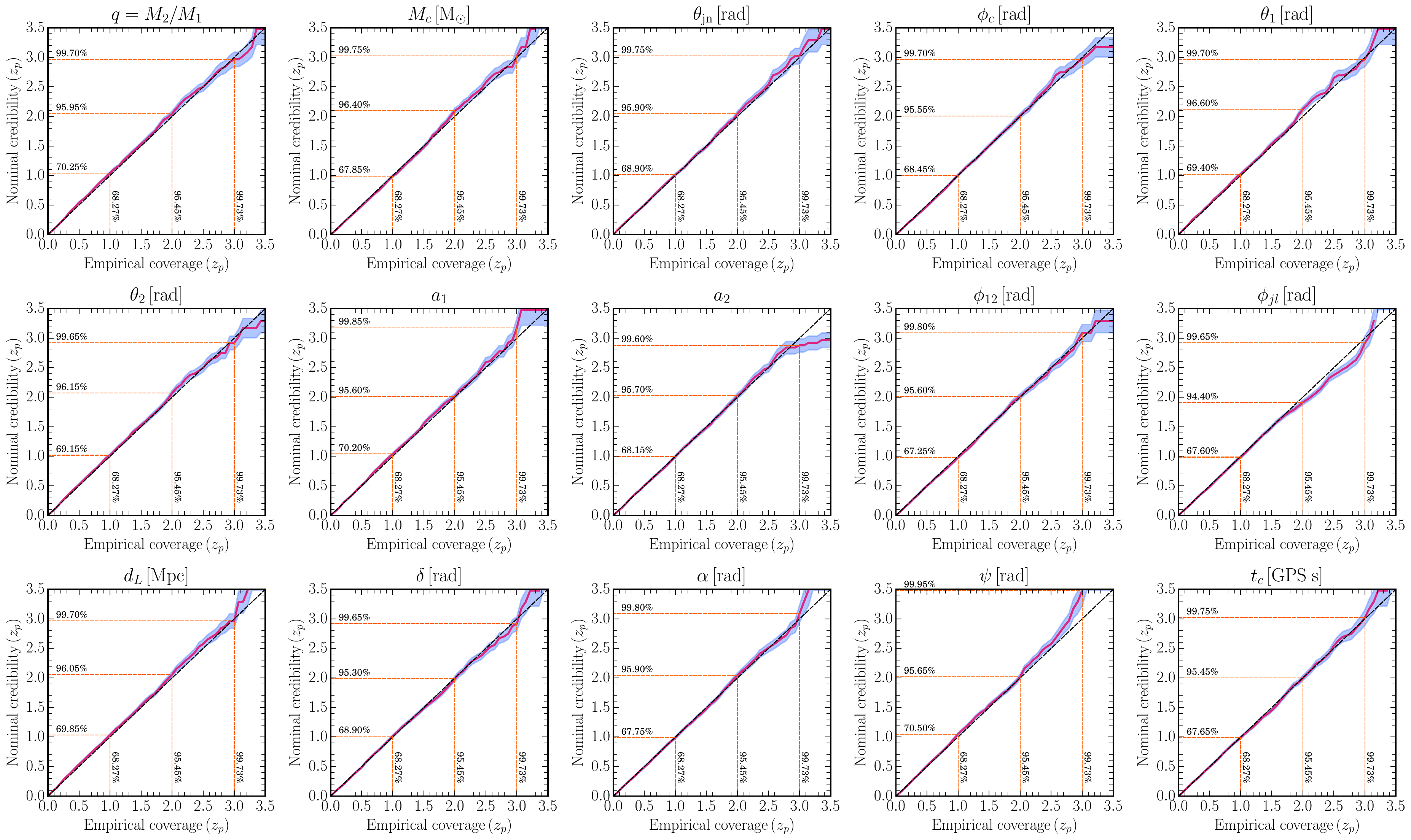}
    \caption{Coverage results for the low SNR case study (\textbf{C1}) for all parameters. This is the same information as Fig.~\ref{fig:pp_highsn}, but with $z_p$ defined by $p = \int_{-z_p}^{z_p}{\mathrm{d}z\,1/\sqrt{2\pi} \exp(-z^2 / 2)}$. This places more emphasis on the behaviour of the posteriors in the tail regions. The pink line indicates the average coverage, while the blue contour indicates the $68$\% confidence interval on the coverage.}
    \label{fig:zz_lowsn}
\end{figure*}

\begin{figure*}[t]
    \centering
    \includegraphics[width=\linewidth,trim={0.1cm 0.1cm 0.1cm 0.1cm},clip]{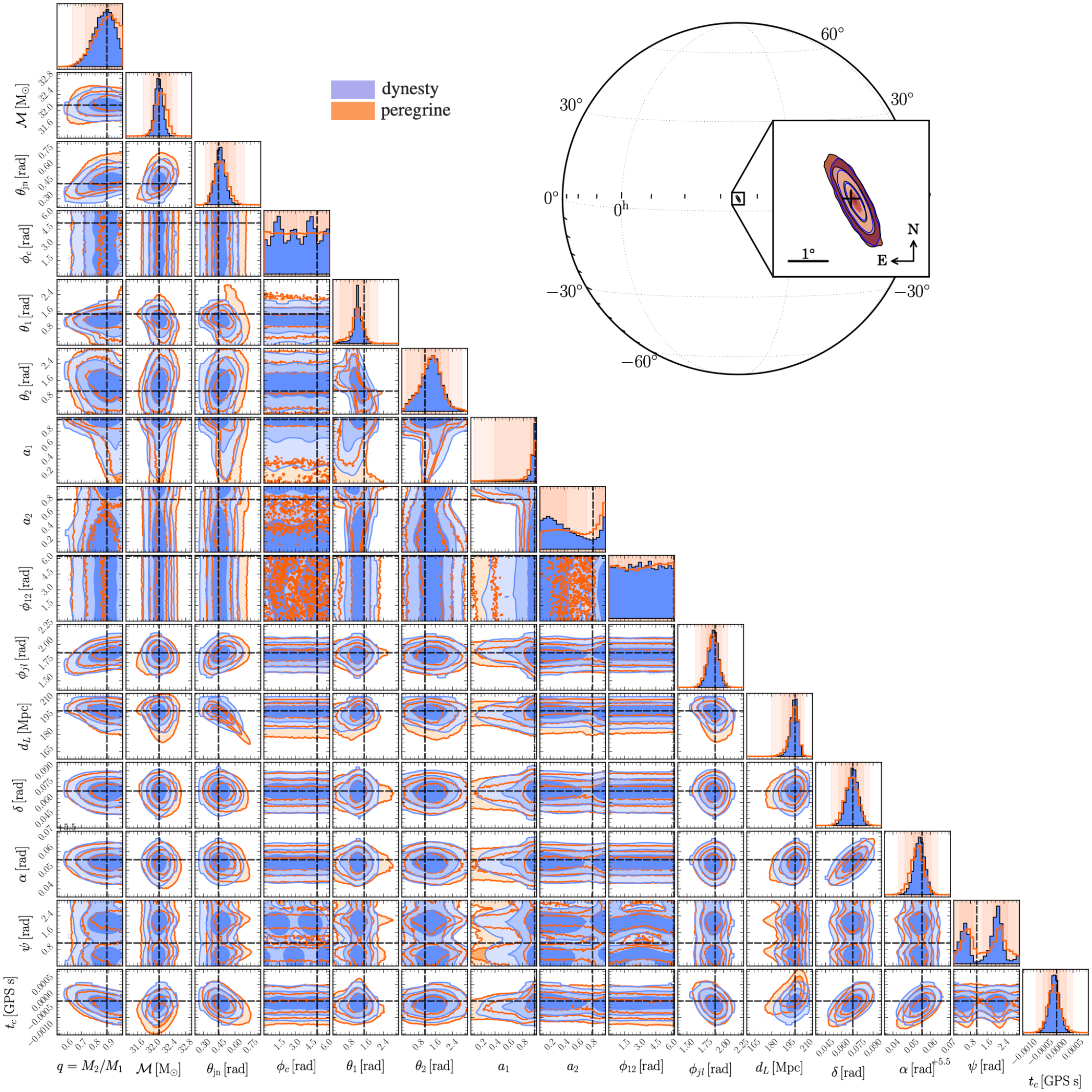}
    \caption{Full set of 2d marginal posteriors in the high SNR case study (\textbf{C2}). The results from the TMNRE analysis are shown in orange, whilst the corresponding \texttt{dynesty} analysis is shown in blue. The darker and subsequently lighter contours in the 2d marginals indicate the $1\sigma$, $2\sigma$ and $3\sigma$ confidence intervals respectively. The sky map shows the ($\alpha$,$\delta$) contours centered at the injection value.}
    \label{fig:2d_highsn}
\end{figure*}

\begin{figure*}[t]
    \centering
    \includegraphics[width=0.92\linewidth,trim={0.1cm 0.1cm 0.1cm 0.1cm},clip]{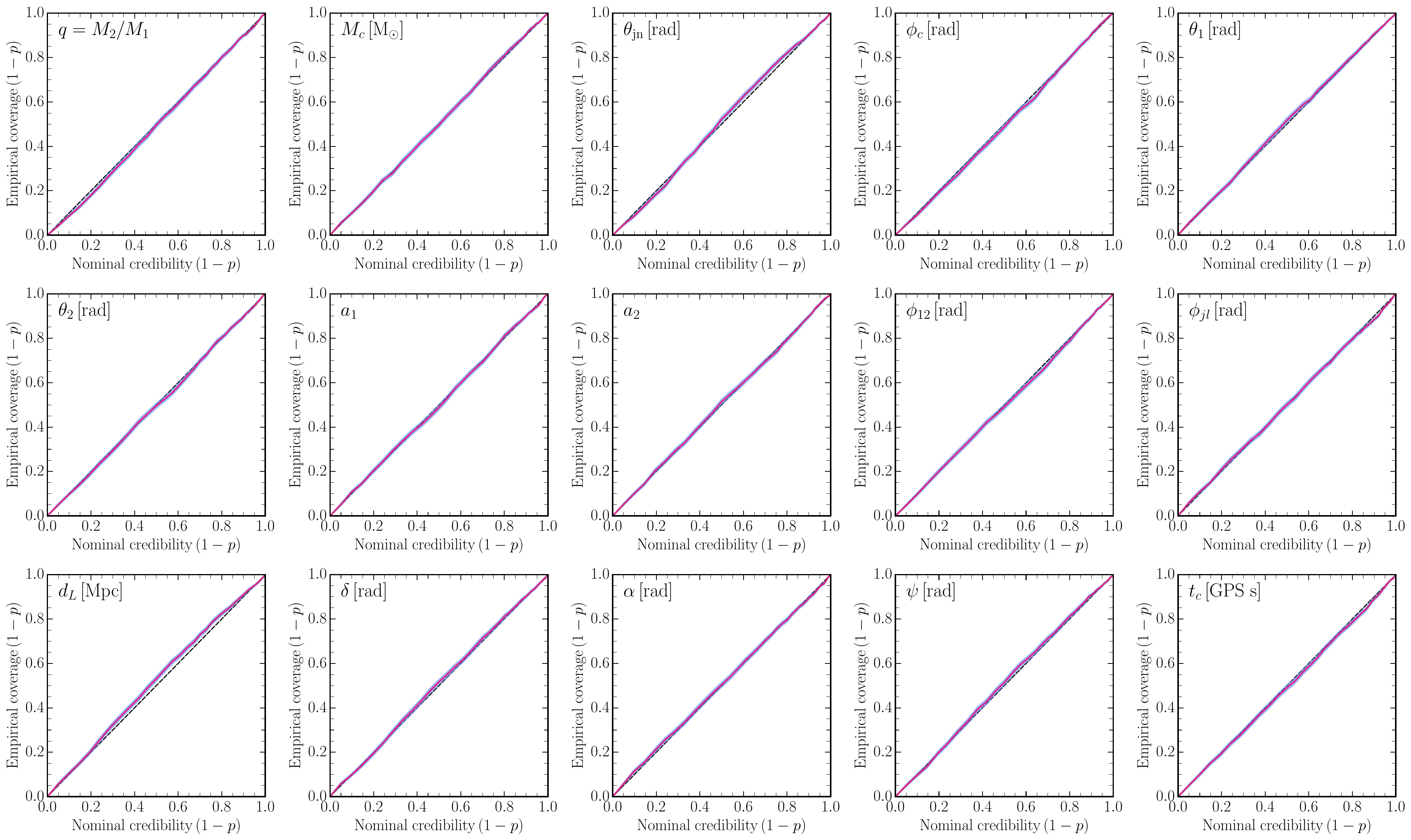}
    \caption{Coverage results for the high SNR case study (\textbf{C2}) for all parameters. This compares the expected coverage of the true value as a percentage on the $x$-axis against the actual coverage of our ratio estimator on the $y$-axis. The pink line indicates the average coverage, while the blue contour indicates the $68$\% confidence interval on the coverage.}
    \label{fig:pp_highsn}
\end{figure*}

\begin{figure*}[t]
    \centering
    \includegraphics[width=0.92\linewidth,trim={0.1cm 0.1cm 0.1cm 0.1cm},clip]{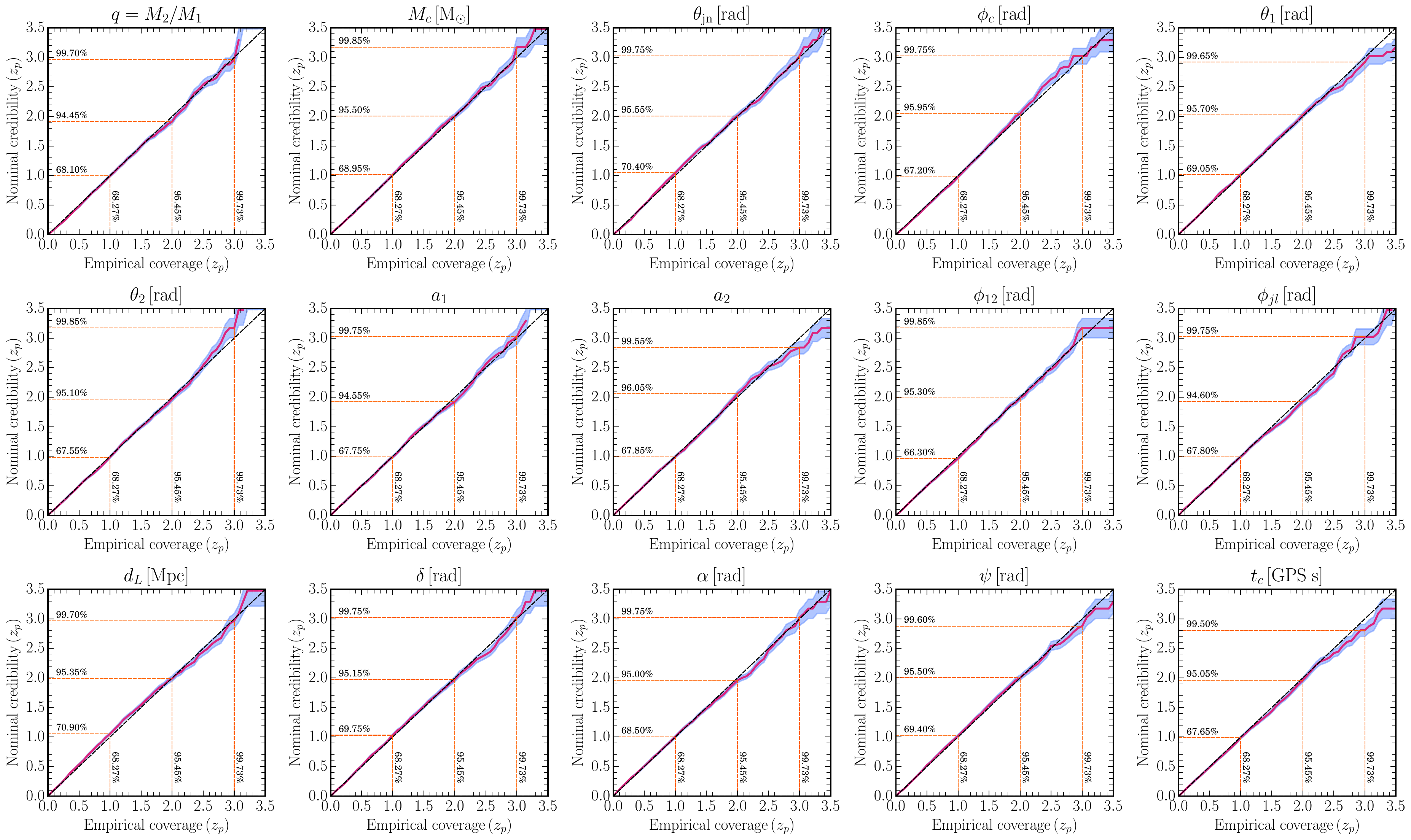}
    \caption{Coverage results for the high SNR case study (\textbf{C2}) for all parameters. This is the same information as Fig.~\ref{fig:pp_highsn}, but with $z_p$ defined by $p = \int_{-z_p}^{z_p}{\mathrm{d}z\,1/\sqrt{2\pi} \exp(-z^2 / 2)}$. This places more emphasis on the behaviour of the posteriors in the tail regions. The pink line indicates the average coverage, while the blue contour indicates the $68$\% confidence interval on the coverage.}
    \label{fig:zz_highsn}
\end{figure*}

\end{document}